\UseRawInputEncoding
\documentclass[aps,prc,amsfonts,preprintnumbers,superscriptaddress,showpacs,nofootinbib,10pt]{revtex4-1}
\usepackage{epsfig}
\usepackage{graphics}
\usepackage{amsmath}
\usepackage{amssymb}
\usepackage{mathtools}
\usepackage{dsfont}
\usepackage{bm}
\usepackage[colorlinks=true,linkcolor=blue,urlcolor=blue,citecolor=blue]{hyperref}

\begin{document}
\title{Renormalization of nuclear chiral effective field theory 
with non-perturbative leading order interactions}

\author{A.~M.~Gasparyan}
\email[]{Email: ashot.gasparyan@rub.de}
\affiliation{Ruhr-Universit\"at Bochum, Fakult\"at f\"ur Physik und
        Astronomie, Institut f\"ur Theoretische Physik II,  D-44780
        Bochum, Germany}
      
\author{E.~Epelbaum}
\email[]{Email: evgeny.epelbaum@rub.de}
\affiliation{Ruhr-Universit\"at Bochum, Fakult\"at f\"ur Physik und
        Astronomie, Institut f\"ur Theoretische Physik II,  D-44780
        Bochum, Germany}

\begin{abstract}
We extend the renormalizability study of the formulation of
chiral effective field theory with a finite cutoff,
applied to nucleon-nucleon scattering, by taking into account
non-perturbative effects.
We consider the nucleon-nucleon interaction up to next-to-leading
order in the chiral expansion.
The leading-order interaction is treated non-perturbatively.
In contrast to the previously considered case when the 
leading-order interaction was assumed to be perturbative, 
new features related to the renormalization of the effective
field theory are revealed.
In particular, more severe constraints on the leading-order potential
are formulated, which can enforce the renormalizability and the correct power counting
for the next-to-leading order amplitude.
To illustrate our theoretical findings, several partial waves in the nucleon-nucleon
scattering, $^3P_0$, $^3S_1-{^3D_1}$ and $^1S_0$
are analyzed numerically.
The cutoff dependence and the convergence of the chiral expansion
for those channels are discussed.
\end{abstract}

\maketitle

\section{Introduction}
Over the last decades, the effective field theory (EFT) approach
has become a standard tool
in studies of the nucleon-nucleon (NN), few-nucleon and many-nucleon systems
due to the possibility to perform
systematically improvable calculations in accordance with the
chiral power counting. 
The chiral power counting implies an expansion of observables in terms of the 
ratio of the soft and the hard scales $Q=q/\Lambda_b$.
The soft scale is given by the pion mass $M_\pi$ and the
external particle 3-momenta $|\vec p\,|$, whereas the hard scale $\Lambda_b$
is the breakdown scale of the EFT expansion of the order of
the $\rho$-meson mass.

Starting with the seminal work by Weinberg~\cite{Weinberg:1990rz,Weinberg:1991um},
a lot of progress has been achieved in this field, see Refs.~\cite{Bedaque:2002mn,Epelbaum:2008ga,Machleidt:2011zz,
Epelbaum:2012vx,Epelbaum:2019kcf,Hammer:2019poc} for reviews. 

In realistic calculations, one has to deal with regularization of an infinite
number of divergent Feynman diagrams originating from the field theoretic treatment
of non-perturbative amplitudes. 
One of the most practical approaches is related to introducing a finite 
(of the order of the hard scale $\Lambda_b$) cutoff $\Lambda$ in momentum space
(or a corresponding short distance cutoff in coordinate space).
The success of such a scheme is reflected in very accurate calculations 
at high orders in the chiral expansion, see Refs.~\cite{Reinert:2017usi,Entem:2017gor,Reinert:2020mcu}
for recent applications.

A justification of such an approach from the fundamental point of view
is complicated by the issue of renormalization and power counting violation
due to the appearance of positive powers of the cutoff in the amplitude.
Such contributions are generated by loop momenta of the order of the cutoff $\Lambda$.
There exists a qualitative understanding in the literature~\cite{Lepage:1997cs,Gegelia:1998iu,Gegelia:2004pz,Epelbaum:2006pt} 
that such positive powers of $\Lambda$ in the leading-order (LO) amplitude get compensated by the negative
powers of the scale $\Lambda_V$ stemming from the LO potential, which
is also regarded to be of the order of the hard scale $\Lambda_b$: $\Lambda_V\sim\Lambda\sim\Lambda_b$.
Further, one believes that at higher chiral orders, the power counting breaking terms
can be absorbed by a renormalization (shift) of lower order contact interactions ~\cite{Lepage:1997cs}.
However, until recently, a rigorous treatment of these problems and a systematic analysis
of conditions under which the renormalization program can be carried out 
has been missing. Such a rigorous treatment is extremely important 
within the EFT approach, where systematic power counting is utilized 
to estimate theoretical uncertainties.

We addressed this issues in our study in Ref.~\cite{Gasparyan:2021edy}.
In particular, we considered the LO potential consisting of the long-range one-pion-exchange
term and a set of contact interactions that are momentum-independent or quadratic in momenta.
The LO potential was regularized by various types of the form factors in momentum space,
including local and non-local regulators both power-like and Gaussian.
This covers most of the schemes considered in the literature.

In Ref.~\cite{Gasparyan:2021edy}, it was assumed that the 
iterations of the leading-order potential $V_0$ 
can be treated perturbatively. More precisely, the series
of the LO and the next-to-leading-order (NLO) amplitude 
in powers of $V_0$ were assumed to be convergent.
However, the convergence rate of the expansion in $V_0$ might
still be slower than the convergence rate of the chiral EFT expansion,
which makes it necessary to sum up all (or many) iterations of $V_0$.
On the other hand, the NLO potential needs not be iterated
in the NLO amplitude.
In the physical case of the NN scattering, such a perturbative 
regime is realized in most of the partial waves.
The prominent exceptions are the $^1S_0$, $^3S_1-{^3D_1}$ and
$^3P_0$ channels.

Under the above rather general assumptions, we proved
the following statements:
\begin{itemize}
 \item[-] The LO amplitude satisfies the dimensional power counting at each order in $V_0$ and is of chiral order $O(Q^0)$.
 If necessary, contact interactions quadratic in momenta can be promoted to leading order.
 \item[-] The NLO amplitude in $P$- and higher waves satisfies the dimensional power counting 
 at each order in $V_0$ and is of chiral order $O(Q^2)$.
  \item[-] The unrenormalized NLO amplitudes in the $S$-waves (including the $^3S_1-{^3D_1}$ channel)
 violate the power counting and are of order $O(Q^0)$. 
 To absorb the power-counting breaking terms, we employed
 the Bogoliubov-Parasiuk-Hepp-Zimmermann (BPHZ) renormalization procedure
and performed the overall subtractions in the diagrams as well as subtractions
 in all nested subdiagrams. As a result, the renormalized NLO amplitude
 was shown to satisfy the dimensional power counting and being of chiral order $O(Q^2)$
 up to corrections logarithmic in the cutoff at each order in $V_0$.
\end{itemize}

In the current work, we extend our analysis to the non-perturbative case, i.e. 
to the situation when the series in the LO potential $V_0$ do not converge
for the LO and/or NLO amplitude. This will
allow us to consider the above mentioned non-perturbative channels in NN scattering.
Our analysis is based on the application of the Fredholm method of solving integral equations,
which enables us to match the perturbative and non-perturbative regimes.

Our paper is organized as follows.
In Sec.~\ref{Sec:formalism}, we briefly  describe our formalism based on the effective 
Lagrangian, the corresponding effective potential and the way the
amplitude is constructed in the non-perturbative case.
In Sec.~\ref{Sec:LO_Fredholm}, we explain the application of the Fredholm method
for the LO Lippmann-Schwinger equation.
In Sec.~\ref{Sec:Pwaves}, we demonstrate the renormalization of the
nucleon-nucleon interaction in $P$-waves and higher.
The renormalization in the $S$-waves is addressed in Sec.~\ref{Sec:Subtractions}.
Numerical results that illustrate our formal considerations
are presented in Sec.~\ref{Sec:results}. 
The paper ends with a summary.
Bounds on the effective potential and various integrals are collected in Appendix.

\section{Formalism}
\label{Sec:formalism}
\subsection{Effective Lagrangian and potential}
In this section we briefly describe the formalism of chiral EFT
used in our analysis. Some details are omitted and can be found
in Ref.~\cite{Gasparyan:2021edy}.

The starting point is the effective chiral Lagrangian
represented as a series
of all possible terms
consistent with the symmetries of the underlying theory \cite{Weinberg:1978kz}.
The expansion of the Lagrangian is performed in terms of  the quark masses 
and field derivatives.
The effective Lagrangian contains
purely pionic terms, single nucleon terms, two-nucleon interactions,
etc.:
\begin{align}
\mathcal{L}_{\rm eff} = \mathcal{L}_{\pi}^{(2)} + 
\mathcal{L}_{\pi}^{(4)} + \mathcal{L}_{\pi N}^{(1)} +
\mathcal{L}_{\pi N}^{(2)} +
\mathcal{L}_{NN}^{(0)} +\mathcal{L}_{NN}^{(2)}+\dots \,, 
\label{Eq:effective_Lagrangian}
\end{align}
where the superscripts denote chiral orders.

The chiral expansion of the NN amplitude in terms of the small parameter $Q$
is performed according to 
the Weinberg power counting~\cite{Weinberg:1991um} 
(with possible modifications based on phenomenological arguments,
e.g. promotion of certain higher order contributions
to lower orders).
The power of $Q$ for a potential (i.e. two-nucleon-irreducible) contribution
is determined by a sum over all vertices $i$ in the diagram:
\begin{align}
 D=2L+\sum_{i}\left(d_i+\frac{n_i}2-2\right)\,,
 \label{Eq:Weinberg_PC}
\end{align}
where $L$ is the number of loops, $n_i$ is the number of nucleon lines 
at vertex $i$ and $d_i$ is the number of derivatives and pion-mass
insertions at vertex $i$. The chiral order of a 2N-reducible diagram
is equal to the sum of the orders of its irreducible components.

Since the LO contributions appear at order $O(Q^0)$,
the corresponding potential terms have to be iterated an infinite
number of times. To implement this procedure on a formal level
and to regularize multiple-loop integrals,
it is convenient to reformulate the effective Lagrangian
of two-nucleon interactions
in Eq.~\eqref{Eq:effective_Lagrangian}
in terms of the non-local regularized potential contributions of the form
(see Ref.~\cite{Gasparyan:2021edy} for details)
\begin{align}
\mathcal{L}_V(x)&=-\int  d\vec y \,d\vec y\,'
\frac{1}{2} N^{\dagger}_{j_1}(x_0,\vec x-\vec y\,'/2) 
N^{\dagger}_{j_2}(x_0,\vec x+\vec y\,'/2)
V(\vec y\,',\vec y)_{j_1,j_2;i_1,i_2}
N_{i_2}(x_0,\vec x+\vec y/2)N_{i_1}(x_0,\vec x-\vec y/2)\,,
\label{Eq:L_V}
\end{align}
where $i_1,i_2,j_1,j_2$ are the combined spin and isospin indices of the 
corresponding nucleons.
This formulation is customary for the few-body and nuclear physics.

The full potential is organized as a series according to the chiral expansion:
\begin{align}
 V=V^{(0)}+V^{(2)}+V^{(3)}+V^{(4)}+\dots.
\end{align}
Bare potentials $V^{(i)}$ are split into the renormalized parts $V_i$ and 
the counter terms $\delta V_i$:
\begin{align}
V^{(i)}=V_i+\delta V_i\,, 
\quad \delta V_i=\delta V_i^{(2)}+\delta V_i^{(3)}+\delta V_i^{(4)}+\dots.
 \label{Eq:counter_terms}
\end{align}
The counter terms $\delta V_i^{(j)}$ ($j>i$) absorb the
divergent and the power counting violating terms
appearing at order ${O}(Q^j)$. 

The LO potential $V_0$ is regulated 
(the details are given in Sec.~\ref{Sec:LO_NLO_potentials} and in Appendix~\ref{Sec:LOpotential})
using a cutoff $\Lambda$ to make the iterations of $V_0$ finite.
We regard the cutoff value
$\Lambda$ (the largest cutoff among all cutoffs used in the LO potential) 
to be of the order of the hard scale $\Lambda\sim\Lambda_b$.
Higher order potentials can be considered either regulated or unregulated depending on 
a particular scheme, which will be discussed in the subsequent sections.

Note that to make some intermediate expressions mathematically well defined,
one might need to introduce additional cutoffs
that drop out from the final results after performing certain
subtractions. Such cutoffs can be chosen to be much larger than $\Lambda$  (or even infinity large).

To make the formulation of the theory in terms of non-local (on the Lagrangian level)
regularized potential contributions completely equivalent to the
original formulation in terms of local interactions,
the regulator corrections $\delta_\Lambda V$ have to be 
taken into account:
\begin{align}
 \delta_\Lambda V=\sum_i\delta_\Lambda V^{(i)},\quad \delta_\Lambda V^{(i)}:=V^{(i)}_{\Lambda=\infty}-V^{(i)}_\Lambda,
\end{align}
where $V^{(i)}_{\Lambda=\infty}$ is the unregulated potential at the chiral order $i$.
One possibility, often implicitly used in practical calculations, is to
expand $\delta_\Lambda V$ in powers of $1/\Lambda$ and absorb the resulting terms
by higher order contact interactions.
This is possible if the potential does not contain
non-locally regularized long-range contributions.
Another approach suggested in Ref.~\cite{Gasparyan:2021edy} is
to keep the terms with $\delta_\Lambda V$ explicitly and consider those as perturbation.
This allows us to reduce the cutoff dependence and extend the range 
of possible values of $\Lambda$, especially to smaller ones.

\subsection{LO and NLO potentials and regulators}
\label{Sec:LO_NLO_potentials}
Our treatment of the LO and NLO potentials is identical to Ref.~\cite{Gasparyan:2021edy}.

Weinberg's power counting in Eq.~\eqref{Eq:Weinberg_PC} implies that the leading-order $O(Q^0)$ potential $V_0(\vec p\,',\vec p\,)$  
is represented by the sum of the regulated static
one-pion-exchange potential and the short-range part:
\begin{align}
 &V_0(\vec p\,',\vec p \, )=V^{(0)}_{1\pi,\Lambda}(\vec p\,',\vec p\,)+V^{(0)}_{\text{short},\Lambda}(\vec p\,',\vec p\,),
\end{align}
where the short-range part $V^{(0)}_{\text{short},\Lambda}$
may contain momentum-independent contact terms as well as the
contact terms quadratic in momentum.
The latter are formally of order $O(Q^2)$, as follows from Eq.~\eqref{Eq:Weinberg_PC}.
Nevertheless, it is known that in some channels, e.g., $^1S_0$ and $^3P_0$,
their promotion to leading order can be motivated by phenomenological arguments, see, e.g., 
Refs.~\cite{Epelbaum:2015sha,Long:2013cya,Nogga:2005hy,Birse:2007sx}.

For the sake of generality, we allow for different forms of regulators:
power-like local, power-like non-local, Gaussian local and Gaussian non-local regulators
as well as all possible combinations of those.
In Ref.~\cite{Gasparyan:2021edy}, we argued that for a local part of the LO potential $V_{0,\text{local}}(\vec q\,)$,
the regulator (if it is also local) can be rather ``mild''. 
If the regulated LO potential behaves as
\begin{align}
 V_{0,\text{local}}(\vec q\,)\sim \frac{1}{|\vec q\,|^2},\qquad \text{for } |\vec q|\to\infty,
 \label{Eq:local_q_minus2}
\end{align}
both LO and NLO amplitudes turn finite after renormalization even if the NLO potential
is not regulated. The reason for that is a 
milder ultraviolet behavior of local structures after performing subtractions.
Such a mild regulator cannot be chosen for the non-local parts of the LO potential.

Equation~\eqref{Eq:local_q_minus2} implies
that in the spin-triplet channels the one-pion-exchange potential
can be regulated by a dipole form factor, 
\begin{align}
F_{q,1\pi,\Lambda,1}=\frac{\Lambda^2-M_\pi^2}{q^2+\Lambda^2},
\end{align}
whereas for the spin-singlet channels it can even be left unregulated.

Although in practical calculations one typically
implements Gaussian or even sharper regulators
to guarantee the finiteness of all integrals,
we consider separately the above mentioned situation
with a local part of the LO potential having the ultraviolet asymptotics
as in Eq.~\eqref{Eq:local_q_minus2} and say that such a potential
has a ``mild'' regulator in contrast to  ``standard'' regulators, i.e. 
all other cases. 
This is done to keep the analysis general and to clarify the difference between perturbative and non-perturbative regimes.
Moreover, such an analysis is useful to understand the cutoff dependence of the NN amplitude:
the milder regulator can be chosen, the weaker cutoff dependence should be expected.

For completeness, we provide the explicit expressions for the LO potential and the 
corresponding regulators in Appendix~\ref{Sec:LOpotential}.

The next-to-leading-order potential $V_2(\vec p\,',\vec p\,)$ contains the short-range part, the two-pion-exchange potential and
the regulator corrections to the leading-order potential:
\begin{align}
 &V_2(\vec p\,',\vec p\,)=V^{(2)}_{2\pi}(\vec p\,',\vec
   p)+V^{(2)}_{\text{short}}(\vec p\,',\vec p\,)
   % \quad
   +\delta_\Lambda V^{(0)}(\vec p\,',\vec p\,)\,.
\end{align}
In Ref.~\cite{Gasparyan:2021edy}, we found that one does not need to regularize the NLO potential 
to perform the renormalization of the NLO amplitude. Or, equivalently, one can introduce a cutoff $\Lambda_\text{NLO}\gg\Lambda$.
On the other hand in practical calculations, one can choose $\Lambda_\text{NLO}\sim\Lambda_b$
if it improves efficiency of a computational scheme.
Both approaches are formally equivalent because the regulator corrections $\delta_\Lambda V^{(2)}$ appear at order $O(Q^4)$
in accordance with the dimensional power counting.

It turns out, that the situation is slightly different in the general non-perturbative case,
where for the choice of the ``mild'' LO regulator we need to keep $\Lambda_\text{NLO}$ finite.
It can still be larger than $\Lambda$, but not arbitrarily large, see discussion in Sec.~\ref{Sec:Subtractions}.

The explicit expressions for the NLO potential can be found in Appendix~\ref{Sec:NLOpotential}.

\subsection{NN amplitudes and contour rotation}
In the present study we work predominantly 
in the partial wave $lsj$ basis, which makes the analysis
of the non-perturbative effects more efficient.
In the $lsj$ basis,
the potential and the amplitude are 
$n_\text{PW}\times n_\text{PW}$  
matrices, where $n_\text{PW}=1$ ($n_\text{PW}=2$) for the uncoupled (coupled)
partial waves.
The series for the partial wave LO amplitude and for the unrenormalized NLO amplitude
are given by
\begin{align}
&T_0=\sum_{n=0}^{\infty}T_0^{[n]},\qquad  T_0^{[n]}=V_0 K^n= \bar K^n V_0, \label{Eq:T0}\\
&T_2=\sum_{m,n=0}^{\infty}T_2^{[m,n]},\qquad  T_2^{[m,n]}=\bar K^m V_2 K^n,\label{Eq:T2}
\end{align}
where $G$ is the free two-nucleon propagator and
\begin{align}
  K=G V_0, \qquad\bar K= V_0 G.
\end{align}

In the non-perturbative case these equations generalize to
\begin{align}
&T_0=V_0 R=\bar R V_0,\label{Eq:T0_NP}\\
&T_2=\bar R V_2 R,\label{Eq:T2_NP}
\end{align}
where $R$ ($\bar R$) is the resolvent of the Lippmann-Schwinger equation (LSE)
\begin{align}
 R=\frac{1}{\mathds{1}-K},\qquad \bar R=\frac{1}{\mathds{1}-\bar K}.
 \label{Eq:resolvent}
\end{align}

The renormalized expression for the NLO amplitude $\mathds{R}(T_2)$ is obtained by adding
the relevant counter term, see Sec.~\ref{Sec:Subtractions} for details:
\begin{align}
&\mathds{R}(T_2)=\bar R \left(V_2 +\delta V_0^{(2)}  \right)R.\label{Eq:T2_renormalized}
\end{align}

The explicit form of the LSE, $T_0=V_0+V_0GT_0$, reads
\begin{align}
\left(T_{0}\right)_{l' l}(p',p;p_\text{on})&=
\sum_{l''}
\int \frac{p''^2 dp''}{(2\pi)^3}
\left(V_0\right)_{l' l''}(p',p'')
G(p'';p_\text{on})
\left(T_{0}\right)_{l'' l}(p'',p;p_\text{on}),\nonumber\\
G(p''; p_\text{on})&=\frac{m_N}{p_\text{on}^2-p''^2+i \epsilon}.
\label{Eq:LS_equation}
\end{align}
The indices $l$, $l'$, $l''$ denote the orbital angular momentum of the NN system,
$p_\text{on}$ is the on-shell c.m.~nucleon momentum and $p$ ($p'$) are the initial (final)
off-shell c.m. momenta.

It turns out useful to modify the integration path over the
off-shell momentum $p''$ and rotate the contour into the complex plane \cite{Hetherington:1965zza,Aaron:1966zz,Cahill:1971ddy}.
The new integration contour $\mathcal{C}$ is defined by
$p''=|p''|e^{-i\alpha_{\mathcal{C}}}$.
Our choice for the rotation angle $\alpha_{\mathcal{C}}$ is determined by the
location of singularities of the LO potential in the complex plane \cite{Gasparyan:2021edy}:
\begin{align}
 \alpha_{\mathcal{C}}=\frac{1}{2}\arctan\frac{M_\pi}{\left(p_\text{on}\right)_\text{max}},
\end{align}
where  $\left(p_\text{on}\right)_\text{max}$ is the maximal considered on-shell momentum.

The contour rotation enables us to perform direct estimations of the bounds on the partial wave amplitudes
avoiding principal value integrals.

\subsection{Bounds on the potentials and the NN propagator}
\label{Sec:Bounds_text}
By analogy with Ref.~\cite{Gasparyan:2021edy},
we use certain upper bounds for the potentials and
the NN propagator that are valid for off-shell momenta lying on
the complex contour $\mathcal{C}$ and for the allowed 
real on-shell momenta.
These bounds allow us to estimate the nucleon-nucleon LO and NLO amplitudes
and to verify the corresponding power counting.

Following Ref.~\cite{Gasparyan:2021edy}, in the bounds considered below,
we introduce dimensionless constants named $\mathcal{M}_i$: $\mathcal{M}_{V_0}$,
$\mathcal{M}_G$, etc., which are supposed to be of order one.
Analogous constants appear in our final estimates for the amplitudes.

Some of the inequalities should be modified compared to Ref.~\cite{Gasparyan:2021edy}
to be better suited for the non-perturbative analysis.
In particular, for the LO potential $V_0(p',p)$, we need bounds that are separable
in momenta $p$ and $p'$.

The inequalities listed below are meant to hold for all matrix elements
of the partial wave potentials $V_0(p',p)$ and $V_2(p',p)$ in $l\,,l'$ space.
Their derivation can be found in 
Appendices~\ref{Sec:bounds_plane_wave} and~\ref{Sec:PW_bounds}.

The LO partial-wave potential obeys the following bounds:
\begin{align}
&\left|V_0(p',p)\right|\le \mathcal{M}_{V_0}
V_{0,\text{max}}\, g(p') h(p),\nonumber\\
&\left|V_0(p',p)\right|\le \mathcal{M}_{V_0}
V_{0,\text{max}}\, h(p') g(p),
\label{Eq:bounds_V0_l_0_text}
\end{align}
with
\begin{align}
 V_{0,\text{max}}=\frac{8\pi^2 }{m_N \Lambda_V}\,,
\end{align}
where 
the exact form of the functions $g$ and $h$ (and the value of $\mathcal{M}_{V_0}$) 
depends on the partial wave and on the
form of a regulator.
For  $l=0$ (for the coupled partial waves, we mean by $l$ the lowest possible orbital angular momentum),
$g$ and $h$ are given by 
\begin{align}
 g(p)=\lambda_\text{log}(p/\Lambda)\,, \ h(p)=1 \,,
 \label{Eq:g_h_local_n1_text}
\end{align}
for the ``mild'' regulator, and by
\begin{align}
 g(p)=\left[\lambda(p/\Lambda)  \right]^2\,, \ h(p)=\left[\lambda(p/\Lambda)  \right]^{-1}\,, 
  \label{Eq:g_h_local_n2_text}
\end{align}
for the ``standard'' regulators with the functions $\lambda$ and $\lambda_\text{log}$ defined as
\begin{align}
\lambda(\xi)&=\theta(1-|\xi|)+\theta(|\xi|-1)\frac{1}{|\xi|^2}\,,\nonumber\\
\lambda_\text{log}(\xi)&=\theta(1-|\xi|)+\theta(|\xi|-1)\frac{1+\ln|\xi|}{|\xi|^2}.
\label{Eq:lambda_lambdalog_text}
\end{align}
For higher partial waves, $l \ge 1$, we adopt the bounds
\begin{align}
 g(p)=\lambda_\text{log}(p/\Lambda)/|p|\,, \ h(p)=|p|.
 \label{Eq:g_h_higherPWs_text}
\end{align}
Notice that while in the latter case one could use a stronger bound
and replace $\lambda_\text{log}$ with $\lambda$
for the ``standard'' regulator, this would not affect our conclusions.
Therefore, we prefer to employ this unified bound.

For spin-singlet partial waves without a short-range LO contribution,
one can improve the above bounds and
replace in Eq.~\eqref{Eq:g_h_higherPWs_text} 
$\lambda_\text{log}(p/\Lambda)$ with $\lambda_\text{log}(p/M_\pi)$.
However, in all such channels 
the perturbative regime for the LO potential is realized, which has already been
analyzed in Ref.~\cite{Gasparyan:2021edy} and will not be considered here.

Note that for $|p|\le\Lambda$, and, in particular, for the on-shell momentum $|p|=p_\text{on}$,
we have in all cases $g(p)=h(p)=1$.

It is convenient also to introduce the functions 
\begin{align}
&v_0(p',p)=V_0(p',p)\left[\mathcal{M}_{V_0}
V_{0,\text{max}}\, h(p') g(p)  \right]^{-1} \,,\nonumber\\
&\bar v_0(p',p)=V_0(p',p)\left[\mathcal{M}_{V_0}
V_{0,\text{max}}\, g(p') h(p)  \right]^{-1},
\label{Eq:v0_text}
\end{align}
for which the following bounds hold:
\begin{align}
&\left|v_0(p',p)\right|\le 1\,,\qquad
\left|\bar v_0(p',p)\right|\le 1\,.
\label{Eq:v0_bound_text}
\end{align}

For the unregulated NLO potential, we adopt the bounds from Ref.~\cite{Gasparyan:2021edy}.
In particular, for $l=0$: 

\begin{align}
&\left|V_2(p',p)\right|\le \mathcal{M}_{V_2,0}
\left(|p|^2+|p'|^2\right)\tilde f_\text{log}(p',p),
\label{Eq:bounds_V2_l_0_text}
\end{align}
with
\begin{align}
&\tilde f_\text{log}(p',p)=\frac{8\pi^2}{m_N \Lambda_V \Lambda_{b}^2} f_\text{log}(p',p)\,,\nonumber\\
& f_\text{log}(p',p)=\theta(|p|-M_\pi)\ln\frac{|p|}{M_\pi}
+\theta(|p'|-M_\pi)\ln\frac{|p'|}{M_\pi}+1\,,
\label{Eq:f_log_text}
\end{align}
where we have dropped the $\log \Lambda/M_\pi$ term in the definition of $f_\text{log}$, 
which is unnecessary and was introduced in 
Ref.~\cite{Gasparyan:2021edy} for convenience.

Note that in Ref.~\cite{Gasparyan:2021edy}, the NLO potential $V_2$
was split into two parts
\begin{align}
 V_2( p\,', p)=\hat V_2( p\,', p)+\tilde V_2( p\,', p),
 \label{Eq:V2_tilde_text}
\end{align}
with 
\begin{align}
\hat V_2( p\,', p)=V_2(0,0)\,,\quad \tilde V_2( p\,', p)=V_2( p\,', p)-V_2(0,0),
\end{align}
and the inequality in Eq.~\eqref{Eq:bounds_V2_l_0_text} is, strictly speaking, valid for $\tilde V_2$.
However, in the present work, we use most of the time the scheme with $V_2(0,0)=0$.
Therefore, in what follows, we will always assume that $\tilde V_2=V_2$
unless specified otherwise.
For alternative schemes, we also provide the bound for $\hat V_2$:
\begin{align}
 \left|\hat V_2(p', p)\right|\le \mathcal{\hat M}_{V_2,0} \frac{8\pi^2 }{m_N \Lambda_V}\frac{M_\pi^2}{\Lambda_{b}^2}.
 \label{Eq:bound_V_2_0_hat_text}
\end{align}

For higher partial waves $l>0$, it is sufficient to implement the $p$-wave bound:
\begin{align}
&\left|V_2(p',p)\right|\le \mathcal{M}_{V_2,1}
|p'||p|\tilde f_\text{log}(p',p).
\label{Eq:bounds_V2_l_2_text}
\end{align}

For the regularized NLO potential with the cutoff $\Lambda_\text{NLO}$,
the bounds in Eq.~\eqref{Eq:bounds_V2_l_0_text} are modified as follows (see Sec.~\ref{Sec:bounds_PW_NLO_Swave}):
\begin{align}
&\left|V_2(p',p)\right|\le \mathcal{M}_{V_2,0}
\left(|p|^2+|p'|^2\right)\tilde f_\text{log}(p',p)  \lambda_\text{log}(p'/\Lambda_\text{NLO}) \,, \text{ or}\nonumber\\
&\left|V_2(p',p)\right|\le \mathcal{M}_{V_2,0}
\left(|p|^2+|p'|^2\right)\tilde f_\text{log}(p',p)  \lambda_\text{log}(p/\Lambda_\text{NLO})\,.
\label{Eq:bounds_V2_l_0_R_0_text}
\end{align}

For the two-nucleon propagator $G(p{;p_\text{on}})=m_N/(p_\text{on}^2-p^2 )$, we use the same bound as
in Ref.~\cite{Gasparyan:2021edy}:
\begin{align}
 |G(p{;p_\text{on}})|\le \mathcal{M}_G \frac{m_N}{|p^2|}\,,
 \label{Eq:bound_on_G}
\end{align}
with  $\mathcal{M}_G=1/\sin(2\alpha_{\mathcal{C}}) $.

\section{Leading-order Lippmann-Schwinger equation}
\label{Sec:LO_Fredholm}
In this section we outline the Fredholm method for
solving integral equations and derive the bounds on
the resolvents of the LSE and on the LO amplitude in the non-perturbative case.
The resolvents $R$ and $\bar R$ of the partial-wave LSE, see Eq.~\eqref{Eq:resolvent},
can be represented by means of the Fredholm formula~\cite{Newton:1982qc, Smithies1958integral} as:
\begin{align}
R&=(\mathds{1}-K)^{-1}=\mathds{1}+\frac{Y}{{D}},\nonumber\\
\bar R &= (\mathds{1}-  \bar K)^{-1}=\mathds{1}+\frac{\bar Y}{{D}},
\label{Eq:Fredholm_decomposition}
\end{align}
where the Fredholm determinant $D$ is a number and depends only on the on-shell momentum $D=D(p_\text{on})$,
whereas the minor $Y$ ($\bar Y$) is a matrix in the $l$, $l'$ space and an operator in the space of 
the off-shell momenta: $Y=Y_{ji}(p',p;p_\text{on})$.
The quantities $Y$, $\bar Y$ and $D$ can be expanded into convergent series in powers of the LO potential $V_0$:
\begin{align}
&  Y=\sum_{n=1}^\infty  Y^{[n]}, \qquad \bar Y=\sum_{n=1}^\infty  \bar Y^{[n]},\qquad
{D} = \sum_{n=0}^\infty  {D}^{[n]}.
\end{align}

In what follows, we will consider the resolvent $R$ and the minor $Y$. 
The results are trivially generalized for $\bar R$ and $\bar Y$.

The terms ${D}^{[n]}$ and $Y^{[n]}$ can be written as~\cite{Newton:1982qc, Smithies1958integral}
\begin{align}
{D}^{[n]}(p_\text{on})=\frac{(-1)^{n}}{n !} 
\sum_{i_{1}, \ldots, i_n}\int \prod_{k=1}^n \frac{p_k^2 dp_k}{(2\pi)^3}
\left[ \text{det}_{D,n} (K) \right]_{i_1,\dots\,i_n}(p_1,\dots,p_n;p_\text{on}),
\label{Eq:D_n_det}
\end{align}
and
\begin{align}
Y^{[n+1]}_{i' i}(p',p;p_\text{on})=&\frac{(-1)^{n}}{n !}
\sum_{i_{1}, \ldots, i_n}
\int \prod_{k=1}^n \frac{p_k^2 dp_k}{(2\pi)^3}\left[ \text{det}_{Y,n+1} (K) \right]_{i,i_1,\dots\,i_n,i'}(p,p_1,\dots,p_n,p';p_\text{on}),
\label{Eq:Y_n_det}
\end{align}
where the matrix indices $i,i_1,..i_n$ and $i'$ correspond to the orbital angular momentum $l=j\pm1$
for coupled partial waves and $l=j$ for uncoupled partial waves.
In the above equations, the determinants for an operator $X$ with matrix elements $X(p',p;p_\text{on})$
(or $X(p',p)$ if it is independent of $p_\text{on}$) are defined as:
\begin{align}
 \left[ \text{det}_{D,n} (X) \right]_{i_1,\dots,i_n}(p_1,\dots,p_n;p_\text{on})=
 \left|\begin{array}{lll}
X_{i_1,i_1}(p_1,p_1;p_\text{on}) & \cdots & X_{i_n,i_1}(p_1,p_n;p_\text{on}) \\
\ldots & \ldots& \ldots\\
X_{i_1,i_n}(p_n,p_1;p_\text{on}) & \cdots & X_{i_n,i_n}(p_n,p_n;p_\text{on})
\end{array}\right|,
\end{align}
and
\begin{align}
 \left[ \text{det}_{Y,n+1} (X) \right]_{i,i_1,\dots,i_n,i'}(p,p_1,\dots,p_n,p';p_\text{on})=
 \left|\begin{array}{llll}
X_{i' i}(p',p;p_\text{on}) & X_{i_{1} i}(p_1,p;p_\text{on}) & \cdots & X_{i_{n} i}(p_n,p;p_\text{on}) \\
X_{i' i_{1}}(p',p_1;p_\text{on}) & X_{i_{1} i_{1}}(p_1,p_1;p_\text{on}) & \cdots & X_{i_{n} i_{1}}(p_n,p_1;p_\text{on}) \\
\cdots & \cdots & \cdots & \cdots  \\
X_{i' i_{n}}(p',p_n;p_\text{on}) & X_{i_{1} i_{n}}(p_1,p_n;p_\text{on}) & \cdots & X_{i_{n} i_{n}}(p_n,p_n;p_\text{on})
\end{array}\right|.
\end{align}

Rescaling $V_0$ as in Eq.~\eqref{Eq:v0_text}, we obtain:
\begin{align}
&K_{i'i}(p',p;p_\text{on})=(v_0)_{i'i}(p',p)\mathcal{M}_{V_0}
V_{0,\text{max}}\, g(p') h(p)G(p';p_\text{on}),
\end{align}
so that
\begin{align}
 {D}^{[n]}(p_\text{on})=\frac{(-1)^{n}}{n !} 
\left( \mathcal{M}_{V_0} V_{0,\text{max}} \right)^n
\sum_{i_{1}, \ldots, i_n}
\int \left[\prod_{k=1}^n \frac{p_k^2 dp_k}{(2\pi)^3}g(p_k) h(p_k)G(p_k;p_\text{on})\right] 
\left[ \text{det}_{D,n} (v_0) \right]_{i_1,\dots,i_n}(p_1,\dots,p_n),
\label{Eq:D_n}
\end{align}
and
\begin{align}
Y^{[n+1]}_{i' i}(p',p;p_\text{on})&=\frac{(-1)^{n}}{n !}
\left( \mathcal{M}_{V_0} V_{0,\text{max}} \right)^{n+1}
g(p') h(p)G(p';p_\text{on}) \nonumber\\
&\times\sum_{i_{1}, \ldots, i_n}
\int \left[\prod_{k=1}^n \frac{p_k^2 dp_k}{(2\pi)^3}g(p_k) h(p_k)G(p_k;p_\text{on})\right] 
\left[ \text{det}_{Y,n+1} (v_0) \right]_{i,i_1,\dots,i_n,i'}(p,p_1,\dots,p_n,p').
\label{Eq:Y_n}
\end{align}

\subsection{Upper bounds for the Fredholm determinant}
\label{Sec:bound_Frehdolm_determinant}
First, we analyze the series for the Fredholm determinant $D$.
Since the matrix elements $v_{0;ji}(p',p)$
are bounded by (see Eq.~\eqref{Eq:v0_bound_text})
\begin{align}
\left|v_{0;ji}(p',p)\right|\le 1\,,
\label{Eq:bound_on_kappa}
\end{align}
the  Hadamard's inequality for determinants gives~\cite{Newton:1982qc, Smithies1958integral} 
\begin{align}
 \left|\text{det}_{D,n} (v_0) \right|\le n^{n/2}.
 \label{Eq:Hadamard_Dn}
\end{align}
Therefore, using Stirling's formula, we can estimate $ D^{[n]}$ as follows:
\begin{align}
\left|D^{[n]}\right|  &\leq 
\frac{1}{n !}
 \Sigma^n n^{n/2}  \le \frac{1}{\sqrt{2\pi n}}
\left(\frac{e \Sigma}{\sqrt{n}}\right)^n
=\frac{1}{\sqrt{2\pi} e\Sigma}
\left(\frac{e \Sigma}{\sqrt{n}}\right)^{n+1}\nonumber\\
&=\frac{1}{\sqrt{2\pi} e\Sigma}
\exp\bigg[-(n+1)\ln\frac{\sqrt{n}}{e\Sigma}\bigg]
\eqqcolon\mathcal{M}_{{D,n}}.
\label{Eq:bounds_Delta_n}
\end{align}
where $\Sigma$ is defined as
\begin{align}
\mathcal{M}_{V_0} V_{0,\text{max}}n_\text{PW} \left|\int\frac{p^2 dp}{(2\pi)^3}g(p) h(p)G(p;p_\text{on})\right|\le 
\frac{\mathcal{M}_{V_0}\mathcal{M}_G}{\Lambda_V} n_\text{PW}\int\frac{d|p|}{\pi}g(p) h(p)
\eqqcolon\Sigma\,.
\end{align}
Since $g(p)$ and $h(p)$ depend only on the ratio $p/\Lambda$, 
we can write
\begin{align}
 \Sigma=\mathcal{M}_\Sigma\frac{\Lambda}{\Lambda_V}\,,
\end{align}
where the numerical value of the constant $\mathcal{M}_\Sigma$ depends on a particular 
form of $g(p)$ and $h(p)$.

If we assume $\Lambda\sim\Lambda_V$, then $\Sigma\sim1$ up to a numerical factor.
The situation when $\Sigma<1$ corresponds to a convergent series for the LO amplitude in terms of $V_0$.
In contrast, for the non-perturbative regime that we consider, we have $\Sigma\ge 1$.

The maximal value of $ {D}^{[n]}$ is achieved at some $n=n_{D_\text{max}}$ and can be estimated by differentiating
Eq.~\eqref{Eq:bounds_Delta_n} with respect to $n$:
\begin{align}
&n_{D_\text{max}}\approx e\Sigma^2\,,\nonumber\\
&| {D}^{[n]}|\le \mathcal{M}_{ {D}^{[n]},\text{max}}\approx\frac{e^{e\Sigma^2/2}}{\sqrt{2\pi e }\Sigma}\,,
 \label{Eq:Delta_n_max}
\end{align}
which is formally a number of order one, but it grows very rapidly with $\Sigma$.

The whole series for $D$ is also bounded by a constant of order one:
\begin{align}
 |D|\le  \mathcal{M}_{D},
 \label{Eq:MD}
 \end{align}
which can be estimated by replacing the sum with an integral and using
Laplace's method:
\begin{align}
 \mathcal{M}_{D}&= \sum_{n=0}^\infty \mathcal{M}_{{D,n}}\approx\int_0^\infty d t \mathcal{M}_{{D,t}}
\approx 
 \sqrt{2\pi}
\left(-\frac{\partial^2 \ln\mathcal{M}_{{D,t}}}{\partial t^2}\right)^{-1/2}\mathcal{M}_{{D,t}}\Big|_{t=n_{D_\text{max}}}
\approx\sqrt{2}e^{e\Sigma^2/2}\,,
 \label{Eq:MD_estimate}
 \end{align}
which agrees rather well with the series summed numerically
(see Eq.~\eqref{Eq:bounds_Delta_n}).
For example, for $\Sigma=1$, both results give
$\mathcal{M}_{D}\approx 5$.

The bounds~\eqref{Eq:bounds_Delta_n} and~\eqref{Eq:MD_estimate} are 
rather weak and very conservative.
If $\Sigma$ is not close to one, the numerical values for $\mathcal{M}_{D}$
become very large. However, in realistic calculations, we can see that 
$D$ does actually not exceed the values of order one. Clearly, one can always perform a numerical check
in order to verify whether our approach to the renormalizability of the NN amplitude
based on the Fredholm method is reliable.
Note also that for the $^1S_0$ and $^3S_1-{^3D_1}$  NN channels,
one can expect $\Sigma$ to be close to one
(ignoring the fine-tuning between attractive and repulsive forces)
because the first (quasi) bound states in these channels are very shallow.
This is roughly confirmed by an analysis of the 
Weinberg eigenvalues in Ref.~\cite{Reinert:2017usi}.

There are particular cases when the estimate in Eq.~\eqref{Eq:bounds_Delta_n}
can be readily improved.
For example, for purely local LO potentials,
the quantities ${D}$ and $Y$ correspond to the Jost function and the regular solution
of the Schr\"odinger equation in configuration space and 
the terms in their expansion, $ D^{[n]}$ and $Y^{[n]}$,
decrease as $1/n!$. 
On the other hand, if the LO consists of only a short-range separable potential
(or is dominated by such a contribution),
the series for ${D}$ and $Y$ contain a finite number of terms.
However, in our general discussion, we will simply assume that 
Eq.~\eqref{Eq:MD} holds.

We will also need an estimate for the series remainder:
\begin{align}
 \delta_n D= \sum_{k=n+1}^\infty  D^{[n]}.
\label{Eq:remainder_D}
\end{align}
From Eq.~\eqref{Eq:bounds_Delta_n}, we can conclude
that for sufficiently large $n$,
\begin{align}
 n>n_0\equiv \mathcal{\tilde M}_{\delta D},
\end{align}
the terms $ D^{[n]}$ and, therefore, also the remainder $\delta_n \Delta$
decrease faster than exponential 
\begin{align}
 \delta_n D \le e^{-\mathcal{M}_{\delta D}\,n},
 \label{Eq:deltaD_exponential}
\end{align}
with any $\mathcal{M}_{\delta D}$, which we will use in our further estimates.
The value $\mathcal{\tilde M}_{\delta D}$ depends on $\mathcal{M}_{\delta D}$ and on $\Sigma$.
Based on Eq.~\eqref{Eq:bounds_Delta_n}, we can conclude that the exponential decrease starts only
for
\begin{align}
 \mathcal{\tilde M}_{\delta D}> (e\Sigma)^2,
\end{align}
which, being formally a number of order one, becomes extremely large unless $\Sigma\approx 1$.
However, as follows from the discussion above, in realistic calculations,
such an exponentially suppressed regime can be reached much earlier. In fact in the numerical calculation
presented in Sec.~\ref{Sec:results}, the relative error $\delta_n D/D$ becomes less than
one percent in most cases for $n=3$ or $4$.

\subsection{\texorpdfstring{Bounds for the minor $Y$}{Bounds for the minor Y}}
By analogy with the Fredholm determinant $D$, we can perform the same analysis for
the minor $Y$ starting from the definition in Eq.~\eqref{Eq:Y_n}.
Using again the Hadamard's inequality,
\begin{align}
  \left|\text{det}_{Y,n} (v_0) \right|\le n^{n/2},
   \label{Eq:Hadamard_Yn}
\end{align}
we get the bound for $Y^{[n]}$:
\begin{align}
\left|Y^{[n]}_{j i}(p',p;p_\text{on})\right|  
&\leq \mathcal{M}_{Y,n}
\left|G(p',p_\text{on})\right| \frac{8\pi^2 \mathcal{M}_{V_0}}{m_N \Lambda_V}
 g(p') h(p) \nonumber\\
\end{align}
with
\begin{align}
\mathcal{M}_{Y,n}&=\frac{1}{(n-1) !}
 \Sigma^{n-1} n^{n/2}\le\frac{e}{\sqrt{2\pi}}
\left(\frac{e \Sigma}{\sqrt{n}}\right)^{n-1}.
\end{align}
Further, taking into account the bound for the propagator in Eq.~\eqref{Eq:bound_on_G},
we obtain
\begin{align}
\left|Y^{[n]}_{j i}(p',p;p_\text{on})\right|  
&\le \mathcal{M}_{Y,n}
\frac{8\pi^2 \mathcal{M}_{V_0} \mathcal{M}_G}{\Lambda_V |p'|^2}
 g(p') h(p)\eqqcolon
 \frac{8\pi^2\mathcal{M}_Y}{\Lambda_V |p'|^2}\mathcal{M}_{Y,n}\, g(p') h(p).
\label{Eq:bounds_Y_n}
\end{align}

Analogously to Eq.~\eqref{Eq:MD_estimate},
the whole series for $Y$ can be estimated to be
\begin{align}
\left|Y_{ji}(p',p;p_\text{on})\right|  
 \le \frac{8\pi^2\mathcal{M}_Y}{\Lambda_V |p'|^2} Y_\text{max}\, g(p') h(p)
 \eqqcolon\frac{8\pi^2\mathcal{M}_{Y_\text{max}}}{\Lambda_V |p'|^2} \, g(p') h(p),
 \label{Eq:bound_Y_Y_max}
\end{align}
where
\begin{align}
Y_\text{max} =\sum_{k=0}^\infty\mathcal{M}_{Y,n} \le \sqrt{2}e \, \Sigma \, e^{e\Sigma^2/2}.
\label{Eq:Y_max_sum}
 \end{align}

The remainder $\delta_n Y_\text{max}$, defined as
\begin{align}
 \delta_n Y_\text{max}= \sum_{k=n+1}^\infty\mathcal{M}_{Y,n},
\end{align}
can be bounded similarly to $\delta_n D$ by an exponent with an arbitrary base:
\begin{align}
\delta_n Y_\text{max} \le e^{-\mathcal{M}_{\delta Y}\, n},\qquad
\text{for } n>\mathcal{\tilde M}_{\delta Y},
 \label{Eq:deltaY_exponential}
\end{align}
with some $\mathcal{\tilde M}_{\delta Y}$.
As in the case of $\delta_n D$, the estimated value of $\mathcal{\tilde M}_{\delta Y} \sim (e\Sigma)^2$
becomes very large for $\Sigma$ significantly larger than one. However, in the actual calculations,
its numerical value is typically much more natural, see the discussion in the previous subsection.
The same comment applies also to the bound in Eq.~\eqref{Eq:Y_max_sum} for $Y_\text{max}$.

The remainder $\delta_n Y(p',p;p_\text{on})$ follows from Eq.~\eqref{Eq:deltaY_exponential}:
\begin{align}
\left|\delta_n Y_{ji}(p',p;p_\text{on})\right|  
=\left|\sum_{k=n}^\infty Y^{[n]}_{ji}(p',p)\right| 
 \le \frac{8\pi^2\mathcal{M}_Y}{\Lambda_V |p'|^2} \delta_n Y_\text{max}\, g(p') h(p)
 \eqqcolon\frac{8\pi^2\mathcal{N}_{\delta_n Y}}{\Lambda_V |p'|^2} \, g(p') h(p)\,.
 \label{Eq:deltaY_max}
\end{align}

The bounds for $\bar Y(p',p;p_\text{on})$ are obtained from Eqs.~\eqref{Eq:bound_Y_Y_max} and~\eqref{Eq:deltaY_max} by interchanging $p\leftrightarrow p'$.

\subsection{Bounds for the LO amplitude}
After these preparations, we are finally in the position to deduce the bounds for the on-shell LO amplitude,
which can be represented as
\begin{align}
 T_0=V_0R=\frac{N_0}{D},\qquad N_0=V_0 D+V_0 Y.
 \label{Eq:N_0_def}
\end{align}
First, consider the quantity $N_0$ defined explicitly as follows:
\begin{align}
 (N_0)_{ji}(p_\text{on})&=(V_0)_{ji}(p_\text{on},p_\text{on})D(p_\text{on})
 \nonumber\\
 &+\sum_{i'}
 \int\frac{p'^2 dp'}{(2\pi)^3}
 (V_0)_{ji'}(p_\text{on},p')Y_{i'i}(p',p_\text{on};p_\text{on}).
  \end{align}
Applying the bounds from Eqs.~\eqref{Eq:bounds_V0_l_0_text},~\eqref{Eq:MD}~and~\eqref{Eq:bound_Y_Y_max} , we obtain
  \begin{align}
 \left|(N_0)_{ji}(p_\text{on})\right|&\le
 \mathcal{M}_{V_0} V_{0,\text{max}}
 \left[ \mathcal{M}_D+\frac{n_\text{PW}\mathcal{M}_{Y_\text{max}}}{\Lambda_V }
 \int\frac{d|p|}{\pi}g(p)h(p)\right]\nonumber\\
 &\le \mathcal{M}_{V_0} V_{0,\text{max}}
 \left( \mathcal{M}_D+\frac{\mathcal{M}_{Y_\text{max}}\Sigma}{\mathcal{M}_{V_0}\mathcal{M}_G}\right)
 \eqqcolon \mathcal{M}_{N_0} V_{0,\text{max}}.
 \label{Eq:bound_N0}
  \end{align}
  
Now, we can analyze the bounds for the LO amplitude $T_0$.
Since $T_0$ is the ratio of $N_0$ and $D$, it is important
how the Fredholm determinant $D$ is bounded from below.
From the definition in Eq.~\eqref{Eq:D_n_det}, it follows 
that all terms $ D^{[n]}$ should be in general of order $O(Q^0)$.
However, in a realistic situation, there might be certain cancellations
among terms in the series, and the actual numerical 
value of $D(p_\text{on})$ might turn out to be very small.
This can happen when there is a shallow bound or quasibound state, which 
leads to an enhancement of the amplitude at threshold.
Such a situation only takes place in the $^1S_0$ of NN scattering.
Therefore, in our analysis for higher partial waves with $l\ge 1$,
we regard the Fredholm determinant as being ``natural'':
\begin{align}
 |D(p_\text{on})|&\ge \mathcal{M}_{D,\text{min}},
 \label{Eq:MDmin_P_waves}
\end{align}
where $\mathcal{M}_{D,\text{min}}$ is a constant of order one.
From Eqs.~\eqref{Eq:bound_N0} and~\eqref{Eq:MDmin_P_waves}, we conclude
that for $l\ge 1$, the LO amplitude is bounded by 
\begin{align}
 |(T_0)_{ji}|\le  \mathcal{M}_{T_0} V_{0,\text{max}},
 \label{Eq:T_0_natural}
\end{align}
and satisfies the same power counting as $V_0$, i.e. is of order $O(Q^0)$.

For the $S$-wave channels,
we allow for the real part of $D$ to be small,
while still bounded from below
at least at threshold.
Moreover, we assume that the imaginary part of $D$, which is proportional to $p_\text{on}$,
is not a subject to additional cancellations.
In particular, we exclude the situation when both $N$ and $D$ are equal to 
zero, i.e. the presence of a  Castillejo-Dalitz-Dyson (CDD) pole \cite{Castillejo:1955ed,Johnson:1979jy}.
We combine these conditions into the following constraint:
\begin{align}
 |D(p_\text{on})|&\ge \mathcal{M}_{D,\text{min}} \left(\kappa+\frac{p_\text{on}}{\Lambda_V}\right),
 \label{Eq:MDmin_S_waves}
\end{align}
where $\kappa>0$ is not necessarily of order one, but can be numerically small.
The factor $1/\Lambda_V$ in front of $p_\text{on}$ follows from the upper bound for the
imaginary part of $D$.

The LO amplitude $T_0$ is enhanced compared to $V_0$, which can be written as
\begin{align}
 |(T_0)_{ji}|\le  \mathcal{M}_{T_0} \kappa^{-1} V_{0,\text{max}},
 \label{Eq:T_0_enhanced}
\end{align}
or
\begin{align}
 |(T_0)_{ji}|\le  \mathcal{M}_{T_0} \frac{\Lambda_V}{p_\text{on}} V_{0,\text{max}},
 \label{Eq:T_0_enhanced2}
\end{align}
depending on the value of the on-shell momentum $p_\text{on}$.
The latter bound is in fact a unitary limit for the LO amplitude up to a numerical factor of order one,
which justifies the coefficient $1/\Lambda_V$ in Eq.~\eqref{Eq:MDmin_S_waves},
see the definition of $V_{0,\text{max}}$ in Eq.~\eqref{Eq:V0max_text}.
Equation~\eqref{Eq:T_0_enhanced2} means that the LO amplitude becomes effectively of order $O(Q^{-1})$
in agreement with findings of Refs.~\cite{Kaplan:1998tg,Kaplan:1998we}.

To summarize, we have applied the Fredholm method to decompose the resolvent of the LS equation
and derived the bounds for the Fredholm determinant $D$, the minor $Y$ and the on-shell LO amplitude.
The bounds involve undetermined dimensionless constants of order one, which can be calculated
for each particular situation.

\section{\texorpdfstring{Next-to-leading order amplitude in the non-perturbative case. $P$- and higher partial waves}
{Next-to-leading order amplitude in the non-perturbative case. P-waves and higher}}
\label{Sec:Pwaves}
In this section we consider the on-shell ($p=p'=p_\text{on}$)
NLO amplitude $T_2$ for orbital angular momenta
$l\ge 1$ and derive the corresponding bounds in the non-perturbative regime.
We represent the amplitude $T_2$ using the Fredholm 
decomposition of the resolvent in Eq.~\eqref{Eq:Fredholm_decomposition} as follows:
\begin{align}
  &T_2 = \bar R V_2 R=V_2+T_{2,Y}/D+T_{2,\bar{Y}}/D+T_{2,\bar{Y}Y}/D^2
  \eqqcolon\frac{N_2}{D^2},
  \label{Eq:decomposition_N2_pwaves}
\end{align}
with
\begin{align}
  &T_{2,Y}= V_2 Y\,,\qquad T_{2,\bar{Y}}= \bar Y V_2\,,\qquad T_{2,\bar{Y}Y}=\bar Y V_2 Y\,,
\end{align}
or more explicitly:
\begin{align}
 &T_{2,Y}(p',p;p_\text{on})=
\int\frac{p_1^2 d p_1}{(2\pi)^3}V_2(p',p_1)Y(p_1,p;p_\text{on})\,,\nonumber\\
 &T_{2,\bar{Y}}(p',p;p_\text{on})=\int\frac{p_1'^2 d p'_1}{(2\pi)^3}
 \bar{Y}(p',p_1';p_\text{on})V_2(p_1',p)\,,\nonumber\\ 
 &T_{2,\bar{Y}Y}(p',p;p_\text{on})=\int\frac{p_1^2 d p_1}{(2\pi)^3}\frac{p_1'^2 d p'_1}{(2\pi)^3}
 \bar{Y}(p',p_1';p_\text{on})V_2(p_1',p_1)Y(p_1,p;p_\text{on}).
\end{align}

First, consider $T_{2,Y}$.
The bounds for $V_2$ and $Y$ in
Eqs.~\eqref{Eq:bounds_V2_l_2_text} and~\eqref{Eq:bound_Y_Y_max} give
\begin{align}
 &\left|T_{2,Y}(p',p;p_\text{on})\right|\le
 \int\frac{|p_1|^2 d |p_1|}{(2\pi)^3}|V_2(p',p_1)||Y(p_1,p;p_\text{on})|\nonumber\\
 &\le \mathcal{M}_{V_2,1}n_\text{PW} \frac{8\pi^2 \mathcal{M}_{Y_\text{max}}}{\Lambda_V} |p'|h(p)
 \int\frac{|p_1| d |p_1|}{(2\pi)^3}
\tilde f_\text{log}(p',p_1) g(p_1) .
\end{align}
The functions $g$ and $h$ for $P$- and higher partial waves
are given in Eq.~\eqref{Eq:g_h_higherPWs_text}, which 
results in the following inequality:
\begin{align}
 &\left|T_{2,Y}(p',p;p_\text{on})\right|\le
  \mathcal{M}_{V_2,1}n_\text{PW} \frac{8\pi^2 \mathcal{M}_{Y_\text{max}}}{\Lambda_V} |p'||p|
 \int\frac{d |p_1|}{(2\pi)^3}
\tilde f_\text{log}(p',p_1) \lambda_\text{log}(p_1/\Lambda) \nonumber\\
&=\mathcal{M}_{V_2,1} n_\text{PW}\frac{8\pi^2 \mathcal{M}_{Y_\text{max}}}{\Lambda_V} 
\frac{8\pi^2}{m_N \Lambda_V \Lambda_{b}^2} |p'||p|
\int\frac{d |p_1|}{(2\pi)^3}
f_\text{log}(p',p_1) \lambda_\text{log}(p_1/\Lambda)  \nonumber\\
&=\frac{\mathcal{M}_{V_2,1}n_\text{PW}\mathcal{M}_{Y_\text{max}} }{\Lambda_V} 
\frac{8\pi^2}{m_N \Lambda_V \Lambda_{b}^2} |p'||p|\left\{
\left[1+\theta(|p'|-M_\pi)\ln\frac{|p'|}{M_\pi}\right]I_{\lambda_\text{log},1}
+I_{\lambda_\text{log},2}  \right\},
\label{Eq:T2Y}
\end{align}
where the typical integrals $I_{\lambda_\text{log},1}$ and 
$I_{\lambda_\text{log},2}$
are defined and estimated in Appendix~\ref{Sec:integrals}
and we have used Eq.~\eqref{Eq:f_log_text}.
Using those estimates, we obtain:
\begin{align}
 \left|T_{2,Y}(p',p;p_\text{on})\right|\le
 \mathcal{M}_{{2,Y}}\frac{8\pi^2 }{m_N\Lambda_V\Lambda_b^2} |p'||p|\frac{\Lambda}{\Lambda_V}
 \left[1+\theta(|p'|-M_\pi)\ln\frac{|p'|}{M_\pi}+\ln\frac{\Lambda}{M_\pi}  \right],
 \label{Eq:bound_T2Y_offshell}
\end{align}
which reduces to
\begin{align}
 \left|T_{2,Y}(p_\text{on})\right|\le
 \mathcal{M}_{{2,Y;\text{on}}}\frac{8\pi^2 }{m_N\Lambda_V\Lambda_b^2} \frac{\Lambda}{\Lambda_V} p_\text{on}^2
 \ln\frac{\Lambda}{M_\pi}, 
 \label{Eq:bound_T2Y_onshell}
\end{align}
for the on-shell momenta $p=p'=p_\text{on}$.
The bounds for $T_{2,\bar Y}$ are the same as for $T_{2,Y}$.

Next, we analyze $T_{2,\bar{Y}Y}$:
\begin{align}
\left|T_{2,\bar{Y}Y}(p',p;p_\text{on})\right|&\le
 \int\frac{|p_1|^2 d |p_1|}{(2\pi)^3}|\frac{|p_1'|^2 d |p_1'|}{(2\pi)^3}|\bar Y(p',p_1';p_\text{on})||V_2(p_1',p_1)||Y(p_1,p;p_\text{on})|\nonumber\\
 &\le \mathcal{M}_{V_2,1}n_\text{PW}^2 \left(\frac{8\pi^2 \mathcal{M}_{Y_\text{max}}}{\Lambda_V}\right)^2
 h(p')h(p)
 \int\frac{|p_1| d |p_1|}{(2\pi)^3}\frac{|p_1'| d |p_1'|}{(2\pi)^3}
\tilde f_\text{log}(p',p_1) g(p_1') g(p_1) \,.
\end{align}
The integrals over $p_1$ and $p_1'$ factorize, giving rise to the same set of integrals as in $T_{2,Y}$.
The analog of Eq.~\eqref{Eq:bound_T2Y_onshell} for $T_{2,\bar{Y}Y}$ in the on-shell kinematics is given by
\begin{align}
 \left|T_{2,\bar{Y}Y}(p_\text{on})\right|\le
 \mathcal{M}_{{2,\bar{Y}Y;\text{on}}}\frac{8\pi^2 }{m_N\Lambda_V\Lambda_b^2} \frac{\Lambda^2}{\Lambda_V^2} p_\text{on}^2
 \ln\frac{\Lambda}{M_\pi}. 
 \label{Eq:bound_T2barYY_onshell}
\end{align}

Combining the bounds for  $V_{2}$, $T_{2,Y}$, $T_{2,\bar{Y}}$ and $T_{2,\bar{Y}Y}$
and setting $\Lambda\sim\Lambda_V$, we obtain
\begin{align}
 |T_2(p_\text{on})|\le
 \mathcal{\tilde M}_2\frac{8\pi^2 }{m_N\Lambda_V\Lambda_b^2}p_\text{on}^2\ln\frac{\Lambda}{M_\pi}
 \left[ 1+D(p_\text{on})^{-1}+D(p_\text{on})^{-2} \right]\,.
 \label{Eq:T2_bound1}
\end{align}
Since we assume that for the $P$- and higher partial waves the Fredholm determinant
is bounded from below by a constant of order one, see Eq.~\eqref{Eq:MDmin_P_waves},
equation~\eqref{Eq:T2_bound1} takes the form
\begin{align}
 |T_2(p_\text{on})|\le
  \mathcal{M}_2\frac{8\pi^2 }{m_N\Lambda_V\Lambda_b^2}p_\text{on}^2\ln\frac{\Lambda}{M_\pi}.
 \label{Eq:T2_bound2}
\end{align}
Thus, the NLO amplitude is of order $O(Q^2)$ up to a factor $\ln\Lambda/M_\pi$,
which agrees with the dimensional power counting.
This result reproduces the one obtained in Ref.~\cite{Gasparyan:2021edy}
for the case of a perturbative LO interaction.

\subsection{Promoting a contact term to leading order}
\label{Sec:Contact_term_promotion}
In this subsection we consider separately the scenario with
promoting leading $P$-wave contact terms to the LO potential.
As already discussed in Sec.~\ref{Sec:LO_NLO_potentials},
phenomenological arguments may require a promotion of contact interactions
quadratic in momenta to the LO potential, even though they are formally 
of order $O(Q^2)$.
A typical example is the $^3P_0$ partial wave, where the promotion of the contact interaction to 
leading order is often considered as necessary.

Below, we discuss the subtlety related to the freedom of choosing the renormalization condition,
i.e., deciding what part of the considered contact interaction should be included into the LO potential
and what part of it should be left in the NLO potential.

The LO partial wave contact interaction in the $P$-wave channel $i$
is given by
\begin{align}
&V^{(0)}_{\text{short},\Lambda,i}(p\,', p)=
C_i \, V_{C_i,\Lambda}(p\,',p),
\label{Eq:short_range_LO}
\end{align}
where $V_{C_i,\Lambda}(p\,',p)$ is the partial wave projection of the regulated 
contact term (see Appendix~\ref{Sec:LOpotential}) relevant for the considered channel.
The corresponding NLO contact interaction has the same structure:
\begin{align}
&V^{(2)}_{\text{short},\Lambda,i}(p\,', p)=
C_{2,i} \, V_{C_i,\Lambda}(p\,',p).
\label{Eq:short_range_NLO}
\end{align}

In our estimates, we always assume that 
the LO low energy constants (LECs) are of natural size,
\begin{align}
&C_i=\frac{\mathcal{M}_{C_i}}{\Lambda_b^2} \frac{8\pi^2 }{m_N \Lambda_V},
\label{Eq:natural_C_i}
\end{align}
see Appendix of Ref.~\cite{Gasparyan:2021edy} (the factor of $4\pi$ 
corresponds to the partial-wave basis), so that
the contact interactions quadratic in momenta are of order 
$\sim p^2/\Lambda^2\sim O(Q^2)$ and are suppressed for small momenta.
As a consequence, the regulator corrections to 
the contact interactions quadratic in momenta
are effects of order $O(Q^4)$ and can be neglected in the present study.
This is why we adopt the same regulator for $V^{(0)}_{\text{short},\Lambda,i}$
and $V^{(2)}_{\text{short},\Lambda,i}$ even though,
in principle, one could employ a larger cutoff for the NLO terms or
even use the unregulated potential.
Nevertheless, if the contact interactions quadratic in momenta
are promoted to leading order, 
their contribution
in the iterations of the LO potential at momenta
$p\sim \Lambda$
is of the same order as
those of the momentum-independent contact interactions and
of the one-pion-exchange potential as long as we treat $\Lambda \sim \Lambda_b$.

The freedom to choose the renormalization scheme manifests itself
schematically as follows:
if we perform the transformation 
\begin{align}
 C_i\to C_i+\delta C_i, \qquad   C_{2,i}\to C_{2,i}-\delta C_i, \qquad \delta C_i\ll C_i\,,
\end{align}
and expand the LO and NLO amplitudes in Eqs.~\eqref{Eq:T0_NP} and~\eqref{Eq:T2_NP} in $\delta C$,
then the linear in $\delta C$ terms cancel:
\begin{align}
 &\delta T_0 \approx -\delta T_2\approx\delta C \bar R V_{C_i,\Lambda} R,
\end{align}
where we have neglected higher order effects, such as the
terms proportional simultaneously to $\delta C$ and the NLO potential.

As was shown in this section, there are no power counting breaking contributions
in $P$-waves at NLO stemming from the iterations of the LO potential.
This means that $C_{2,i}$ is the renormalized quantity, where we assume that the divergent
contributions to the two-pion-exchange diagrams are subtracted within some scheme,
e.g. as is done for our choice of the non-polynomial two-pion-exchange contribution, 
see Eq.~\eqref{Eq:two_pion_exchange}.
Then, one obvious choice for the renormalization condition is
\begin{align}
 C_{2,i}=0.
\end{align}

However, at higher orders, power counting breaking terms will appear also in $P$-waves,
and one will have to absorb them by performing renormalization of the same contact interaction.
Therefore, to be consistent with our subtraction scheme for the $S$-waves,
we impose the renormalization condition on $C_i$ and $C_{2,i}$
by requiring that the NLO amplitude in channels with $l=1$ vanishes
at threshold faster than $p_\text{on}^2$:
\begin{align}
 (T_2)_{11}(p_\text{on})/p_\text{on}^2\Big|_{p_\text{on}=0}=0.
 \label{Eq:condition_Pwave}
\end{align}
Instead of the threshold point $p_\text{on}=0$, one can also take another 
renormalization point below or above threshold within the applicability of our approach.

A potential problem related to the above renormalization condition 
was discussed in great detail in Ref.~\cite{Gasparyan:2022isg} when studying schemes with 
large or infinite cutoffs.
It arises near ``exceptional'' cutoff values for which the contribution of the contact interaction to the NLO amplitude
is unnaturally small:
\begin{align}
\big(\bar R V_{C_i,\text{short},\Lambda} R\big)(p_\text{on})/p_\text{on}^2\Big|_{p_\text{on}=0}\approx 0,
\label{Eq:Pwave_zero}
\end{align}
which, in turn, leads to an unnaturally large value of $C_{2,i}$.
In such a case, the power counting is violated unless the zero of the 
function on the left-hand side of Eq.~\eqref{Eq:Pwave_zero} is factorizable (i.e., it appears at all energies).
The condition in Eq.~\eqref{Eq:Pwave_zero} can take place, e.g., in the spin-triplet channels with 
attractive one-pion-exchange potential such as $^3P_0$ if the adopted cutoff value is too large.
Then one starts to feel the singular nature of the one-pion-exchange potential, 
which is reflected in oscillations of the scattering wave function at short distances.
Note that this effect does not directly correspond to the appearance of spurious bound states,
although the two issues are related to each other.

In Ref.~\cite{Gasparyan:2022isg}, several particular cases were discussed when 
the condition in Eq.~\eqref{Eq:Pwave_zero} can be avoided
or the corresponding zero is factorizable.
However, we are interested in the general case, in which the practical solution
of the problem would be to explicitly verify that the LO potential is chosen in such a way that 
the condition in Eq.~\eqref{Eq:Pwave_zero} is not fulfilled.
In such a case, the NLO amplitude will satisfy the expected power counting.
In fact, for the regulators mentioned in
the discussion in Sec.~\ref{Sec:results} and many other choices tested by us,
if the cutoff value is of the order of the hard scale,
Eq.~\eqref{Eq:Pwave_zero} is never fulfilled.
A simple indication that the cutoff of the LO potential 
is not ``exceptional'' is the naturalness of the renormalized NLO 
low energy constants.

To summarize, we have shown that the $P$-wave NLO amplitudes formally satisfy the dimensional power counting
in the non-perturbative regime.
This holds also for the case when a contact interaction quadratic in momenta is promoted to
LO if one makes sure that a certain condition on the LO potential is satisfied.

\newpage
\section{\texorpdfstring{Non-perturbative renormalization of the amplitude at NLO. $S$-waves.}
{Non-perturbative renormalization of the amplitude at NLO. S-waves.}}
\label{Sec:Subtractions}
In this section we consider the renormalization of the NLO amplitude in the non-perturbative
regime for $S$-waves. As in the perturbative case considered in Ref.~\cite{Gasparyan:2021edy},
subtractions have to be made in order to absorb contributions that violate power counting.
We will start with generalizing the perturbative result of Ref.~\cite{Gasparyan:2021edy}
and then analyze under which conditions a particular power counting can be established.
\subsection{General formula}
Analogously to the situation discussed in Sec.~\ref{Sec:Contact_term_promotion},
there is freedom to choose the momentum-independent part of the NLO potential
\begin{align}
 \hat V_2(p',p)=V_2(0,0),
 \label{Eq:V2_hat_0}
\end{align}
because it can be partly or completely absorbed by the LO potential.
In the perturbative case, the NLO amplitude corresponding to
$\hat V_2$ does not contain any power counting breaking contributions in contrast
to the remaining part $\tilde T_2$ that is generated by 
\begin{align}
\tilde V_2(p',p)=V_2(p',p)-\hat V_2(p',p). 
\end{align}
In what follows, we will mostly consider the scheme with $ \hat V_2(p',p)=0$,
which is well suited for compensating possible threshold enhancement of the LO amplitude
due to non-perturbative effects. Alternative schemes will be briefly discussed separately.
Therefore, when using the results of Ref.~\cite{Gasparyan:2021edy}, we will assume
\begin{align}
 \tilde V_2=V_2,\quad \tilde T_2=T_2.
\end{align}

First, we recall some notation from Ref.~\cite{Gasparyan:2021edy}.
For an operator $X=X_{l'l}(p',p;p_\text{on})$, 
where $l$($l'$) is the initial (final) orbital angular momentum,
we define the subtraction operation $\mathds{T}$:
\begin{align}
 \mathds{T}(X)=X_{00}(0,0,0) V_\text{ct},
 \label{Eq:T_operation}
\end{align}
where the contact term is given by
\begin{align}
& V_\text{ct}=|\chi\rangle\langle\chi|,\nonumber\\
 &\langle p,lsj|\chi\rangle = \delta_{l,0}.\\
\end{align}
We assume that the counter term is unregulated or regulated with some $\Lambda_\text{ct}\gg\Lambda$.
Analogously, we introduce the subtraction operation $\mathds{T}^{m_i,n_i}$ for 
subdiagrams $(m_i,n_i)$ of the diagram $(m,n)$
corresponding to $T_2^{[m,n]}$.
We follow the Bogoliubov-Parasiuk-Hepp-Zimmermann (BPHZ)
subtraction scheme~\cite{Bogoliubov:1957gp,Hepp:1966eg,Zimmermann:1969jj}
and represent the renormalized amplitude via the forest formula:
\begin{align}
\mathds{R}( T_2^{[m,n]})= T_2^{[m,n]}+\sum_{U_k\in \mathcal{F}^{m,n}}
 \bigg(\prod_{(m_i,n_i)\in U_k} -\mathds{T}^{m_i,n_i}\bigg)  T_2^{[m,n]}\,,
\label{Eq:R_operation1}
\end{align}
where $\mathcal{F}^{m,n}$ represents the set of all forests, i.e, 
the set of all possible distinct sequences of nested subdiagrams $(m_i,n_i)$:
\begin{align}
&U_k=((m_{k;1},n_{k;1}),(m_{k;2},n_{k;2}),\dots)\,,\nonumber\\
& m\ge m_{k;i+1}\ge m_{k;i}\ge 0\,, \quad n\ge n_{k;i+1}\ge n_{k;i}\ge 0\,,
\quad n+m>0.
\label{Eq:R_operation2}
\end{align}
In Ref.~\cite{Gasparyan:2021edy}, it was proved that
each term in the expansion in $V_0$ of the
renormalized NLO amplitude satisfies the dimensional power counting
and is bounded by
\begin{align}
&\left|\mathds{R}( T_2^{[m,n]})(p_\text{on})\right|\le \frac{8\pi^2 \mathcal{M}_{T_2}}{m_N\Lambda_V} 
\Sigma_{2,0}^{m+n}
\frac{p_\text{on}^2}{\Lambda_{b}^2}
\, \ln{\frac{\Lambda}{M_\pi}}\,,
\label{Eq:Power_Counting_T2_m_n}
\end{align}
where 
\begin{align}
 \Sigma_{2,0}=2\mathcal{M}_{\text{max}}\frac{{\Lambda}}{\Lambda_V}
\end{align}
is a quantity of order one ($\Sigma_{2,0}\ge 1$ in the non-perturbative case).

To resum the series 
\begin{align}
 \mathds{R}( T_2)(p_\text{on})=\sum_{m,n=0}^\infty \mathds{R}( T_2^{[m,n]})(p_\text{on}),
\end{align}
we perform some rearrangement of Eq.~\eqref{Eq:R_operation1}, as explained below.

It is convenient to introduce the following notation:
\begin{align}
  &|\bar\psi\rangle =\bar R|\chi\rangle,\qquad
 \langle\psi| =\langle\chi|R,\nonumber\\
\psi_{l}(p;p_\text{on})&=\langle\psi|p,lsj\rangle = \langle p,lsj |\bar\psi\rangle.
\end{align}
For on-shell momenta $p=p_\text{on}$, the explicit form of $\psi_{l}$ reads
\begin{align}
 \psi_{l}(p_\text{on})\coloneqq \psi_{l}(p_\text{on};p_\text{on}) = \delta_{l,0}+\int \frac{p^2 dp}{(2\pi)^3} G(p;p_\text{on}) (T_{0})_{0,l}(p, p_\text{on};p_\text{on}),
  \label{Eq:psi_explicit}
\end{align}
and it coincides with the scattering wave function at the origin ($r=0$).

Now, consider the sum of all unrenormalized diagrams:
\begin{align}
  T_2 = \bar R V_2 R,
\end{align}
and perform first all single overall subtractions:
\begin{align}
\delta  T_2^{(1),\text{overall}}= - \mathds{T}( T_2) 
=- (T_2)_{00}(0,0;0)  |\chi\rangle  \langle\chi|,
\end{align}
where the superscript $(1)$ denotes the number of subtractions.

If we add all possible rescatterings with the LO potential,
we will obtain all terms with single subtractions in subdiagrams:
\begin{align}
& \delta T_2^{(1)} =\bar R\delta  T_2^{(1),\text{overall}}R=
 -  (T_2)_{00}(0,0;0)  \bar R|\chi\rangle  \langle\chi|R
=-  (T_2)_{00}(0,0;0) \rangle |\bar\psi\rangle  \langle\psi|\,.
\end{align}
Analogously, the sum of all double nested subtractions (one of which is an overall subtraction)
is given by
\begin{align}
\delta  T_2^{(2),\text{overall}}= - \mathds{T}\left(\delta T_2^{(1)} - \delta  T_2^{(1),\text{overall}} \right)
= (T_2)_{00}(0,0;0) 
\left[\psi_0(0)^2-1\right]|\chi\rangle  \langle\chi|
=-\left[\psi_0(0)^2-1\right]\delta  T_2^{(1),\text{overall}},
\end{align}
where the constant term $\delta  T_2^{(1)} $ was already subtracted in the previous step and should be excluded.
All terms with double nested subtractions in subdiagrams are obtained by adding the
rescattering contributions:
\begin{align}
& \delta T_2^{(2)}=\bar R\delta  T_2^{(2),\text{overall}}R=-\left[\psi_0(0)^2-1\right]\delta  T_2^{(1)}\,.
\end{align}
Continuing with further multiple nested subtractions, we obtain recursion relations:
\begin{align}
\delta  T_2^{(n+1),\text{overall}} &= - \mathds{T}\left( \delta T_2^{(n)} - \delta  T_2^{(n),\text{overall}} \right) \nonumber\\
 &=-\left[\psi_0(0)^2-1\right]\delta  T_2^{(n),\text{overall}},
\end{align}
and
\begin{align}
&\delta  T_2^{(n+1)} =-\left[\psi_0(0)^2-1\right]\delta  T_2^{(n)},
\end{align}
where the superscripts $(n)$ and $(n+1)$ denote the number of nested subtractions.
The terms $T_2^{(n)}$ can be summed up to
\begin{align}
 \delta  T_2 &= \sum_{n=1}^\infty \delta  T_2^{(n)}
 =\delta  T_2^{(1)}\sum_{n=0}^\infty \left[1-\psi_0(0)^2\right]^n\nonumber\\
 &=\delta  T_2^{(1)}\frac{1}{\psi_0(0)^2}.
\end{align}
Finally,
\begin{align}
\mathds{R}( T_2)   =  
T_2+\delta T_2=
T_2 - \frac{ (T_2)_{00}(0,0;0)}{\psi_0(0)^2}
|\bar\psi\rangle  \langle\psi|\,.
\label{Eq:nonperturbative_subtraction}
\end{align}
Taking the on-shell matrix elements of $\mathds{R}( T_2) $, we obtain:
\begin{align}
\mathds{R}( T_2)_{l'l}(p_\text{on})   =  (T_2)_{l'l}(p_\text{on})
+\delta C\psi_{l'}(p_\text{on})\psi_l(p_\text{on}),
\label{Eq:nonperturbative_subtraction_ij}
\end{align}
with the counter term constant
\begin{align}
\delta C=
-  \frac{(T_2)_{00}(0)}{\psi_0(0)^2}\,.
\label{Eq:delta_C_a}
\end{align}
Equation~\eqref{Eq:nonperturbative_subtraction_ij} can also be 
obtained directly without referring to the perturbative result from 
the renormalization condition:
\begin{align}
\mathds{R}( T_{2})_{l'l}(0)=0.
\label{Eq:renormalization_condition_Swave}
\end{align}
Therefore, the perturbative and non-perturbative results
match in the regime where both are applicable.

Similarly to the analysis of higher partial waves in Sec.~\ref{Sec:Pwaves},
we use the Fredholm decomposition of the resolvent of 
the LS equation and introduce
the quantities $N_2$ and $\nu$,
\begin{align}
 & (T_2)_{l'l}(p_\text{on})=\frac{ (N_2)_{l'l}(p_\text{on})}{D(p_\text{on})^2},\qquad
 \psi_{l}(p_\text{on})=\frac{\nu_{l}(p_\text{on})}{D(p_\text{on})}.
 \label{Eq:definitions_N2_n}
\end{align}
The counter term constant can be expressed as
\begin{align}
\delta C=
-  \frac{(N_2)_{00}(0)}{\nu_0(0)^2}.
\label{Eq:delta_C}
\end{align}
Then, the renormalized amplitude $\mathds{R}( T_{2})$ reads
\begin{align}
\mathds{R}( T_{2})_{l'l}(p_\text{on})
&= \frac{1}{D(p_\text{on})^2}
\Big[
 (N_2)_{l'l}(p_\text{on})
+\delta C\,\nu_{l'}(p_\text{on})\nu_l(p_\text{on})\Big]\nonumber\\
 &=\frac{\mathds{R}(N_2)_{l'l}(p_\text{on})}{D(p_\text{on})^2}
 =\frac{\mathds{R}(\tilde N_2)_{l'l}(p_\text{on})}{D(p_\text{on})^2\,\nu_0(0)^2}
 ,
\label{Eq:R_T2_nonperturbative}
\end{align}
where, for convenience, the following quantities have been introduced:
\begin{align}
\mathds{R}(N_2)_{l'l}(p_\text{on})&=(N_2)_{l'l}(p_\text{on})
 +  \delta C\,\nu_{l'}(p_\text{on})\nu_l(p_\text{on}) ,\label{Eq:R_N2}\\
\mathds{R}(\tilde N_2)_{l'l}(p_\text{on})
&=  (N_2)_{l'l}(p_\text{on})\nu_0(0)^2
 -  (N_2)_{00}(0)\nu_{l'}(p_\text{on})\nu_l(p_\text{on}).
\label{Eq:R_N2_tilde}
\end{align}

\subsection{\texorpdfstring{Power counting with the naturalness condition for $\nu_0(0)$}
{Power counting with the naturalness condition for nu0(0)}}
\label{Sec:natural_case}
In this subsection we analyze the expression for the 
renormalized NLO amplitude $\mathds{R}( T_{2})$ in Eq.~\eqref{Eq:R_T2_nonperturbative}
and determine what power counting it satisfies under which conditions.
Considering different constraints on various quantities entering
$\mathds{R}( T_{2})$, we can understand to 
what extent the renormalizability of the amplitude 
depends on details of the short-range dynamics.

We assume that the Fredholm determinant 
$D(p_\text{on})$ satisfies the bound in Eq.~\eqref{Eq:MDmin_S_waves},
which includes also the case of a shallow (quasi-) bound state.
For the function $D(p_\text{on})^2$, we can write
\begin{align}
 |D(p_\text{on})^2|&\ge \mathcal{M}_{D,\text{min}} ^2\kappa^2,
 \label{Eq:Dsquare1}
\end{align}
or, if $\kappa$ is very small,
\begin{align}
 |D(p_\text{on})^2|&\ge \mathcal{M}_{D,\text{min}} ^2\frac{p_\text{on}^2}{\Lambda_V^2}.
 \label{Eq:Dsquare2}
\end{align}
We will also need the upper bound for the quantity $\nu_l(p_\text{on})$, see Eq.~\eqref{Eq:bound_n_l}:
 \begin{align}
  \nu_l(p_\text{on})\le\mathcal{M}_{\nu}.
  \label{Eq:bound_n_l_text}
 \end{align}

First, we consider the ``natural'' case when the quantity $\nu_0(0)$
is bounded not only from above as in Eq.~\eqref{Eq:bound_n_l_text},
but also from below by some constant of order one:
 \begin{align}
  \nu_0(0)\ge\mathcal{M}_{\nu,\text{min}},
  \label{Eq:bound_n_0_min}
 \end{align}
which also implies the natural value of the counter term constant $\delta C$, see Eq.~\eqref{Eq:delta_C},
similarly to the condition of the absence of ``exceptional'' cutoffs in Sec.~\ref{Sec:Contact_term_promotion}.
Then as follows from Eq.~\eqref{Eq:R_T2_nonperturbative},
to analyze the power counting that the renormalized amplitude $\mathds{R}( T_{2})$
satisfies, it is sufficient to 
find bounds for $\mathds{R}(\tilde N_2)$.

As we show in Appendix~\ref{Sec:Bounds_N2tilde_nu},
the quantity $\mathds{R}(\tilde N_2)$
can be expanded into a convergent series in terms of $V_0$:
\begin{align}
 \mathds{R}(\tilde N_{2})(p_\text{on})&=\sum_{m,n=0}^\infty \left[ \mathds{R}(\tilde N_{2})(p_\text{on}) \right]^{[m,n]}\nonumber\\
 &=\sum_{m,n=0}^{n_\text{max}}\left[ \mathds{R}(\tilde N_{2})(p_\text{on}) \right]^{[m,n]}
 + \delta_{n_\text{max}} \left[\mathds{R}(\tilde N_2)(p_\text{on})\right]\nonumber\\
 &\eqqcolon S_{\tilde N_2, n_\text{max}}(p_\text{on})
 + \delta_{n_\text{max}} \left[\mathds{R}(\tilde N_2)(p_\text{on})\right],
 \label{Eq:series_N2_tilde }
\end{align}
and the remainder $\delta_n \left[\mathds{R}(\tilde N_2)(p_\text{on})\right] $
decreases faster than exponential with any base $\mathcal{M}_{\delta \tilde N_2}$
starting with some $n=\mathcal{\tilde M}_{\delta \tilde N_2}$
(see Eq.~\eqref{Eq:delta_tildeN2_exponential}):
 \begin{align}
|\delta_n [\mathds{R}(\tilde N_2)]| 
\le  \frac{8\pi^2}{m_N\Lambda_V}\mathcal{N}_{\tilde N_2} e^{-\mathcal{M}_{\delta \tilde N_2}n},\qquad\text{ for }
 n>\mathcal{\tilde M}_{\delta \tilde N_2}.
 \label{Eq:delta_tildeN2_exponential_text}
\end{align}

The prefactor $\mathcal{N}_{\tilde N_2}$ is given by
\begin{align}
 \mathcal{N}_{\tilde N_2}=\frac{\Lambda^2}{\Lambda_b^2}
\ln\frac{\Lambda}{M_\pi}
\end{align}
in the case of the ``standard'' regulators of the LO potential.
For the ``mild'' regulator, 
it depends also on the regulator of the NLO potential $\Lambda_\text{NLO}$:
\begin{align}
 \mathcal{N}_{\tilde N_2}=\frac{\Lambda\Lambda_\text{NLO}}{\Lambda_b^2}
 \ln\frac{\Lambda_\text{NLO}}{\Lambda}\ln\frac{\Lambda_\text{NLO}}{M_\pi},
\end{align}
and, in contrast to the perturbative regime,
the regulator $\Lambda_\text{NLO}$
cannot be set to infinity (in general) but can be chosen $\Lambda_\text{NLO}\gg\Lambda$.
Note that we do not consider the choice $\Lambda_\text{NLO}\sim\Lambda$
for the ``mild'' LO regulator because in such a case, we would simply
reproduce the variant with the ``standard'' regulators.
The appearance of $\Lambda_\text{NLO}$ in the expression for $\mathcal{N}_{\tilde N_2}$
is an indication of a potentially stronger cutoff dependence
of the NLO amplitude in the non-perturbative regime.

The general conservative estimate for $\mathcal{\tilde M}_{\delta \tilde N_2}$
yields $\mathcal{\tilde M}_{\delta \tilde N_2}\gtrsim (e\Sigma)^2$, which is rather large.
In realistic calculations, it turns out to be much smaller, 
see the discussion in Sec.~\ref{Sec:bound_Frehdolm_determinant}
and the numerical results in Sec.~\ref{Sec:results}.

On the other hand, expanding Eq.~\eqref{Eq:R_T2_nonperturbative} in $V_0$
gives
\begin{align}
 \left[ \mathds{R}(\tilde N_{2})(p_\text{on}) \right]^{[m,n]}&=
 \sum_{m_1=0}^m
 \sum_{m_2=0}^{m-m_1}
 \sum_{n_1=0}^n
 \sum_{n_2=0}^{n-n_1}
 D^{[m-m_1-m_2]}(p_\text{on})D^{[n-n_1-n_2]}(p_\text{on})\nonumber\\
 &\times \nu_0(0)^{[m_2]} \nu_0(0)^{[n_2]}
 \mathds{R}(T_2^{[m_1,n_1]})(p_\text{on}).
\end{align}

Using the perturbative bounds on $\mathds{R}(T_2^{[m,n]})$ in Eq.~\eqref{Eq:Power_Counting_T2_m_n}
and Eqs.~\eqref{Eq:MD},~and~\eqref{Eq:bound_n_l_text}, we obtain
\begin{align}
 \left[ \mathds{R}(\tilde N_{2})(p_\text{on}) \right]^{[m,n]}&\le
 \mathcal{M}_{D}^2 \mathcal{M}_{\nu}^2
 \sum_{m_1=0}^m\sum_{n_1=0}^n
\left|\mathds{R}(T_2^{[m_1,n_1]})(p_\text{on})\right|\nonumber\\
&\le \frac{8\pi^2 \mathcal{M}_{T_2}\mathcal{M}_{D}^2 \mathcal{M}_{\nu}^2}{m_N\Lambda_V} 
\frac{p_\text{on}^2}{\Lambda_{b}^2}
\, \ln{\frac{\Lambda}{M_\pi}}
\sum_{m_1=0}^m\sum_{n_1=0}^n
\Sigma_{2,0}^{m_1+n_1}.
\end{align}
Performing the summation up to $n=n_\text{max}$, we obtain
\begin{align}
 \left|S_{\tilde N_2, n_\text{max}}(p_\text{on})\right|
  &\le\frac{8\pi^2 \mathcal{M}_{T_2}\mathcal{M}_{ {D}^{[n]},\text{max}}^2 }{m_N\Lambda_V} 
 \frac{p_\text{on}^2}{\Lambda_{b}^2}
\, \ln{\frac{\Lambda}{M_\pi}}
\sum_{m,n=0}^{n_\text{max}}\sum_{m_1=0}^m\sum_{n_1=0}^n
\Sigma_{2,0}^{m_1+n_1}\nonumber\\
&\le \frac{8\pi^2 \mathcal{M}_{N_2;2}}{m_N\Lambda_V}\frac{p_\text{on}^2}{\Lambda_{b}^2}
\, \ln{\frac{\Lambda}{M_\pi}} n_\text{max}^4\Sigma_{2,0}^{2n_\text{max}}\nonumber\\
&\eqqcolon \frac{8\pi^2 \mathcal{M}_{S}}{m_N\Lambda_V}\frac{p_\text{on}^2}{\Lambda_{b}^2}\Phi_\text{log}.
\label{Eq:sum_nmax}
\end{align}
Given that the remainder $\delta_n [\mathds{R}(\tilde N_2)] $
can be made arbitrarily small by choosing a sufficiently large $n_\text{max}$,
e.g.
\begin{align}
|\delta_n [\mathds{R}(\tilde N_2)]|\le \frac{8\pi^2}{m_N\Lambda_V}\frac{M_\pi^2 \kappa^2}{\Lambda_b^2} , 
\label{Eq:delta_n_RN2tilde}
\end{align}
whereas the sum in Eq.~\eqref{Eq:sum_nmax} has the bound similar
to the one for the perturbative amplitude up to numerical constants of order one
and possible factors logarithmic in $\Lambda$, $\Phi_\text{log}$, we can conclude that $\mathds{R}(\tilde N_2)$
is bounded as:
\begin{align}
&\left|\mathds{R}(\tilde N_2)(p_\text{on})\right|\le \frac{8\pi^2 \mathcal{M}_{\tilde N_2}}{m_N\Lambda_V} 
\left[\frac{p_\text{on}^2}{\Lambda_{b}^2}\Phi_\text{log}
\, 
+\frac{M_\pi^2}{\Lambda_{b}^2}\kappa^2 \right].
\label{Eq:Power_Counting_N2_tilde}
\end{align}
Whether this picture
is indeed realized for the realistic NN interaction,
i.e., whether $\mathcal{M}_{\tilde N_2}$ is really (and not only formally) of the order of one,
is straightforward to verify by explicit numerical checks of the series for $\mathds{R}(\tilde N_{2})$
as we do partly in Sec.~\ref{Sec:results}.

For completeness, we show below that Eq.~\eqref{Eq:Power_Counting_N2_tilde}
holds formally in the chiral limit, i.e. for the expansion parameter $Q\ll 1$.
What we have to prove is that there exists such a value of $n_\text{max}$
that the remainder $\delta_{n_\text{max}} [\mathds{R}(\tilde N_2)] $
satisfies Eq.~\eqref{Eq:delta_n_RN2tilde}, and, at the same time,
the prefactor 
\begin{align}
\chi=n_\text{max}^4\Sigma_{2,0}^{2 n_\text{max}} 
\end{align}
in Eq.~\eqref{Eq:sum_nmax}
does not contain inverse powers of $Q$ and, therefore, does not destroy the power counting.

The choice
\begin{align}
 n_\text{max}\ge\max(k_0,\bar k_0),
 \end{align}
with
\begin{align}
 k_0=\mathcal{\tilde M}_{\delta \tilde N_2},\qquad
 \bar k_0=-\frac{1}{\mathcal{M}_{\delta \tilde N_2}}\ln \frac{M_\pi^2 \kappa^2}{\Lambda_b^2 \mathcal{N}_{\tilde N_2}},
\end{align}
guarantees that Eq.~\eqref{Eq:delta_n_RN2tilde} holds, as follows from Eq.~\eqref{Eq:delta_tildeN2_exponential_text}.
Note that the inequality $\bar k_0>k_0$ holds only for extremely small $Q=M_\pi/\Lambda_b$.
However, in the actual calculations, this can happen also for physical values of $Q$.

The factor $\chi$ is then given by
\begin{align}
 \chi=\mathcal{\tilde M}_{\delta \tilde N_2}^4 \Sigma_{2,0}^{2\mathcal{\tilde M}_{\delta \tilde N_2}}
\end{align}
if $n_\text{max}=k_0$,
and by
\begin{align}
 \chi=\frac{1}{\mathcal{M}_{\delta \tilde N_2}^4}\left(\ln \frac{M_\pi^2 \kappa^2}{\Lambda_b^2 \mathcal{N}_{\tilde N_2}}\right)^4 
 \left( \frac{M_\pi^2 \kappa^2}{\Lambda_b^2
 \mathcal{N}_{\tilde N_2}} \right)^{-2\frac{\ln\Sigma_{2,0}}{\mathcal{M}_{\delta \tilde N_2}}},
\end{align}
if $n_\text{max}=\bar k_0$.
 In the latter case, if $\mathcal{M}_{\delta \tilde N_2}$ is chosen to be 
$\mathcal{M}_{\delta \tilde N_2}\gg \ln\Sigma_{2,0}$, the factor
$ \left( \frac{M_\pi^2 \kappa^2}{\Lambda_b^2
 \mathcal{N}_{\tilde N_2}} \right)^{-2\frac{\ln\Sigma_{2,0}}{\mathcal{M}_{\delta \tilde N_2}}}$ can be neglected.

Thus, we conclude that Eq.~\eqref{Eq:Power_Counting_N2_tilde} holds with
 \begin{align}
 \Phi_\text{log}=\left\{\begin{array}{ll}\mathcal{M}_\text{log}\ln\frac{\Lambda}{M_\pi},&n_\text{max}=k_0\\[6pt]
  \mathcal{M}_\text{log}\ln\frac{\Lambda}{M_\pi} \left(\ln \frac{M_\pi^2 \kappa^2}{\Lambda_b^2 \mathcal{N}_{\tilde N_2}}\right)^4,
  \qquad&n_\text{max}=\bar k_0\end{array}\right.
 \label{Eq:a_j}
 \end{align}
 
Now we come back to the expression for the renormalized NLO amplitude in Eq.~\eqref{Eq:R_T2_nonperturbative}.
For small on-shell momenta $p_\text{on}$, i.e., when 
 \begin{align}
   \left |S_{\tilde N_2, n_\text{max}}(p_\text{on})\right|\le|\delta_n [\mathds{R}(\tilde N_2)]|,
 \end{align}
Eqs.~\eqref{Eq:Dsquare1},~\eqref{Eq:bound_n_0_min} and~\eqref{Eq:delta_n_RN2tilde} give:
\begin{align}
 \left|\mathds{R}( T_{2})_{l'l}(p_\text{on})\right|&\le
 \frac{8\pi^2\mathcal{M}_{T_2,\text{low}}}{m_N\Lambda_V}\frac{M_\pi^2}{\Lambda_b^2} , 
 \label{Eq:bound_T2_low}
\end{align}
which means that in this energy region, $\mathds{R}( T_{2})$
is of order $O(Q^2)$.

As the on-shell momentum increases, i.e.,
\begin{align}
   \left |S_{\tilde N_2, n_\text{max}}(p_\text{on})\right|\ge|\delta_n [\mathds{R}(\tilde N_2)]|,
\end{align}
we should use Eq.~\eqref{Eq:sum_nmax} instead of Eq.~\eqref{Eq:delta_n_RN2tilde}
to obtain
\begin{align}
 \left|\mathds{R}( T_{2})_{l'l}(p_\text{on})\right|&\le
 \frac{8\pi^2\mathcal{M}_{T_2,\text{high}}}{m_N\Lambda_V}\frac{p_\text{on}^2}{\Lambda_b^2}\frac{\Phi_\text{log}}{\kappa^2} , 
 \label{Eq:bound_T2_high1}
\end{align}
which is enhanced compared to $O(Q^2)$ by a factor $1/\kappa^2$.
In the worst case of the unitary limit, we obtain from Eq.~\eqref{Eq:Dsquare2}:
\begin{align}
 \left|\mathds{R}( T_{2})_{l'l}(p_\text{on})\right|&\le
 \frac{8\pi^2\mathcal{M}_{T_2,\text{high}}}{m_N\Lambda_V}\frac{\Lambda_V^2}{\Lambda_b^2}\Phi_\text{log},
  \label{Eq:bound_T2_high2}
\end{align}
which corresponds effectively to $\mathds{R}( T_{2})\sim O(Q^0)$.
This is still one order higher than the LO amplitude $O(Q^{-1})$, see Eq.~\eqref{Eq:T_0_enhanced2},
but the convergence rate is rather low in this case.
A natural way to reduce the effect of the numerical enhancement of the LO amplitude
and to improve convergence 
is to promote some part of the NLO potential to leading order, which
will make the numerical constant $\mathcal{M}_{T_2,\text{high}}$ smaller.
The simplest recipe would be to promote the contact interactions quadratic in momentum.
As already mentioned, this approach is suggested, e.g., for the $^1S_0$ partial wave.
We will discuss this possibility in Sec.~\ref{Sec:Contact_term_promotion_2}.

\subsection{Local LO potential in a spin-singlet channel and analogous cases}
\label{Sec:Local_LO}
Above, we considered the general case of the LO potential under an additional assumption
on its short-range part formulated in Eq.~\eqref{Eq:bound_n_0_min}
in terms of the naturalness of $\nu_0(0)$.
It is instructive to consider one particular case, when the LO potential in a spin-singlet channel is fully local.
Then, this condition is satisfied automatically. 
Moreover, for a local LO potential, the following identity holds:
\begin{align}
 \nu_0(p_\text{on})\equiv 1,
\end{align}
which follows from the fact that 
the scattering wave function at the origin 
$\psi_{p_\text{on}}$ coincides with 
the inverse of the Jost function $f(p_\text{on})$
and the inverse of the Fredholm determinant \cite{Newton:1982qc}:
\begin{align}
 \psi(p_\text{on})=f(p_\text{on})^{-1}=D(p_\text{on})^{-1},
\end{align}
and the definition~\eqref{Eq:definitions_N2_n}.
Therefore, we have (see the definitions in Eqs.~\eqref{Eq:R_N2_tilde} and~\eqref{Eq:R_N2})
\begin{align}
\mathds{R}(\tilde N_{2})(p_\text{on})=\mathds{R}(N_{2})(p_\text{on})=
\Delta N_{2}(p_\text{on}) = N_{2}(p_\text{on}) -  N_{2}(0).
\label{Eq:RN2_local}
\end{align}

The whole discussion in the previous subsection applies for the case of a local LO potential,
except the absence of the additional condition~\eqref{Eq:bound_n_0_min}.
In the general case, when the constraint in Eq.~\eqref{Eq:bound_n_0_min} is not satisfied,
we still can have a situation similar to the local single-channel potential
if we assume that the series for $\mathds{R}(N_2)$ (not for $\mathds{R}(\tilde N_2)$)
converges and the bound analogous to Eq.~\eqref{Eq:Power_Counting_N2_tilde} holds:
\begin{align}
&\left|\mathds{R}(N_2)(p_\text{on})\right|\le \frac{8\pi^2 \mathcal{M}_{\tilde N_2}}{m_N\Lambda_V} 
\left[\frac{p_\text{on}^2}{\Lambda_{b}^2}\Phi_\text{log}
\, 
+\frac{M_\pi^2}{\Lambda_{b}^2}\kappa^2 \right].
\label{Eq:Power_Counting_N2}
\end{align}
This is possible if the smallness of $\nu_0(0)$ in the denominator of $\mathds{R}(N_2)$
is compensated by a corresponding small factor in the numerator, see~Eq.~\eqref{Eq:R_N2}.
Whether this indeed takes place can be verified numerically 
in any particular case.
From Eq.~\eqref{Eq:Power_Counting_N2}, we can deduce the same bounds for the
renormalized NLO amplitude as in Eqs.~\eqref{Eq:bound_T2_low},~\eqref{Eq:bound_T2_high1}~and~\eqref{Eq:bound_T2_high2}.

We made this comment to emphasize that the naturalness constraint on $\nu_0(0)$
is not necessary to guarantee renormalizability of the NLO amplitude,
but is the most simple one from the practical point of view.

\subsection{Promoting a momentum dependent contact term to leading order}
\label{Sec:Contact_term_promotion_2}
In this subsection we analyze the situation when it is necessary
to promote the momentum dependent $S$-wave contact term to leading order.
For definiteness, we consider the $^1S_0$ partial wave, 
where such a promotion has been shown to significantly improve the 
convergence of the chiral EFT expansion,
see Refs.~\cite{Epelbaum:2015sha,Long:2012ve}.
Since this is a spin-singlet channel, we omit the $l$, $l'$ indices in this subsection.
We also omit all channel indices.

The whole analysis in the preceding subsections remains valid in this case,
except that similarly to the promotion of the subleading term in the  $P$-waves
considered in Sec.~\ref{Sec:Contact_term_promotion}, 
there is freedom choosing what part of such a contact term
should be included in LO potential $V_0$ and what part remains in the NLO potential $V_2$.

We rewrite Eq.~\eqref{Eq:R_N2} by explicitly separating the part with the contact term quadratic in momenta:
\begin{align}
\mathds{R}(N_{2})(p_\text{on})
&=N_{2}(p_\text{on})+\delta C \nu(p_\text{on})^2\nonumber\\
 &\eqqcolon \Delta N_{2}(p_\text{on})
 +\delta  C \nu(p_\text{on})^2
 +C_{2}N_{C_2}(p_\text{on}),
\label{Eq:R_T2_nonperturbative_2}
\end{align}
with
\begin{align}
 N_{C_2}(p_\text{on})=[\bar R V_{C} R](p_\text{on}) D(p_\text{on})^2.
\end{align}
The potential $V_{C}$ is the contact interaction quadratic in momenta
that projects onto the $^1S_0$ partial wave.
This potential can remain regulated because the regulator corrections
to it are of higher order.

Following our subtraction scheme at $p_\text{on}=0$,
we introduce two renormalization conditions to fix $\delta C$ and $C_2$:
\begin{align}
 \mathds{R}(N_2)(0)&=0,\nonumber\\
 \frac{d^2 \mathds{R}(N_2)(p_\text{on})}{d p_\text{on}^2}\bigg|_{p_\text{on}=0}&=0.
\label{Eq:N2_two_subtractions} 
 \end{align}
Note that $N_2$ is an analytic function of $p_\text{on}^2$ at $p_\text{on}=0$.

Of course, Eq.~\eqref{Eq:N2_two_subtractions} can be also formulated in terms
of the amplitudes:
\begin{align}
 \mathds{R}(T_2)(0)&=0,\nonumber\\
 \frac{d^2 \mathds{R}(T_2)(p_\text{on})}{d p_\text{on}^2}\bigg|_{p_\text{on}=0}&=0.
 \label{Eq:renormaization_2_subtractions}
 \end{align}
Analogously to the situation in $P$-waves, the above renormalization conditions
can lead to a problem for ``exceptional'' cutoffs when Eqs.~\eqref{Eq:N2_two_subtractions} 
become inconsistent, which happens
not only when $\nu(0)=0$ but also when the following equation is satisfied~\cite{Gasparyan:2022isg}:
\begin{align}
\bigg[ \frac{d^2 N_{C_2}(p_\text{on})}{d p_\text{on}^2}
 -2 N_{C_2}(p_\text{on})\nu(p_\text{on})\frac{d^2 \nu(p_\text{on})}{d p_\text{on}^2}\bigg]
 \bigg|_{p_\text{on}=0}=0.
\end{align}
As in the case of the $P$-waves, an indirect indication that the
cutoff is not close to an ``exceptional'' value is the naturalness of the NLO LECs.
In our numerical calculation in Sec.~\ref{Sec:results},
we found no ``exceptional'' cutoffs 
for the cutoff values of the order or below the hard scale.

\subsection{Other subtraction schemes}
In all analyses of the non-perturbative regime,
we have always adopted the prescription to perform subtractions at threshold,
see Eq.~\eqref{Eq:renormalization_condition_Swave}.
In this subsection we briefly discuss other possibilities.
Choosing different subtraction points, e.g., the deuteron pole position for the
$^3S_1-{^3D_1}$ channel, is equivalent to setting, in contrast to Eq.~\eqref{Eq:V2_hat_0},
$\hat V_2\ne 0$:
\begin{align}
 \hat V_2(p', p)= \hat\kappa^2 \frac{8\pi^2 }{m_N \Lambda_V}\frac{M_\pi^2}{\Lambda_{b}^2},
 \end{align}
where $\hat\kappa$ is a constant of order one, see Eq.~\eqref{Eq:bound_V_2_0_hat}.
Since this potential is just an $S$-wave contact term,
the corresponding NLO amplitude is given by 
\begin{align}
(\hat T_2)_{l'l}(p_\text{on})   =  
\hat\kappa^2 \frac{8\pi^2 }{m_N \Lambda_V}\frac{M_\pi^2}{\Lambda_{b}^2}
\psi_{l'}(p_\text{on})\psi_l(p_\text{on})
=  \hat\kappa^2 \frac{8\pi^2 }{m_N \Lambda_V}\frac{M_\pi^2}{\Lambda_{b}^2}
\frac{\nu_{l'}(p_\text{on})\nu_l(p_\text{on})}{D(p_\text{on})^2}.
\label{Eq:hat_T2}
\end{align} 
From Eqs.~\eqref{Eq:Dsquare1}~and~\eqref{Eq:bound_n_l_text},
we obtain the following bound:
\begin{align}
|(\hat T_2)_{l'l}(p_\text{on})|\le
   \frac{8\pi^2 }{m_N \Lambda_V}
  \frac{\mathcal{M}_{\nu}^2}{\mathcal{M}_{D,\text{min}} ^2}
  \frac{M_\pi^2}{\Lambda_{b}^2}\,
\frac{\hat\kappa^2}{\kappa^2}.
\label{Eq:hat_T2_bound}
\end{align} 
For the perturbative case considered in Ref.~\cite{Gasparyan:2021edy}
and for the case without an enhancement of the LO amplitude,
the amplitude $\hat T_2$ satisfies the dimensional power counting: $T_2\sim O(Q^2)$.
However, in the situation when the LO amplitude is enhanced, the additional factor
$\hat\kappa^2/\kappa$ in Eq.~\eqref{Eq:hat_T2_bound} relative to Eq.~\eqref{Eq:T_0_enhanced}
spoils convergence even at threshold.
We will have the worst situation in the unitary limit with $\kappa\ll 1$.

Thus, we conclude that for a reasonable convergence in the case of
an enhanced LO amplitude, one should choose a subtraction scheme
not much different from ours, i.e., such that $\hat\kappa/\kappa\sim 1$.

To summarize, we have shown that renormalization of the NLO amplitude
for the $S$-waves can be done explicitly also in the non-perturbative regime
by analyzing the Fredholm decomposition of the amplitudes.
In contrast to the perturbative case discussed in Ref.~\cite{Gasparyan:2021edy},
additional constraints on the LO potential have to be fulfilled
to ensure renormalizability and convergence of the chiral expansion.
Then, the power counting works also in the situation when the 
LO amplitude is enhanced at threshold, although to make the 
scheme more efficient, it might be necessary to promote
certain contributions to leading order.

\section{Numerical results}
\label{Sec:results}
In this section we illustrate our theoretical findings
by explicit numerical calculation of the NLO NN amplitude
in the three channels where the LO interaction should be
treated non-perturbatively: $^3P_0$, $^3S_1-{^3D}_1$ and $^1S_0$.
The results for other channels were presented in Ref.~\cite{Gasparyan:2021edy}.

We adopt the same values for the numerical constants as in Ref.~\cite{Gasparyan:2021edy}:
the pion decay constant $F_\pi=92.1$ MeV,
the isospin average nucleon and pion masses
$m_N=938.9$~MeV, $M_\pi=138.04$~MeV and 
the effective nucleon axial coupling constant $g_A = 1.29$.
The calculations have been performed using \emph{Mathematica} \cite{mathematica12.0}.

For the regularization of the LO and NLO potentials, we
adopt the scheme similar to the one used in realistic calculation in Ref.~\cite{Reinert:2017usi}
at fifth order in the chiral expansion,
which allows us to have a direct interpretation of the numerical 
values of the cutoffs.
In particular, we use the local Gaussian regulator for the one-pion-exchange potential
and the non-local Gaussian regulator for all contact interactions
with the same cutoff $\Lambda$, see Appendix~\ref{Sec:LOpotential}.
For the sake of simplicity, we also employ
the local Gaussian regulator in the form of the overall factor $F_{\Lambda_\text{NLO},\text{exp}}(q)$
for the two-pion-exchange potential.
As in Ref.~\cite{Gasparyan:2021edy}, the cutoff value $\Lambda_\text{NLO}$ 
is set to the hard scale $\Lambda_\text{NLO}=600$~MeV.
This choice for the chiral expansion breakdown scale is consistent
with the recent studies
in the few-nucleon sector \cite{Epelbaum:2014efa,Furnstahl:2015rha,Epelbaum:2019zqc,Epelbaum:2019wvf}.

The momentum-independent contact interactions at NLO are 
included without a regulator in accordance with our power counting.
The contact interactions quadratic in momenta
are regulated with the same cutoff
$\Lambda_\text{NLO}$ at LO and at NLO
in contrast to our choice in Ref.~\cite{Gasparyan:2021edy},
where, for simplicity,
we left the corresponding NLO contact terms unregulated.
Both options are legitimate since the regulator corrections to
the contact interactions quadratic in momenta is an effect of a higher order, $O(Q^4)$ .
By the same reason, the regulator corrections to the 
LO contact interactions quadratic in momenta are not taken into account.

The cutoff values for the one-pion-exchange potential and for the momentum-independent
LO contact interactions are varied in the regions below and above $\Lambda=450$~MeV,
which was found to be the optimal cutoff value in Ref.~\cite{Reinert:2017usi}.
The lower region corresponds to extremely soft cutoffs, 
where explicit regulator corrections to the LO potential
are likely to be important.
The upper region contains relatively hard (of the order of the hard scale) cutoffs
as well as cutoffs above $\Lambda_b$, for which we expect
slower convergence in terms of the Fredholm expansion
and, therefore, potential problems with interpretation
within our renormalization scheme.

The free parameters are determined by a fit to the empirical phase shifts
from the Nijmegen partial wave analysis~\cite{Stoks:1993tb} up to
$E_\text{lab}=150$~MeV. 
The phase shifts and the mixing parameters are calculated through the following 
unitarization procedure. First, the non-unitary NLO $T$-matrix is transformed to
the $S$-matrix via
\begin{align}
 S_{l'l}(p_\text{on})=1-i\frac{m_N p_\text{on}}{8\pi^2} T_{l'l}(p_\text{on}).
 \label{Eq:Smatrix}
\end{align}
The diagonal phase shifts in the Stapp parametrization of the $S$-matrix~\cite{Stapp:1956mz}
are determined as (modulo $\pi$)
\begin{align}
 \delta_{ll}=\frac{1}{2}\arg(S_{ll}),
\end{align}
whereas the mixing parameter $\epsilon_{l+1}$ is obtained from the off-diagonal element
of the $S$-matrix:
\begin{align}
 S_{l+2,l}=i\sin(2\epsilon_{l+1})\exp(i\delta_{l}+i\delta_{l+2}).
\end{align}
The dependence of the results on a particular unitarization scheme
is a higher-order effect, provided the chiral expansion for the amplitude is convergent.

The numerical analysis we perform does not aim at achieving 
a perfect description of the data as 
we work only at next-to-leading order in the chiral expansion.
Rather, we are interested in the convergence and renormalization
issues.
In particular, we make sure that for the cutoff values
we employ, no spurious bound states appear
and no ``exceptional'' cutoffs discussed in Secs.~\ref{Sec:Pwaves}~and~\ref{Sec:Subtractions}
lie within this range.
The latter fact manifests itself in the natural values of 
the fitted next-to-leading-order LECs.
The natural values of the NLO LECs are also an indication of
the ``naturalness'' of the quantity $\nu_0(0)$,
which is the simplest condition for the renormalizability of
the $S$-wave NLO amplitudes, see Sec.~\ref{Sec:natural_case}.
The natural size is roughly given by 
\begin{align}
\frac{8\pi^2 }{m_N \Lambda_b},
\label{Eq:natural_C_0}
\end{align}
for the LECs accompanying momentum-independent contact terms
and by
\begin{align}
 \frac{8\pi^2 }{m_N \Lambda_b^3},
\label{Eq:natural_C_2}
\end{align}
for the LECs of contact terms quadratic in momenta.
Obviously, naturalness is not a mathematically strict criterion.
However a sign of potential problems would be a rapid growth with cutoff
of one or several LECs.

Understanding the power counting for the renormalized
amplitudes in terms of the 
convergence of the Fredholm expansion
is demonstrated by looking at the convergence
of the Fredholm determinant expanded in terms of the LO potential.
Convergence of other elements of the Fredholm formulas for the LO and the NLO 
amplitudes can be analyzed in a similar manner.
Their convergence rates are typically comparable with 
the one for the Fredholm determinant.
An absolute value of Fredholm determinant much larger than $1$
is also a problem for our interpretation of the power counting,
especially for the channels with the enhanced LO amplitude.
In such a case, the numerators in the Fredholm formulas $N_0$, $N_2$
will also be very large, contradicting the power counting that we suggest.
On the contrary, we expect the absolute value of the Fredholm determinant for those channels to be smaller
than $1$.

\subsection{\texorpdfstring{$^3P_0$ channel}{3P0 channel}}
\begin{figure}[tb]
\includegraphics[width=\textwidth]{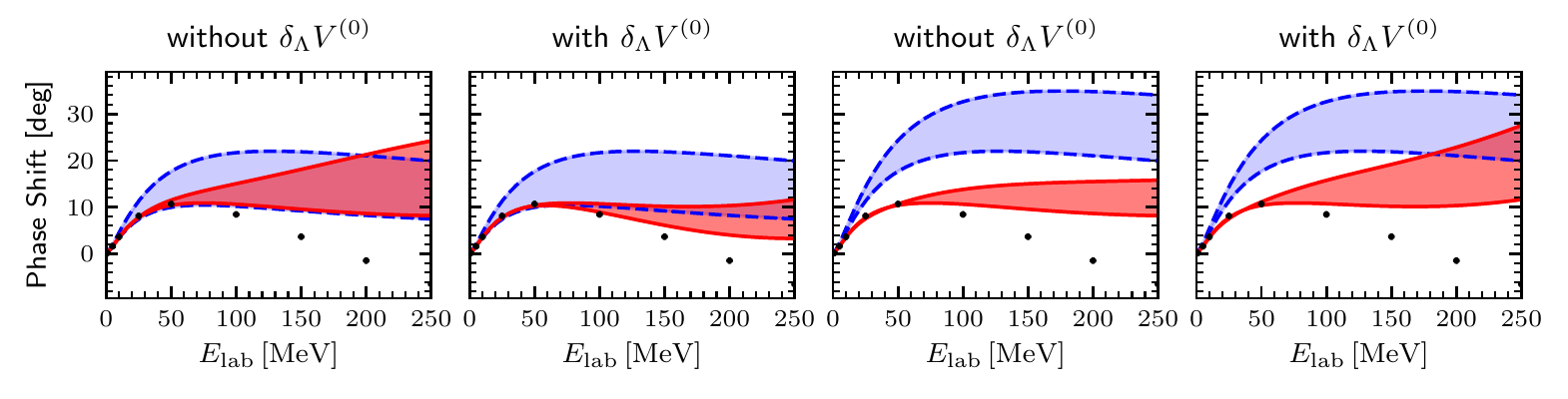}
\caption{The results of the leading-order (blue dashed lines)
and next-to-leading-order (red solid lines)  calculations for  
the $^3P_0$ partial wave without promoting the contact interaction.
The bands indicate the variation of the one-pion-exchange cutoff 
within the range $\Lambda_{1\pi}\in(300,450)$~MeV for two left plots
and within the range $\Lambda_{1\pi}\in(450,600)$~MeV for two right plots.
The second and fourth plots correspond to the NLO potential with the regulator
correction $\delta_\Lambda V^{(0)}$, while the results in the first and third plots
are obtained without this term. The empirical phase shifts shown
by black solid dots are from Ref.~\cite{Stoks:1993tb}.
The plots were created using Matplotlib \cite{Hunter:2007}.} \label{Fig:3P0_perturbative_LambdaC2_600MeV}
\end{figure}

We begin our discussion with the $^3P_0$ partial wave
and first follow the dimensional power counting.
That means that at leading order, we include only the one-pion-exchange potential
and no further terms are promoted.
At next-to-leading order, there is one free parameter $C_{2,^3P_0}$ that determines the strength of the
NLO contact interaction. 
The results for the LO and NLO calculations
are presented in Fig.~\ref{Fig:3P0_perturbative_LambdaC2_600MeV}.
In contrast to other plots in this section, we restrict ourselves to the 
values of the cutoff $\Lambda\le600$~MeV because for larger cutoffs, 
the calculated phase shifts deviate too strongly from the data points.

For soft cutoffs values below $\Lambda=450$~MeV, the convergence of the chiral EFT expansion
and the description of the data are reasonable.
Moreover for such cutoffs, the LO amplitude can be regarded as perturbative,
in the sense that the series in $V_0$ converges very rapidly,
and already a single iteration of the LO potential provides an accuracy of one percent.
Therefore, the analysis of Ref.~\cite{Gasparyan:2021edy} can be applied.
One can also see that the band for next-to-leading order corresponding to the variation of the 
cutoff gets considerably narrower if the regulator correction to the one-pion-exchange
potential is taken into account explicitly.
Further discussion of the fully perturbative approach in the $^3P_0$ channel
can be found in Refs.~~\cite{Birse:2005um,Wu:2018lai}.

As one increases the cutoff value, the convergence of expansion of the amplitude in powers of $V_0$
becomes much slower.
This is not problem for our formalism as we formulated the power counting
in the non-perturbative case in Sec.~\ref{Sec:Pwaves}.
However, as one can see in Fig.~\ref{Fig:3P0_perturbative_LambdaC2_600MeV},
the disagreement with the data gets more severe
and the convergence of the chiral EFT expansion deteriorates.
In fact, such a strong deviation of the LO phase shifts from the data
leads to a strong violation of unitarity.
Another indication of the inefficiency of the resulting EFT expansion is
a rather small value of the Fredholm determinant.
At threshold, it equals $D\sim 0.4$ for $\Lambda=600$~MeV compared
to $D\sim 1$ for $\Lambda=300-450$~MeV.

Large contributions from higher orders makes it more efficient to promote the NLO contact interaction to 
leading order, see also Refs.~\cite{Birse:2007sx}~and~\cite{Nogga:2005hy}.
In fact, the case of very soft cutoffs considered above, which shows a reasonable convergence of the chiral expansion,
can also be viewed as a modification of the short-range part of the LO potential analogous
to promotion of a contact interaction.
Note that our motivation for promoting the NLO contact term is 
not the requirement of the existence of an infinite cutoff limit as advocated, e.g., in Ref.~\cite{Nogga:2005hy},
but rather a large strength of the LO one-pion-exchange potential
in this channel.
Specifically, we demand that the difference between the LO results and empirical
values of the phase shifts can be corrected by a perturbative inclusion of 
higher-order interactions.

\begin{figure}[tb]
\includegraphics[width=\textwidth]{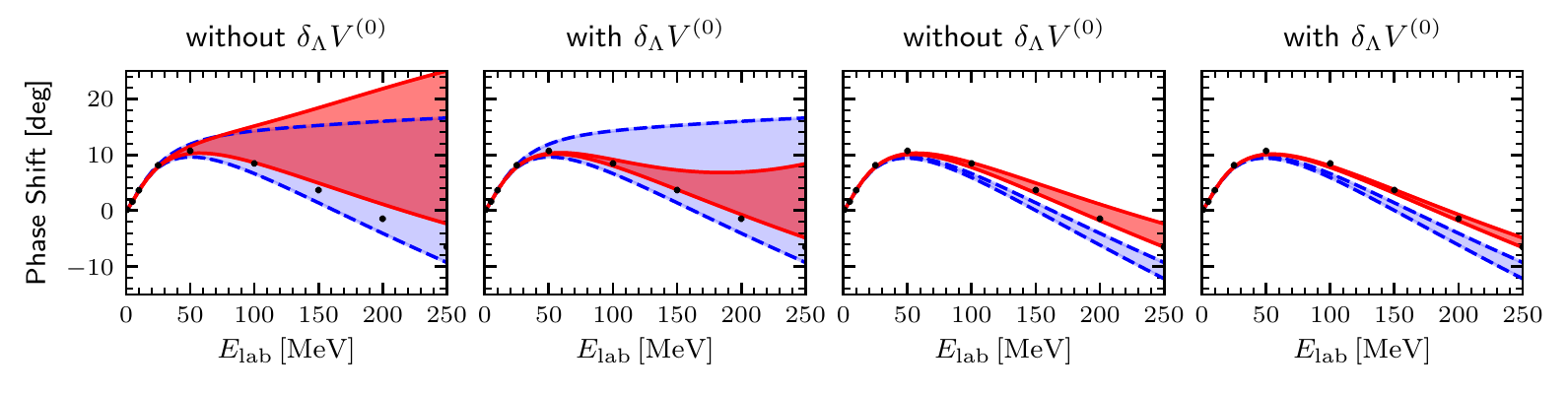}
\caption{The results of the leading-order (blue dashed lines)
and next-to-leading-order (red solid lines)  calculations for  
the $^3P_0$ partial wave with the contact term promoted to leading order.
The bands indicate the variation of the one-pion-exchange cutoff 
within the range $\Lambda_{1\pi}\in(300,450)$~MeV for two left plots
and within the range $\Lambda_{1\pi}\in(450,800)$~MeV for two right plots.
The second and fourth plots correspond to the NLO potential with the regulator
correction $\delta_\Lambda V^{(0)}$, while the results in the first and third plots
are obtained without this term. The empirical phase shifts shown
by black solid dots are from Ref.~\cite{Stoks:1993tb}.} \label{Fig:3P0_nonperturbative_LambdaC2_600MeV}
\end{figure}

In the scheme with a contact term at LO, there is also one free parameter to be determined from the fit,
namely $C_{^3P_0}$,
whereas the NLO constant $C_{2,^3P_0}$ is fixed by the renormalization condition in Eq.~\eqref{Eq:condition_Pwave}.
The corresponding results are shown in Fig.~\ref{Fig:3P0_nonperturbative_LambdaC2_600MeV}.
As one can see, the convergence pattern when going from LO to NLO becomes much better.
Taking into account the regulator correction to the one-pion-exchange potential
$\delta_\Lambda V^{(0)}$ explicitly leads to narrower cutoff-variation bands at NLO, especially for soft cutoffs.

The expansion of the Fredholm determinant in powers of $V_0$
converges rather rapidly for the cutoffs $\Lambda\le 600$~MeV:
at order $(V_0)^3$, a one-percent accuracy is achieved. 
For $\Lambda\sim800$~MeV, the same accuracy requires expansion up to order $(V_4)^4$.
The absolute value of the Fredholm determinant varies within the range $0.7-2.3$
increasing for higher values of the cutoff.
The numerical values of the constant $C_{2,^3P_0}$ in the units of Eq.~\eqref{Eq:natural_C_2}
is reasonably natural for the choice of the hard scale $\Lambda_b=600$~MeV
at least for lower $\Lambda$ values. Specifically,
$C_{2,^3P_0}\sim2$ for $\Lambda\sim450$~MeV
but increases to $C_{2,^3P_0}\sim30$ for $\Lambda\sim800$~MeV.

Combining the above results, we conclude that for the cutoffs below or of the order of the hard scale,
the renormalization of the NLO amplitude can be understood within the approach 
developed in this paper. For higher values of the cutoff, the renormalizability
of the theory becomes questionable.

\subsection{\texorpdfstring{$^3S_1-{^3D_1}$ channel}{3S1-3D1 channel}}
Next, we consider the system of the coupled $^3S_1-{^3D_1}$ partial waves.
The LO potential is obviously non-perturbative due to the presence of the
shallow deuteron bound state.
The enhancement of the LO amplitude at threshold is not as strong as, e.g.,
in the $^1S_0$ channel. Therefore, we assume that 
within the renormalization scheme specified in Eq.~\eqref{Eq:renormalization_condition_Swave},
the dimensional power counting should work.
That means that the LO potential contains only the one-pion-exchange 
and the momentum-independent contact term contributions.

There are three parameters to be determined from the fit: 
the LO constant $C_{^3S_1}$, the NLO constant at the diagonal contact term quadratic in momenta,
$C_{2,^3S_1,p^2}$, and the NLO constant accompanying the off-diagonal contact term $C_{2,\epsilon_1}$.
The NLO momentum-independent contact term with the constant $C_{2,^3S_1}$
is fixed from the renormalization condition in Eq.~\eqref{Eq:renormalization_condition_Swave}.
The above mentioned three parameters are determined by fitting
the phase shifts in the diagonal $^3S_1$ channel and 
the mixing parameter $\epsilon_1$, i.e., the channels with contact terms in the potential.
The $^3D_1$ phase shift comes out as a prediction.

\begin{figure}[tb]
\includegraphics[width=\textwidth]{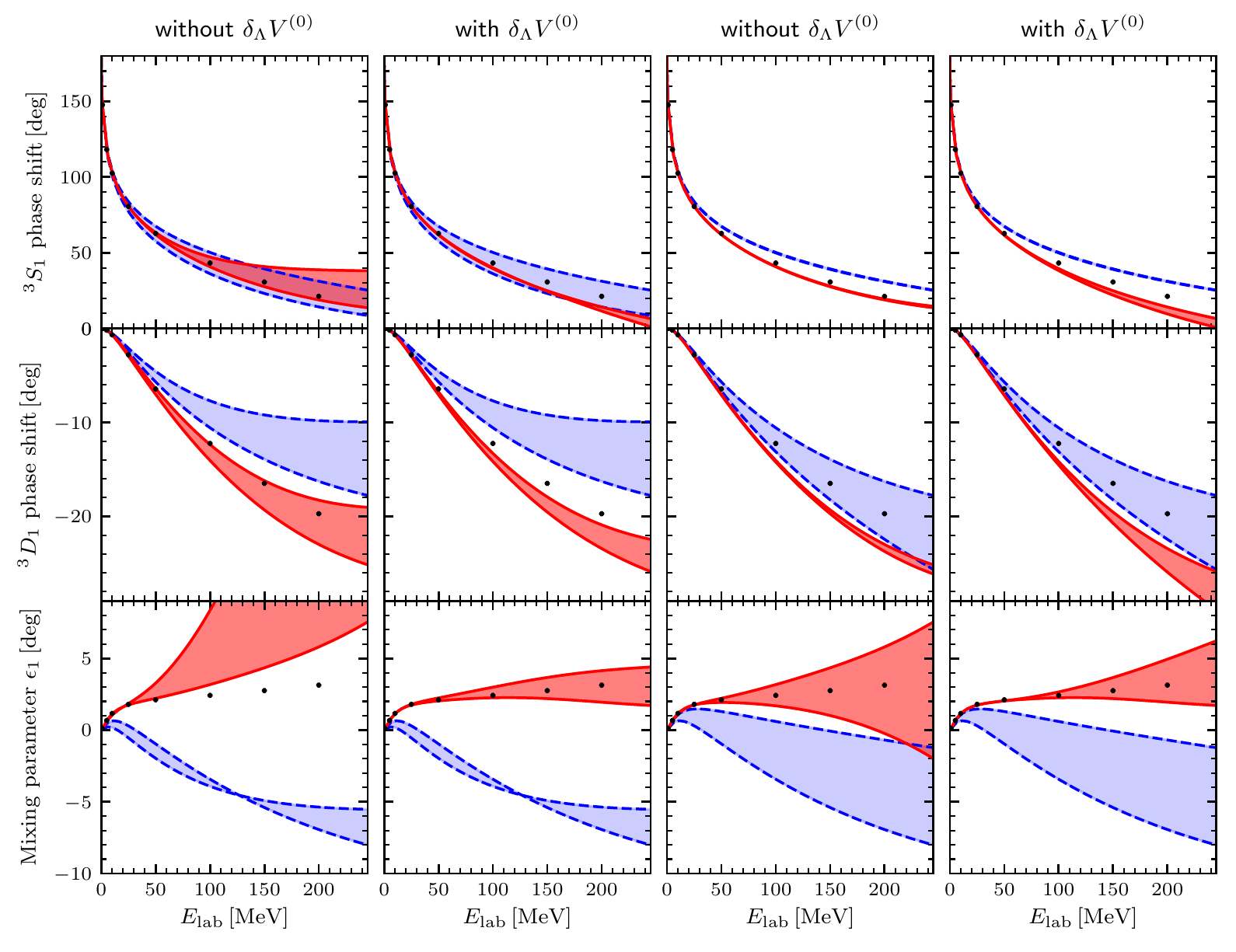}
\caption{The results of the leading-order (blue dashed lines)
and next-to-leading-order (red solid lines)  calculations for  
the $^3S_1-{^3D_1}$ channels with the contact term promoted to leading order.
The bands indicate the variation of the cutoff of the LO potential
within the range $\Lambda\in(300,450)$~MeV for two left columns
and within the range $\Lambda\in(450,800)$~MeV for two right columns.
The second and fourth columns correspond to the NLO potential with the regulator
correction $\delta_\Lambda V^{(0)}$, while the results in the first and third columns
are obtained without this term. 
The data are as in Figs.~\ref{Fig:3P0_perturbative_LambdaC2_600MeV}~and~\ref{Fig:3P0_nonperturbative_LambdaC2_600MeV}.} \label{Fig:3S1_3D1}
\end{figure}

The results of the fit for various cutoffs are shown in Fig.~\ref{Fig:3S1_3D1}.
In general, we observe a reasonable convergence
of the chiral expansion except for the $\epsilon_1$ channel where
the LO as well as the full contributions are rather small.

As expected for soft cutoffs $\Lambda\le450$~MeV, 
taking into account the explicit regulator corrections $\delta_\Lambda V^{(0)}$
for the one-pion-exchange potential and the leading contact term
significantly reduces the cutoff dependence at next-to-leading order.

Given the relatively large number of free parameters and possible fine-tuning,
it is necessary to explicitly verify the renormalizability criteria specified above.

First, we check the naturalness of the NLO LECs in the units specified in Eqs.~\eqref{Eq:natural_C_0}~and~\eqref{Eq:natural_C_2}.
The absolute values of the constants $C_{2,^3S_1,p^2}$ and $C_{2,\epsilon_1}$  do not exceed $12$
for all considered values of the cutoffs.
The maximal absolute value of the constant $C_{^3S_1,p^2}$ is about $6$ for $\Lambda\le600$~MeV,
but it starts rising very fast and reaches the value of $C_{^3S_1,p^2}\sim20$ for $\Lambda=800$~MeV
(and continues rising rapidly).

The Fredholm determinant converges with a one-percent accuracy
at orders $(V_0)^3-(V_0)^5$ for $\Lambda\le600$~MeV and at order $(V_0)^6$
for $\Lambda=800$~MeV.
The absolute value of the Fredholm determinant at threshold (at $E_\text{lab}=250$~MeV)
varies in the range $0.6-0.8$ ($1.8-3.6$) for $\Lambda\le600$~MeV
and is as large as $1.6$ ($7.5$ ) for $\Lambda=800$~MeV.

Summarizing the above observations, our numerical results confirm the
renormalizability of the NLO amplitude in the $^3S_1-{^3D_1}$ channels
for the cutoffs below or of the order of the hard scale.
For higher values of the cutoffs, the renormalizability in the sense discussed in 
the present paper is not guaranteed.

\subsection{\texorpdfstring{$^1S_0$ channel}{1S0 channel}}
\begin{figure}[tb]
\includegraphics[width=\textwidth]{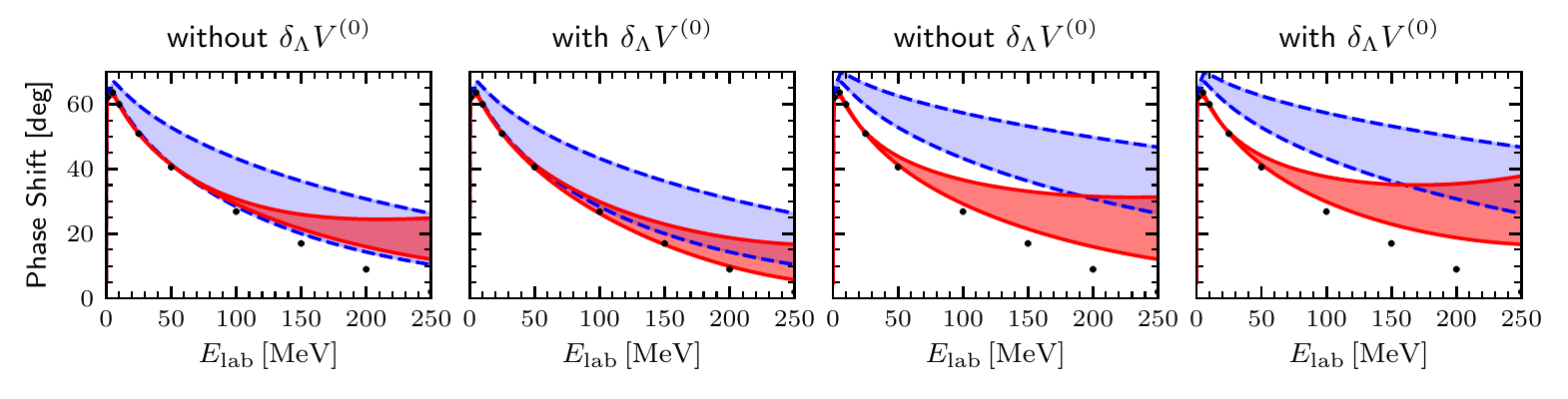}
\caption{The results of the leading-order (blue dashed lines)
and next-to-leading-order (red solid lines)  calculations for  
the $^1S_0$ partial wave without promoting the contact interaction
quadratic in momentum.
The bands indicate the variation of the cutoff of the LO potential
within the range $\Lambda\in(300,450)$~MeV for two left plots
and within the range $\Lambda\in(450,800)$~MeV for two right plots.
The second and fourth plots correspond to the NLO potential with the regulator
correction $\delta_\Lambda V^{(0)}$, while the results in the first and third plots
are obtained without this term. The data are as in Figs.~\ref{Fig:3P0_perturbative_LambdaC2_600MeV}~and~\ref{Fig:3P0_nonperturbative_LambdaC2_600MeV}.} \label{Fig:1S0_perturbativeC2_LambdaC2_600MeV}
\end{figure}

Finally, we discuss the $^1S_0$ partial wave.
The enhancement of the LO amplitude due to the extremely shallow
quasibound state is very strong.
Nevertheless, we start with trying to adopt the dimensional power counting
and do not promote any additional contact interaction to leading order.
Therefore, the LO potential consists of the one-pion-exchange contribution and the
leading contact term.
Two parameters are determined from the fit:
the LO constant $C_{^1S_0}$ and the NLO constant $C_{2,^1S_0,p^2}$
corresponding to the contact term quadratic in momenta.
The NLO constant $C_{2,^1S_0}$ is fixed from the renormalization condition 
in Eq.~\eqref{Eq:renormalization_condition_Swave}.
The results are shown in Fig.~\ref{Fig:1S0_perturbativeC2_LambdaC2_600MeV}.
As in the case of the $^3P_0$ partial wave, 
the convergence of the chiral expansion is 
acceptable only for small values of the cutoff $\Lambda\le450$~MeV.
For larger values of the cutoffs, the LO contribution is too large 
compared to the data, which leads to a strong violation of unitarity.
The regulator corrections
to the one-pion-exchange potential and the leading contact term
practically do not affect the size of the bands corresponding
to the variations of the cutoff, which is also a sign of a slow convergence.
As the cutoff increases, the Fredholm determinant at threshold changes from $0.7$ to $0.3$.
Therefore, the slow convergence of the chiral expansion for the NLO amplitude
is expected from our analysis in Sec.~\ref{Sec:Subtractions}.
Nevertheless, the series for the Fredholm determinant
converges rapidly: the one-percent accuracy is obtained at order $(V_0)^3$.
The naturalness of the NLO LECs in the units of Eqs.~\eqref{Eq:natural_C_0}~and~\eqref{Eq:natural_C_2} 
is also reasonably fulfilled: the absolute value of the constant $C_{2,^1S_0}$ does not exceed $2$,
and the absolute value of the constant $C_{2,^1S_0,p^2}$ is below $25$.

A large deviation of the LO results from the data is a motivation for 
promoting the subleading contact interaction to leading order
(as the simplest solution), see Refs.~\cite{Epelbaum:2015sha,Long:2013cya}.
As we argued in the discussion of the $^3P_0$ partial wave,
adopting soft values of the cutoff $\Lambda\le450$~MeV in the scheme with
one contact term at leading order is a sizable modification of the
short-range part of the LO potential and is, to some extent, equivalent
to the promotion of an additional contact term.

\begin{figure}[tb]
\includegraphics[width=\textwidth]{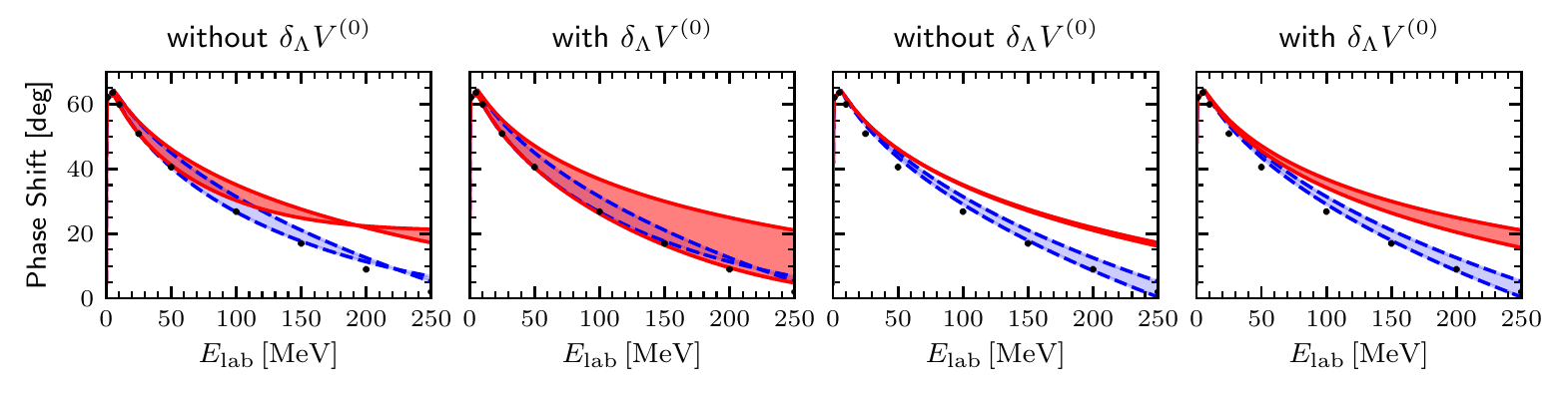}
\caption{The results of the leading-order (blue dashed lines)
and next-to-leading-order (red solid lines)  calculations for  
the $^1S_0$ partial wave with the contact interaction
quadratic in momentum promoted to leading order.
The bands indicate the variation of the cutoff of the LO potential 
within the range $\Lambda\in(300,450)$~MeV for two left plots
and within the range $\Lambda\in(450,800)$~MeV for two right plots.
The second and fourth plots correspond to the NLO potential with the regulator
correction $\delta_\Lambda V^{(0)}$, while the results in the first and third plots
are obtained without this term. The data are as in Figs.~\ref{Fig:3P0_perturbative_LambdaC2_600MeV}~and~\ref{Fig:3P0_nonperturbative_LambdaC2_600MeV}.} \label{Fig:1S0_LambdaC2_600MeV}
\end{figure}

Now, we consider the scheme with the contact interaction quadratic in momenta 
being promoted to the LO potential.
There are still two parameters to fit: $C_{^1S_0}$ and $C_{^1S_0,p^2}$.
The constants $C_{2,^1S_0}$ and $C_{2,^1S_0,p^2}$ are fixed from the renormalization conditions
in Eq.~\eqref{Eq:renormaization_2_subtractions}.
The results for the scheme with two contact terms in the LO potential are presented in Fig.~\ref{Fig:1S0_LambdaC2_600MeV}.
For higher $\Lambda$ values,
the convergence pattern for the EFT expansion in this scheme is significantly better
than in the scheme without  
promotion of the momentum-dependent contact term.
The cutoff dependence is weak for the cutoff values $\Lambda\ge450$~MeV.
For soft cutoffs, it may seem that explicit regulator corrections
makes the cutoff dependence stronger.
However, this is probably accidental because, 
as one can see, the cutoff dependence for the case without
regulator corrections is nonlinear and varies nontrivially with momentum.
This is caused by various cancellations due to the fine-tuning of two contact terms.

The absolute value of the Fredholm determinant at threshold is $D\sim0.1$
for all considered cutoffs, which is in agreement with our expectations
for the strongly enhanced LO amplitude.
The expansion of the Fredholm determinant in powers of the LO potential
approaches an accuracy of one percent at order $(V_0)^3$ for the cutoffs
below or equal to the hard scale and at order $(V_0)^4$ for $\Lambda=800$~MeV.

For all analyzed cutoffs, the naturalness constraint for the NLO constants is
reasonably well satisfied without an obvious tendency to its violation, which 
can be explained by a regular behaviour of the spin-singlet one-pion-exchange potential at
short distances.

To summarize,  the numerical calculations for the channels $^3P_0$, $^3S_1-{^3D_1}$ and $^1S_0$
are in agreement with our theoretical analysis of the renormalization of the NLO amplitude
with a finite cutoff.
We observed a reasonable convergence of the chiral EFT expansion.
However, for the $^3P_0$ and $^1S_0$ partial waves a more efficient
scheme within the considered EFT formulation 
is obtained when the subleading contact interactions are promoted to leading order,
as has already been discussed in the literature.
The naturalness constraints on the NLO LECs and on the 
value of the Fredholm determinants are fulfilled for the cutoff values 
below or of the order of the hard scale. 
The convergence rate of the Fredholm determinants in 
powers of the LO potential also appears to be sufficiently
rapid for such values of the cutoffs.
This allows us to interpret the renormalizability of the NLO amplitude
within the method developed in the current paper.
When the cutoff approaches the value $\Lambda\sim800$~MeV or higher,
the renormalizability constraints are not clearly fulfilled anymore,
even though the convergence of the amplitude might still be reasonable.

Thus, we conclude that the preferable choice of the cutoff values is roughly $\Lambda\lesssim 600$~MeV.
For very soft cutoffs $\Lambda=300-450$~MeV,
the regulator corrections to the LO potential should be explicitly 
taken into account to remove the regulator artifacts.

\newpage
\section{Summary}\label{Sec:summary}
We have extended our previous study in Ref.~\cite{Gasparyan:2021edy}
and analyzed the renormalization of the nucleon-nucleon amplitude
at NLO in chiral EFT 
in the case when the LO interaction is non-perturbative.
Our scheme is based on the formulation of chiral EFT with a finite cutoff
derived from the effective Lagrangian.

In the previous paper, the power counting for the renormalized  NLO
amplitude was justified for the case when the series for the iterations of the LO potential
are (rapidly enough) convergent, i.e., for the perturbative case.
The corresponding subtractions in the form of the LO $S$-wave
contact terms that absorb the power counting breaking contributions
were identified. Starting from $P$-waves,
the NLO amplitudes were found not to 
require any subtractions in agreement with dimensional power counting.

The method of analysis of the power counting in the non-perturbative
regime relies on the Fredholm formula for the solution of the integral
equations, which represents the numerators and denominators of the amplitudes
as individually convergent series in powers of the LO potential.
To implement the Fredholm decomposition, we first had to derive stronger bounds
on the LO potential compared to the ones used in the perturbative case.
In contrast to the perturbative regime, it turned out that the minimal ``mild'' regulator
can, in general, not be employed if the NLO potential remains unregulated.
This implies a potentially stronger cutoff dependence in the non-perturbative case.

The results for the $P$- and higher partial waves in the NN system reproduce to a large extend 
our previous findings. The dimensional power counting for the LO and NLO amplitudes is formally
satisfied without subtractions unless there is an 
enhancement of the LO amplitude due to the presence of a shallow
(quasi-)bound state, which is not the case for the physical channels.
Nevertheless, in some cases, the promotion of NLO contact terms
to leading order can be motivated by phenomenological arguments as, e.g., in the $^3P_0$ channel.
In the latter situation, however, one has to choose the LO potential in such a way as to 
avoid the appearance of ``exceptional'' cutoffs, for which the renormalization breaks down.
The simplest way to verify that the adopted value of the cutoff is not close to ``exceptional''
is to make sure that the NLO LECs are of a natural size.

For the $S$-waves, we have shown that the series for the subtractions
at next-to-leading order, obtained in the perturbative case,
can be resummed in a closed form.
Such a resummation is equivalent to the condition for the renormalized NLO amplitude
to vanish at threshold.
Using the Fredholm formula allowed us to analyze also the case
when the LO amplitude is enhanced at threshold compared to the
dimensional power counting estimate.
This happens in the  $^3S_1-{^3D_1}$ and $^1S_0$ channels where
shallow bound and quasibound states are present.
The dimensional power counting for the NLO amplitudes is
still valid in those cases if certain additional constraints
on the LO potential are fulfilled.
Again, these constraints eventually reveal themselves in the naturalness
of the NLO LECs.
However, the convergence of the chiral expansion 
in the channels with enhanced LO amplitude may become
significantly slower, especially in the $^1S_0$ channel,
where the enhancement is most pronounced.
To improve the convergence, one can, analogously to the $^3P_0$ partial wave,
promote a subleading contact term to the LO potential with the same
warning regarding ``exceptional'' cutoffs.

Finally, we have illustrated our theoretical findings by numerical calculations
of the NN phase shifts at next-to-leading order by fitting the unknown free parameters
to the empirical data.
We considered three channels with non-perturbative dynamics, namely $^3P_0$, $^3S_1-{^3D_1}$ and $^1S_0$,
and varied the LO cutoff in the range of $\Lambda=300-800$~MeV.
We observed reasonable convergence of the chiral expansion,
especially when the subleading contact terms are promoted in the $^3P_0$ and $^1S_0$ channels.

As criteria for the interpretation of the renormalizability of the NLO amplitude
in terms of the Fredholm expansion, we used the naturalness of the NLO LECs
and of the Fredholm determinant as well as the convergence rate of the expansion
of the latter in powers of the LO potential.
It turns out that all these constraints are fulfilled as long as the cutoff values are 
chosen below or of the order of the hard scale $\Lambda_b\sim600$~MeV.
For particularly soft cutoffs $\Lambda=300-450$~MeV, taking into account explicit
regulator corrections to the LO potential compensates for the regulator artifacts
and reduces the cutoff dependence.

When the cutoff increases beyond the hard scale, the renormalizability constraints
start being violated. Therefore, we conclude that the cutoff values $\Lambda\lesssim\Lambda_b$
are preferable from the point of view of the renormalization of EFT.

Further development of our approach goes in the direction of extending it beyond next-to-leading order in the chiral expansion.
It is also important to generalize the scheme to few-nucleon systems and the processes involving
electro-weak interactions.

\section*{Acknowledgments} 
We would like to thank Jambul Gegelia for helpful discussions and for useful comments 
on the manuscript.
This work was supported by DFG (Grant No. 426661267), by BMBF (contract No. 05P21PCFP1), by ERC AdG
NuclearTheory (grant No. 885150) and by the EU Horizon 2020 research and
innovation programme (STRONG-2020, grant agreement No. 824093).

\newpage
\appendix
\section{Leading-order potential}
\label{Sec:LOpotential}
The short-range part of
the leading-order potential in its general form
can be chosen to include 
the momentum-independent contact interactions and
contact terms quadratic in momenta (altogether $9$ terms), multiplied
by the power-like non-local form factor of an appropriate power $n$:
\begin{align}
&V^{(0)}_{\text{short},\Lambda}(\vec p\,',\vec p\,)=\sum_{i}
C_i \, V_{C_i}\, F_{\Lambda_i, n_i}( p\,', p)\,,
\label{Eq:short_range_nonlocal}
\end{align}
where $V_{C_i}$ is any basis for the contact terms, e.g. the
partial wave basis, and the regulators are given by
\begin{align}
&F_{\Lambda, n}( p\,', p)=F_{\Lambda, n}( p\,')F_{\Lambda, n}(p)
\,, \quad F_{\Lambda, n}(p)=\left[ F_{\Lambda}(p)\right]^n\,,\quad
F_{\Lambda}(p)=\frac{\Lambda^2}{p^2+\Lambda^2}\,.
\label{Eq:nonlocal_formfactor}
\end{align}
One can also introduce a regulator of a Gaussian form by replacing
$F_{\Lambda, n}(p)$ with 
\begin{align}
F_{\Lambda, \text{exp}}(p)=\exp{(-p^2/\Lambda^2)}\,. 
\label{Eq:F_exp}
\end{align}

Alternatively, one could introduce local short-range interactions (for the terms that depend only on $\vec q$, 
except for the spin-orbit term) using the appropriate basis \cite{Gezerlis:2014zia}
and the  local regulator
\begin{align}
&F_{q,\Lambda,n}( \vec p\,', \vec p\, )=[F_\Lambda(q)]^n=
\left(\frac{\Lambda^2}{q^2+\Lambda^2}\right)^{n}\,,
\label{Eq:local_formfactor}
\end{align}
or with the regulator in the Gaussian form $F_{\Lambda,\text{exp}}(q)$.

The long-range part of the LO potential is represent 
by the one-pion-exchange contribution, which is split into the 
triplet, singlet, and contact parts
\begin{align}
V^{(0)}_{1\pi}= -\bigg(\frac{g_A}{2F_\pi}\bigg)^2 \bm\tau_1 \cdot \bm \tau_2\, 
\frac{\vec{\sigma}_1 \cdot\vec{q}\,\vec{\sigma}_2\cdot\vec{q}}{q^2 + M_\pi^2} 
\eqqcolon V^{(0)}_{1\pi,\text{t}}+V^{(0)}_{1\pi,\text{s}}+V^{(0)}_{1\pi,\text{ct}}\,,
\label{Eq:V_1pi_0}
\end{align}
with 
\begin{align}
V^{(0)}_{1\pi,\text{s}}&=\bigg(\frac{g_A}{2F_\pi}\bigg)^2 \bm\tau_1 \cdot \bm \tau_2\,
\frac{(\vec{\sigma}_{1} \cdot\vec{\sigma}_{2}-1 )}{4}\frac{M_{\pi}^{2}}{q^{2}+M_{\pi}^{2}} \,,\nonumber\\
V^{(0)}_{1\pi,\text{ct}}&=-\bigg(\frac{g_A}{2F_\pi}\bigg)^2 
\bm\tau_1 \cdot \bm \tau_2\,  \frac{(\vec{\sigma}_{1} \cdot\vec{\sigma}_{2}-1 )}{4}\,.
\label{Eq:V_1pi}
\end{align}
All three parts, if necessary, are regularized individually.
The contact part $V_{1\pi,\text{ct}}$ can be absorbed by the
leading-order $^1S_0$ contact term and thus needs not be considered separately.
The triplet and singlet potentials can be regularized by means
of the non-local form factor (see Eq.~\eqref{Eq:nonlocal_formfactor}):
\begin{align}
V^{(0)}_{1\pi,\Lambda}(\vec p\,',\vec p \,
  )=V^{(0)}_{1\pi,\text{s}}(\vec p\,',\vec p\, )F_{\Lambda_{\text{s}}, n_\text{s}}( p\,', p)
+V^{(0)}_{1\pi,\text{t}}(\vec p\,',\vec p\,)F_{\Lambda_{\text{t}}, n_\text{t}}( p\,', p)\,,
\label{Eq:V_1pi_nonlocal}
\end{align}
or by means of the local regulator: 
\begin{align}
V^{(0)}_{1\pi,\Lambda}(\vec p\,',\vec p \,
  )=V^{(0)}_{1\pi,\text{s}}(\vec p\,',\vec p \,
  )F_{q,1\pi,\Lambda_{\text{s}}}(\vec p\,',\vec p \, )
+V^{(0)}_{1\pi,\text{t}}(\vec p\,',\vec p \,
  )F_{q,1\pi,\Lambda_\text{t}}(\vec p\,',\vec p \, )\,,
  \label{Eq:V_1pi_local}
\end{align}
with
\begin{align}
 F_{q,1\pi,\Lambda_\text{s}}(\vec p\,',\vec p\, )&=
\left(\frac{\Lambda_\text{s}^2-M_\pi^2}{q^2+\Lambda_\text{s}^2}\right)^{n_\text{s}},\nonumber\\
F_{q,1\pi,\Lambda_\text{t}}(\vec p\,',\vec p\, )&=
\left(\frac{\Lambda_\text{t}^2-M_\pi^2}{q^2+\Lambda_\text{t}^2}\right)^{n_\text{t}}.
\label{Eq:local_formfactor_1pi}
\end{align}
Note that in Ref.~\cite{Gasparyan:2021edy}, a more general form of the local regulator
was considered.

The spin-singlet part of the one-pion-exchange potential can, in principle, be
left unregulated. This is, however, only relevant for the spin-singlet channels
without short-range interactions. All such channels can be regarded as 
having perturbative LO potential and were already analyzed in Ref.~\cite{Gasparyan:2021edy}.
For the spin-singlet channel considered in this work, $^1S_0$, the 
effects of a regulator will be driven by the contact interaction in any case.

To regularize the spin-triplet part of the one-pion-exchange potential
in the LO Lippmann-Schwinger equation,
it is sufficient to introduce a dipole ($n_\text{t}=1$) regulator,
which we refer to as the ``mild'' regulator. All other options, i.e., $n_\text{t}\ge 2$
are referred to as the ``standard'' regulators.

One can also adopt the local Gaussian regulator for the one-pion-exchange potential:
\begin{align}
F_{q,1\pi,\text{exp},\Lambda}(\vec p\,',\vec p \, )=\exp{\left[-(q^2+M_\pi^2)/\Lambda^2\right]}\,. 
\label{Eq:1pi_exchange_local_gaussian_regulator}
\end{align}

\section{Next-to-leading-order potential}
\label{Sec:NLOpotential}

The short-range part of the next-to-leading-order potential is given by the sum of contact terms
analogous to Eq.~\eqref{Eq:short_range_nonlocal}:
\begin{align}
&V^{(2)}_{\text{short}}(\vec p\,',\vec p\,)=\sum_{i}
C_{2,i} \, V_{C_i}\,.
\label{Eq:V2_short_range}
\end{align}
  
The non-polynomial part of the two-pion-exchange potential
is given by (it is equivalent to the one provided in Ref.~\cite{Epelbaum:2004fk} up to polynomial terms)
\begin{align}
V^{(2)}_{2\pi}(\vec p\,',\vec p\,)&= - \frac{\bm \tau_1 \cdot \bm \tau_2}{384 \pi^2 F_\pi^4}\,
\tilde L(q) \, \biggl[4M_\pi^2 (5g_A^4 - 4g_A^2 -1)
+ q^2(23g_A^4 - 10g_A^2 -1)
+ \frac{48 g_A^4 M_\pi^4}{4 M_\pi^2 + q^2} \biggr]\nonumber\\
&+ \frac{\bm \tau_1 \cdot \bm \tau_2}{8 \pi^2 F_\pi^4}\frac{g_A^4 M_\pi^2 q^2}{4 M_\pi^2 + q^2}
-\frac{3 g_{A}^{4}}{64 \pi^{2} F_{\pi}^{4}}\tilde  L(q)
\left[\vec{\sigma}_{1} \cdot \vec{q} \vec{\sigma}_{2} \cdot \vec{q}-q^{2}
 \vec{\sigma}_{1} \cdot \vec{\sigma}_{2}\right]\,,
 \label{Eq:two_pion_exchange}
\end{align}
where
\begin{align}
\tilde L(q) :=L(q)-L(0)=L(q)-1\,, \quad 
L(q)=\frac{1}{q} \sqrt{4 M_{\pi}^{2}+q^{2}} \log \frac{\sqrt{4 M_{\pi}^{2}+q^{2}}+q}{2 M_{\pi}}\,.
\end{align}

The regulator of the NLO potential, not shown explicitly in the above expressions, 
can be a combination of any local or non-local forms.
For the two-pion-exchange potential, one can also employ a spectral function regularization
by introducing a finite upper limit in the dispersion representation
of $\tilde L(q)$:
\begin{align}
 \tilde L(q)=q^2\int_{2M_\pi}^{\Lambda_\rho} \frac{d\mu}{\mu^2}\frac{\sqrt{\mu^2-4M_\pi^2}}{q^2+\mu^2}\,.
\end{align}

\section{Bounds on the plane-wave potential}
\label{Sec:bounds_plane_wave}
\subsection{Bounds on the substructures}
\label{Sec:bounds_substructures}
Below, we list the inequalities for the building blocks of the
LO and NLO potentials obtained in Ref.~\cite{Gasparyan:2021edy}.

The components of the initial and final nucleon c.m. momenta $p$ and $p'$
are defined as
\begin{align}
\vec p = p\begin{pmatrix} 0 \\ 0 \\ 1 \end{pmatrix}\,,\
\vec p\,' = p'\begin{pmatrix} \sin \theta \cos \phi  \\ \sin \theta \sin \phi \\ \cos \theta \end{pmatrix}\,,
\end{align}
where $p$ is either $p=p_\text{on}$ or lies on the complex contour $p\in \mathcal{C}$:
$p=|p|\exp(-i\alpha_C)$, and  
$p'$ is either $p'=p_\text{on}$ or 
$p'=|p'|\exp(-i\alpha_C)$.

For the function
\begin{align}
&f_\mu(p',p,x)=\frac{1}{q^2+\mu^2}=\frac{1}{p'^2+p^2-2p p' x+\mu^2}\,,
\end{align}
with $\mu\ge M_\pi$, the following bounds hold
\begin{align}
\left|f_\mu(p',p,x)  \right| &\le\frac{ \mathcal{M}_f}{|p|^2+|p'|^2-2|p||p'|x+\mu^2}\,,
\label{Eq:estimates_denominator2}\\
\left|q_i q_j f_\mu(p',p,x)\right|&\le \mathcal{M}_f,
\label{Eq:bounds_qi_qj_over_q2}\\
\left|( \vec{k} \times \vec{q})_i \,f_\mu(p',p,x)\right|
 &\le\mathcal{M}_f\left(1-x^2\right)^{-1/2}\,,\quad i,j=1,2,3\,.
 \label{Eq:bounds_qi_kj_over_q2}
\end{align}
The subtraction remainders defined as
\begin{align}
 \Delta_p^{(n)} f(p',p)&= 
f(p',p)-\sum_{i=0}^{n}\frac{\partial^i f(p',p)}{i!(\partial p)^i}\bigg|_{p=0}p^i,\nonumber\\
\Delta_{p'}^{(n)} f(p',p)&= 
f(p',p)-\sum_{i=0}^{n}\frac{\partial^i f(p',p)}{i!(\partial p')^i}\bigg|_{p'=0}(p')^i,
\label{Eq:remainders}
\end{align}
satisfy the following inequalities:
\begin{align}
 \left| \Delta_p^{(n)} f_\mu(p',p)\right| &\le
                \mathcal{M}_{f,n}\left|\frac{p}{p'}\right|^{n+1}
                \left|f_\mu(p',p)\right|,
                \quad
\text{ if }|p'|>|p|,\nonumber\\
 \left| \Delta_{p'}^{(n)} f_\mu(p',p)\right| &\le
                \mathcal{M}_{f,n}\left|\frac{p'}{p}\right|^{n+1}
                \left|f_\mu(p',p)\right|б
                \quad
\text{ if }|p|>|p'|.
\label{Eq:Delta_p_f}
\end{align}

For a more general structure 
\begin{align}
&\Psi_{k,m,\{\mu_i\}}(p',p,x)=Q_k(p',p,x) F_{\Lambda,m}(p',p) f_{\{\mu_i\}}(p',p\,,x),
\label{Eq:Psi}
\end{align}
where the form factor $F_{\Lambda,m}$ is defined in Eq.~\eqref{Eq:nonlocal_formfactor},
$f_{\{\mu_i\}}$ is a product of several $f_\mu$
\begin{align}
& f_{\{\mu_i\}}(p',p,x)=\prod_{i=1,r} f_{\mu_i}(p',p,x),
\label{Eq:product_f_mu_i}
\end{align}
and $Q_k$ is a homogeneous polynomial of degree $k$,
one can deduce the bounds for derivatives:
\begin{align}
\left|p^n\frac{\partial^n \Psi_{k,m,\{\mu_i\}}(p',p,x)}{\partial p^n}\bigg|_{p=0}\right|
& \le\mathcal{M}_{\partial \Psi}^{k,n}\left|p'^k F_{\Lambda,m-\frac{n+1}{2}}(p') 
f_{\{\mu_i\}}(p',0,x)\right|\left|\frac{p}{p'}\right|^{n} ,\nonumber\\
\left|(p')^n\frac{\partial^n \Psi_{k,m,\{\mu_i\}}(p',p,x)}{\partial (p')^n}\bigg|_{p'=0}\right|
&\le\mathcal{M}_{\partial \Psi}^{k,n}\left|p^k F_{\Lambda,m-\frac{n+1}{2}}(p) 
f_{\{\mu_i\}}(0,p,x)\right|\left|\frac{p'}{p}\right|^{n} ,
\qquad n\ge0\,,
\label{Eq:derivatives_Psi}
\end{align}
and for the subtraction remainders:
\begin{align}
\left| \Delta_p^{(n)} \Psi_{k,m,\{\mu_i\}}(p',p,x)\right| &\le \mathcal{M}_{\Psi}^{k,n}\left|\frac{p}{p'}\right|^{n+1} 
\nonumber\\&\times\Big(|\Psi_{k,m,\{\mu_i\}}(p',p,x)|
+\left|p'^k F_{\Lambda,m-\frac{n+1}{2}}(p') f_{\{\mu_i\}}(p',0,x)\right|\Big),\qquad\text{ if }|p'|>|p|,\nonumber\\
\left| \Delta_{p'}^{(n)} \Psi_{k,m,\{\mu_i\}}(p',p,x)\right| &\le \mathcal{M}_{\Psi}^{k,n}\left|\frac{p'}{p}\right|^{n+1} 
\nonumber\\&\times\Big(|\Psi_{k,m,\{\mu_i\}}(p',p,x)|
+\left|p^k F_{\Lambda,m-\frac{n+1}{2}}(p) f_{\{\mu_i\}}(0,p,x)\right|\Big),\qquad\text{ if }|p|>|p'|.
\label{Eq:Delta_p_Psi}
\end{align}

\subsection{Bounds on the plane-wave leading-order potential}
\label{Sec:bounds_LO_plane_wave_potential}
In this section we provide bounds for the leading-order potential.
We will need slightly stronger bounds than those obtained in Ref.~\cite{Gasparyan:2021edy}.
In particular, we will need bounds that factorize in initial and finale momenta in 
the partial wave basis.
In order to obtain them, we will partly keep the angular dependence in 
binding functions.

The derivation is only slightly different from that of Ref.~\cite{Gasparyan:2021edy},
which we demonstrate for the case of the spin-triplet one-pion-exchange potential.

The locally regularized one-pion exchange potential in the spin-triplet channels
can be bounded using equations of Appendix.~\ref{Sec:bounds_substructures} by the following inequality:
\begin{align}
&\left|V^{(0)}_{1\pi,\text{t}}(\vec p\,',\vec p\,)\right|
\le \left|\frac{g_A^2}{4F_\pi^2}\sum_{i,j}\mathcal{M}_{\text{t},ij}\frac{q_iq_j}{q^2+M_\pi^2}
\left(\frac{\Lambda_{\text{t}}^2-M_\pi^2}{q^2+\Lambda_{\text{t},1}^2}\right)^{n_{\text{t}}}
\right|\nonumber\\
&\le\frac{2\pi M_\text{t}}{m_N \Lambda_V}F_{\Lambda,n_{\text{t}}}(|p'|,|p|,x)
\label{Eq:bound_triplet_1pi_local}\,,
\end{align}
where we have introduced the form factors
\begin{align}
F_{\Lambda,n}(|p'|,|p|,x)=\left( F_{\Lambda}(|p'|,|p|,x) \right)^n,\qquad
 F_{\Lambda}(|p'|,|p|,x)=\frac{\Lambda^2}{|p|^2+|p'|^2-2|p||p'|x+\Lambda^2}\,.
\end{align}
In Eq.~\eqref{Eq:bound_triplet_1pi_local}, we replaced 
$\Lambda_{\text{t}}$ with the largest cutoff $\Lambda$ among all regulators in the LO potential,
which is possible due to the inequality:
\begin{align}
 F_{\Lambda_1}(|p'|,|p|,x)<F_{\Lambda_2}(|p'|,|p|,x) \qquad \text{ for }
 \Lambda_1<\Lambda_2\,.
\end{align}

If the triplet one-pion exchange potential is regularized by the
non-local form factor, we obtain
\begin{align}
\left|V^{(0)}_{1\pi,\text{t}}(\vec p\,',\vec p\,)\right|
&\le \left|\frac{g_A^2}{4F_\pi^2}\sum_{i,j}\mathcal{M}_{\text{t},ij}\frac{q_iq_j}{q^2+M_\pi^2}
\left(\frac{\Lambda^2}{p'^2+\Lambda^2}\frac{\Lambda^2}{p^2+\Lambda^2}  \right)^{n_{\text{t}}}\right|\nonumber\\
&\le\frac{2\pi M_\text{t}}{m_N \Lambda_V}F_{\Lambda, n_{\text{t}}}( |p\,'|)F_{\Lambda, n_{\text{t}}}(|p|).
\label{Eq:bound_triplet_1pi_nonlocal}
\end{align}

Analogously, we obtain bounds for 
other LO contributions as in Ref.~\cite{Gasparyan:2021edy}
retaining the angular dependence of local form factors and the powers of the form factors.

Finally, the full leading-order potential satisfies
\begin{align}
&\left|V_0(\vec p\,',\vec p\,)\right|\le \frac{\mathcal{M}_{V_0}}{4\pi}
V_{0,\text{max}}(p',p,x),\qquad
\left|V_0(\vec p\,',\vec p\,)\right|
\le \frac{\mathcal{M}_{V_0}}{4\pi}V_{0,\text{max}}(p,p',x)
\,,
\label{Eq:bound_full_LO}
\end{align}
where we have introduced
\begin{align}
&V_{0,\text{max}}(p',p,x)
=\frac{8\pi^2 }{m_N \Lambda_V}\Big[F_{\Lambda,n}(|p'|,|p|,x)+F_{\Lambda,n}(|p'|)\Big]\,,
\label{Eq:V0max}
\end{align}
with $n$ being the smallest power among all regulators in the LO potential.
The cases of the ``mild'' and the ``standard'' regulators correspond to $n=1$
and $n\ge2$, respectively.
The difference of Eq.~\eqref{Eq:V0max} from an analogous bound in  Ref.~\cite{Gasparyan:2021edy}
is that the powers of both local and non-local form factors are retained 
and the $x$-dependence of the local form factor is kept.
The bounds for the Gaussian regulators can be reduced to
the ones for the power-like regulators as was shown in Ref.~\cite{Gasparyan:2021edy}.

For the spin-singlet channels without a short-range interaction, the
bounds in Eq.~\eqref{Eq:V0max} can be improved by replacing $\Lambda$ with $M_\pi$.
However as mentioned above, those channels were already covered in our previous study.

The remainders $\Delta_p^{(n)} V_0(\vec p\,',\vec p\,)$ for $|p'|>|p|$ can be estimated using 
Eq.~\eqref{Eq:Delta_p_Psi}:
\begin{align}
&\left|\Delta_p^{(n)} V_0(\vec p\,',\vec p\,)\right| \le \frac{\mathcal{M}_{\Delta V_0,n}}{4\pi}
\left|\frac{p}{p'}\right|^{n+1} V_{0,\text{max}}(p',p,x)\qquad\text{ if } |p'|>|p|\,,\nonumber\\
&\left|\Delta_{p'}^{(n)} V_0(\vec p\,',\vec p\,)\right| \le \frac{\mathcal{M}_{\Delta V_0,n}}{4\pi}
\left|\frac{p'}{p}\right|^{n+1} V_{0,\text{max}}(p,p',x)\qquad\text{ if } |p|>|p'|\,.
\label{Eq:Delta_p_V0_constrained}
\end{align}

From Eq.~\eqref{Eq:derivatives_Psi} 
one obtains the estimates for the derivatives of the leading-order potential:
\begin{align}
&\left|p^m\frac{\partial^m V_0(\vec p\,',\vec p\,)}{(\partial p)^m}\bigg|_{p=0}\right|
\le \frac{2\pi \mathcal{M}_{\partial V_0,n}}{m_N \Lambda_V}F_{\tilde\Lambda,n}(|p'|)\left|\frac{p}{p'}\right|^m
\le \frac{\mathcal{M}_{\partial V_0,n}}{4\pi}\left|\frac{p}{p'}\right|^mV_{0,\text{max}}(p',p,x)
\,,\label{Eq:V0_derivatives_a}\\
&\left|p'^m\frac{\partial^m V_0(\vec p\,',\vec p\,)}{(\partial p')^m}\bigg|_{p'=0}\right|
\le \frac{2\pi \mathcal{M}_{\partial V_0,n}}{m_N \Lambda_V}F_{\tilde\Lambda,n}(|p|)\left|\frac{p'}{p}\right|^m
\le \frac{\mathcal{M}_{\partial V_0,n}}{4\pi}\left|\frac{p'}{p}\right|^m V_{0,\text{max}}(p,p',x)\,,
\label{Eq:V0_derivatives_b}
\end{align}
including the case $m=0$,
where we have used that the local form factor satisfies
\begin{align}
 F_\Lambda(p',0,x)=F_\Lambda(p').
\end{align}

Applying Eq.~\eqref{Eq:V0_derivatives_a} (Eq.~\eqref{Eq:V0_derivatives_b})
to the definition of $\Delta_p^{(n)} V_0(\vec p\,',\vec p\,)$
($\Delta_{p'}^{(n)} V_0(\vec p\,',\vec p\,)$) in Eq.~\eqref{Eq:remainders}
for $|p|>|p'|$ ($|p'|>|p|$), and combining it with Eq.~\eqref{Eq:Delta_p_V0_constrained}, we obtain the following bounds for the remainders:
\begin{align}
&\left|\Delta_p^{(n)} V_0(\vec p\,',\vec p\,)\right| \le \frac{\mathcal{M}_{\Delta V_0,n}}{4\pi}
\left|\frac{p}{p'}\right|^{n+1} V_{0,\text{max}}(p',p,x)\,,\nonumber\\
&\left|\Delta_{p'}^{(n)} V_0(\vec p\,',\vec p\,)\right| \le \frac{\mathcal{M}_{\Delta V_0,n}}{4\pi}
\left|\frac{p'}{p}\right|^{n+1} V_{0,\text{max}}(p,p',x)\,,
\label{Eq:Delta_p_V0}
\end{align}
which are valid for all considered $p$ and $p'$.
All above general formulas do not include the case when 
the LO potential contains a locally regulated
spin-orbit short-range interaction such as
\begin{align}
V^{(0)}_{C_5}(\vec p\,',\vec p\,)=C_5 \frac{i}{2}( \vec{\sigma}_1 + \vec{\sigma}_2)
\cdot ( \vec{k} \times \vec{q}\,)\left(\frac{\Lambda_5^2}{q^2+\Lambda_5^2}\right)^{n_5}\,, 
\end{align}
with $n_5>1$ (or with the Gaussian form factor).
Following the arguments provided in Ref.~\cite{Gasparyan:2021edy},
one can formulate the same bounds as in Eqs.~\eqref{Eq:bound_full_LO}~and~\eqref{Eq:Delta_p_V0}
for the quantity $\tilde V^{(0)}_{C_5}$ defined as
\begin{align}
V^{(0)}_{C_5}(\vec p\,',\vec p\,)&=\tilde V^{(0)}_{C_5}(\vec p\,',\vec p\,)
\frac{i}{2}( \vec{\sigma}_1 + \vec{\sigma}_2)\cdot \vec n_\phi/\sin\theta,\nonumber\\
\vec n_\phi&=(-\sin\phi,\cos\phi,0),
\label{Eq:V_C5_tilde}
 \end{align}
which makes it possible, after the partial-wave projection, to treat
this interaction on the same footing as all other LO terms.

\subsection{Bounds on the plane-wave next-to-leading-order potential}
\label{Sec:Bounds_NLO}
For the NLO potential, we use the bounds obtained in Ref.~\cite{Gasparyan:2021edy}.
The NLO potential is split into two parts:
\begin{align}
 V_2(\vec p\,',\vec p\,)=\hat V_2(\vec p\,',\vec p\,)+\tilde V_2(\vec p\,',\vec p\,)\,,
 \label{Eq:V2_tilde}
\end{align}
with 
\begin{align}
\hat V_2(\vec p\,',\vec p\,)=V_2(0,0)\,,\quad \tilde V_2(\vec p\,',\vec p\,)=V_2(\vec p\,',\vec p\,)-V_2(0,0)\,,
\end{align}
which are bound as
\begin{align}
 \left|\hat V_2(\vec p\,',\vec p\,)\right|\le \mathcal{\hat M}_{V_2} \frac{2\pi}{m_N \Lambda_V}\frac{M_\pi^2}{\Lambda_{b}^2}\,,
 \label{Eq:bound_V_2_0_hat_plane_wave}
\end{align}
and
\begin{align}
& \left|\tilde V_2(\vec p\,',\vec p\,) \right|
\le  \frac{2\pi \mathcal{M}_{V_2}}{m_N \Lambda_V} \frac{|p|^2+|p'|^2}{\Lambda_{b}^2}f_\text{log}(p',p)
=\frac{\mathcal{M}_{V_2}}{4\pi}\left(|p|^2+|p'|^2\right)\tilde f_\text{log}(p',p)\,,
 \label{Eq:bounds_V2_full}
\end{align}
with
\begin{align}
&\tilde f_\text{log}(p',p)=\frac{8\pi^2}{m_N \Lambda_V \Lambda_{b}^2} f_\text{log}(p',p)\,,\nonumber\\
& f_\text{log}(p',p)=\theta(|p|-M_\pi)\ln\frac{|p|}{M_\pi}
+\theta(|p'|-M_\pi)\ln\frac{|p'|}{M_\pi}+\ln\frac{\tilde\Lambda}{M_\pi}+1\,.
\label{Eq:f_log}
\end{align}
In the function $f_\text{log}(p',p)$, the term $\ln\frac{\tilde\Lambda}{M_\pi}$
was introduced for convenience so we can omit it (or set $\tilde\Lambda=M_\pi$).

We also allow for a regulator (local or non-local) for the NLO potential.
We can introduce it simply as a factor, so that the bounds in Eq.~\eqref{Eq:bounds_V2_full}
are modified as
\begin{align}
&\left|\tilde V_2(\vec p\,',\vec p\,) \right|
\le \frac{\mathcal{M}_{V_2}}{4\pi}\left(|p|^2+|p'|^2\right)\tilde f_\text{log}(p',p)
\Big[F_{\Lambda_\text{NLO}}(|p'|,|p|,x)+F_{\Lambda_\text{NLO}}(|p'|)\Big]\,,\nonumber\\
\label{Eq:bounds_V2_regulated_full}
\end{align}
where we combined local and non-local regulators into one factor.
The first power ($n=1$) of the form factors is sufficient for our estimates.
The cases of higher powers (or Gaussian cutoffs) are included automatically,
because 
\begin{align}
 F_{\Lambda_\text{NLO},n}(|p|)\le F_{\Lambda_\text{NLO}}(|p|)\,,\qquad
 F_{\Lambda_\text{NLO},n}(|p'|,|p|,x)\le F_{\Lambda_\text{NLO}}(|p'|,|p|,x)\,,
\end{align}
for $n>1$.
If different values of the cutoff are used for different NLO contributions, $\Lambda_\text{NLO}$
can be chosen to be the largest value.

\section{Bounds on the partial-wave potential}
\label{Sec:PW_bounds}
Below, we repeat the arguments of Ref.~\cite{Gasparyan:2021edy}
for deriving the bounds on the partial-wave potential, but
take into account an angular dependence of the binding functions.

The partial-wave potential is obtained from the plain-wave potential via
\begin{align}
 &V_{l', l}^{s j}(p', p) = \sum_{ \lambda_{1}, \lambda_{2},\lambda_{1}^{\prime}, \lambda_{2}^{\prime}}
\int d\Omega\
\langle jl's|\lambda_{1}^{\prime} \lambda_{2}^{\prime}\rangle
\langle \lambda_{1}^{\prime} \lambda_{2}^{\prime}|V(\vec p\,',\vec p\,)| \lambda_{1} \lambda_{2}\rangle
\langle \lambda_{1} \lambda_{2}|jls \rangle
d_{\lambda_{1}-\lambda_{2}, \lambda_{1}^{\prime}- \lambda_{2}^{\prime}}^{j}(\theta)\,,\nonumber\\
&\langle \lambda_{1} \lambda_{2}|jls\rangle
=\left(\frac{2 l+1}{2 j+1}\right)^{\frac{1}{2}} C(l\,, s\,, j ; 0\,, \lambda_{1}-\lambda_{2}) 
C\left(1/2\,, 1/2\,, s ; \lambda_{1},-\lambda_{2}\right),
\end{align}
where $\lambda_i$, $\lambda_i'$ are the helicities of the corresponding nucleons.

Due to unitarity of the transformation, the following 
constraints hold:
\begin{align}
& |\langle \lambda_{1} \lambda_{2}|jls\rangle|\le 1,\qquad
|\langle 1/2\,,s_z|\lambda\rangle|\le 1,\qquad
|d_{\lambda, \lambda^{\prime}}^{j}(\theta)|\le 1\,.
 \nonumber\\
\end{align}
Therefore, if the plain-wave potential 
is bounded by some angle-dependent function $\phi(p',p,x)$:
\begin{align}
&|V(\vec p\,',\vec p\,)|\le M_k \phi(p',p,x)\,,\nonumber\\
\end{align}
then, for the partial-wave potential, we obtain:
\begin{align}
&|V_{l', l}^{s j}(p', p)|\le 2\pi \tilde M_k \int_{-1}^1 dx\, \phi(p',p,x)\,.
\label{Eq:bound_PW_x}
\end{align}

For the special case of the locally regulated spin-orbit contact interaction, 
a bound of the same type can be obtained if one replaces $|V(\vec p\,',\vec p\,)|$
by $|\tilde V(\vec p\,',\vec p\,)|=|V(\vec p\,',\vec p\,)|\sqrt{1-x^2}$,
see Appendix.~\ref{Sec:bounds_LO_plane_wave_potential} and Ref.~\cite{Gasparyan:2021edy}.

\subsection{\texorpdfstring{Bounds on the form factor $F_{\mu,n}(q)$ integrated over $x$}
{Bounds on the form factor Fmu,n(q) integrated over x}}
In this subsection we derive the bounds on the local form factors
\begin{align}
 F_{\mu}(q)=\frac{\mu^2}{q^2+\mu^2}\,,\qquad F_{\mu,2}(q)=F_{\mu}(q)^2,
\end{align}
integrated over the angle variable $x$, which are relevant when 
considering bounds for the partial-wave potentials.
The form factors $F_{\mu,n}(q)$ with $n>2$ satisfy (at least) the same bounds 
as $F_{\mu,2}(q)$, which is sufficient for our estimates.
The same is true for the form factors of the Gaussian form, 
which was analyzed in detail in Ref.~\cite{Gasparyan:2021edy}.

From Eq.~\eqref{Eq:estimates_denominator2}, it follows
\begin{align}
\big|q^2+\mu^2\big|&\ge \mathcal{M}_f^{-1}\left(|p|^2+|p'|^2-2|p||p'|x+\mu^2\right)
=\mathcal{M}_f^{-1}\left[\left(|p'|x-|p|\right)^2+|p'|^2(1-x^2)+\mu^2\right]\nonumber\\
&\ge \mathcal{M}_f^{-1}\left[|p'|^2(1-x)/2+\mu^2 \right].
\label{Eq:denominator_factorized_3}
\end{align}

For $|p'|\ge\mu$, we obtain
\begin{align}
\left|\int_{-1}^1F_{\mu}(q)dx\right|&=
\left|\int_{-1}^1\frac{\mu^2 dx}{q^2+\mu^2}\right|\le
 2\mathcal{M}_f \mu^2
\int_{-1}^1\frac{dx}{|p'|^2(1-x)+2\mu^2 }
=\frac{2\mathcal{M}_f \mu^2}{|p'|^2}\ln\left(1+|p'|^2/\mu^2\right)\nonumber\\
&\le \frac{2\mathcal{M}_f \mu^2}{|p'|^2}\ln\frac{2|p'|^2}{\mu^2}
<\frac{2\mathcal{M}_f \mu^2}{|p'|^2}\left( 1+ \ln\frac{|p'|^2}{\mu^2}\right)\,,
\label{Eq:denominator_integrated_2}
\end{align}
and

\begin{align}
&\left|\int_{-1}^1F_{\mu,2}(q)dx\right|=
\left|\int_{-1}^1\frac{\mu^4 dx}{\left(q^2+\mu^2\right)^2}\right|\le
 4\mathcal{M}_f\mu^4
\int_{-1}^1\frac{dx}{\left[|p'|^2(1-x)+2\mu^2\right]^2 }
=\frac{2\mathcal{M}_f\mu^2}{|p'|^2+\mu^2}<\frac{2\mathcal{M}_f\mu^2}{|p'|^2}\,,
\label{Eq:denominator_integrated_3}
\end{align}
whereas for $|p'|<\mu$, we can simply use
\begin{align}
&\left|\int_{-1}^1 dx F_{\mu,n}(q)\right|\le
 \int_{-1}^1 dx (\mathcal{M}_f)^n
 =2(\mathcal{M}_f)^n\,, \qquad n=1,2\,.
\label{Eq:denominator_integrated_1}
\end{align}

Combining Eq.~\eqref{Eq:denominator_integrated_1} with Eq.~\eqref{Eq:denominator_integrated_2}
or Eq.~\eqref{Eq:denominator_integrated_3} and introducing the functions
\begin{align}
& \lambda(\xi)=\theta(1-|\xi|)+\theta(|\xi|-1)\frac{1}{|\xi|^2}\,,\nonumber\\
& \lambda_\text{log}(\xi)=\theta(1-|\xi|)+\theta(|\xi|-1)\frac{1+\ln|\xi|}{|\xi|^2}\,,
\label{Eq:lambda_lambdalog}
\end{align}
we arrive at the following bounds (obviously symmetric under the interchange $p\leftrightarrow p'$):
\begin{align}
&\left|\int_{-1}^1F_{\mu}(q)dx\right|\le \mathcal{M}_{F,1}\lambda_\text{log}(p'/\mu)\,,\
\text{and the same for }p\leftrightarrow p'\,,
\label{Eq:denominator_integrated_2a}
\end{align}
and
\begin{align}
&\left|\int_{-1}^1F_{\mu,2}(q)dx\right|\le \mathcal{M}_{F,2}\lambda(p'/\mu)\,,\
\text{and the same for }p\leftrightarrow p'\,.
\label{Eq:denominator_integrated_3a}
\end{align}

For the function $F_{\mu,2}(q)$, we can also obtain another bound:
\begin{align}
&\left|\int_{-1}^1F_{\mu,2}(q)dx\right|\le \mathcal{M}_{F,2}\lambda(p'/\mu)^2/\lambda(p/\mu)\,,\
\text{and the same for }p\leftrightarrow p'\,.
\label{Eq:denominator_integrated_4a}
\end{align}

To prove Eq.~\eqref{Eq:denominator_integrated_4a},
we consider three cases.
\begin{enumerate}
 \item $|p'|\le\mu$. In this case, $\lambda(p'/\mu)=1$. Since $\lambda(p/\mu)\le 1$,
 Eq.~\eqref{Eq:denominator_integrated_4a} follows from Eq.~\eqref{Eq:denominator_integrated_3a}.

 \item $|p|\ge|p'|>\mu$. In this case, $\lambda(p'/\mu)\ge\lambda(p/\mu)$
 and Eq.~\eqref{Eq:denominator_integrated_3a} yields Eq.~\eqref{Eq:denominator_integrated_4a}.

 \item $|p|<|p'|$ and $|p'|>\mu$.
 Consider the definition of the subtraction remainder $\Delta_p^{(1)}$ in Eq.~\eqref{Eq:Delta_p_f}:
\begin{align}
 F_{\mu,2}(q)=F_{\mu,2}(p')+
 p\frac{\partial F_{\mu,2}(q)}{\partial p}\bigg|_{p=0}
 +\Delta_p^{(1)} F_{\mu,2}(q).
\end{align}
Now, we estimate the three terms in the last equation individually.
\begin{align}
 \left| \int_{-1}^1F_{\mu,2}(p')dx\right|
 \le \int_{-1}^1\left| F_{\mu,2}(p')\right|dx\le \frac{2\mu^4}{|p'|^4}=2\lambda(p'/\mu)^2\le 
 2 \lambda(p'/\mu)^2/\lambda(p)\,.
\end{align}
From the fact that $\frac{\partial F_{\mu,2}(q)}{\partial p}\bigg|_{p=0}\propto x$, it follows
\begin{align}
 \int_{-1}^1 p\frac{\partial F_{\mu,2}(q)}{\partial p}\bigg|_{p=0}dx=0.
\end{align}
The bound from Eq.~\eqref{Eq:Delta_p_f} gives
\begin{align}
&\left|\int_{-1}^1\Delta_p^{(1)} F_{\mu,2}(q)\right|dx
\le \mathcal{M}_{f,1} \frac{|p|^2}{|p'|^2}\int_{-1}^1\left| F_{\mu,2}(q)\right|dx\,,
\end{align}
which (see Eq.~\eqref{Eq:denominator_integrated_3}) leads to
\begin{align}
&\left|\int_{-1}^1\Delta_p^{(1)} F_{\mu,2}(q)\right|dx
\le 2\mathcal{M}_{f,1}\frac{ |p|^2\mu^2}{|p'|^4}
=2\mathcal{M}_{f,1}\frac{ |p|^2}{\mu^2}\lambda(p'/\mu)^2
\le 2\mathcal{M}_{f,1}\lambda(p'/\mu)^2/\lambda(p/\mu)\,.
\end{align}
Finally,
\begin{align}
\left| \int_{-1}^1F_{\mu,2}(q)dx\right|\le
\left| \int_{-1}^1F_{\mu,2}(p')dx\right|+\left|\int_{-1}^1\Delta_p^{(1)} F_{\mu,2}(q)\right|dx
\le 2(\mathcal{M}_{f,1}+1)\lambda(p'/\mu)^2/\lambda(p/\mu)\,.
\end{align}
Combining all three cases, we obtain Eq.~\eqref{Eq:denominator_integrated_4a}.

\end{enumerate}

\subsection{Bounds on the partial-wave leading-order potential}
We represent the bounds for the partial-wave LO potential in the 
separable form:
\begin{align}
&\left|V_0(p',p)\right|\le \mathcal{M}_{V_0}
V_{0,\text{max}}\, g(p') h(p)\,,\nonumber\\
&\left|V_0(p',p)\right|\le \mathcal{M}_{V_0}
V_{0,\text{max}}\, h(p') g(p)\,,
\label{Eq:bounds_V0_l_0}
\end{align}
with
\begin{align}
 V_{0,\text{max}}=\frac{8\pi^2 }{m_N \Lambda_V}\,,
 \label{Eq:V0max_text}
\end{align}
where the exact form of functions $g$ and $h$ (and the value of $\mathcal{M}_{V_0}$) 
depends on the partial wave and on the
form of a regulator.

Introducing the functions 
\begin{align}
&v_0(p',p)=V_0(p',p)\left[\mathcal{M}_{V_0}
V_{0,\text{max}}\, h(p') g(p)  \right]^{-1} \,,\nonumber\\
&\bar v_0(p',p)=V_0(p',p)\left[\mathcal{M}_{V_0}
V_{0,\text{max}}\, g(p') h(p)  \right]^{-1} \,,
\label{Eq:v0}
\end{align}
we obtain the bounds
\begin{align}
&\left|v_0(p',p)\right|\le 1\,,\qquad
\left|\bar v_0(p',p)\right|\le 1\,.
\label{Eq:v0_bound}
\end{align}

The above inequalities are meant to hold for all matrix elements
of  $V_0(p',p)$ in the $l\,,l'$ space.

\subsubsection{\texorpdfstring{$S$-wave}{S-wave}}                                                                                                            
\label{Sec:LO_bounds_Swaves}
Using the bounds for the plane-wave leading-order potential
in Eq.~\eqref{Eq:bound_full_LO}
and performing the partial-wave projection according to 
Eqs.~\eqref{Eq:bound_PW_x},~\eqref{Eq:denominator_integrated_2a},~\eqref{Eq:denominator_integrated_4a}, 
we obtain
for $l=0$ (for the coupled partial waves, we mean by $l$ the lowest orbital angular momentum):                                                                                                                                                      
\begin{align}
 g(p)=\lambda_\text{log}(p/\Lambda), \qquad h(p)=1,
 \label{Eq:g_h_local_n1}
\end{align}
for the ``mild'' regulator, and
\begin{align}
 g(p)=\left[\lambda(p/\Lambda)  \right]^2, \qquad h(p)=\left[\lambda(p/\Lambda)  \right]^{-1}, 
  \label{Eq:g_h_local_n2}
\end{align}
for the ``standard'' regulators.

Note that for $|p|\le\Lambda$, in particular, for the on-shell momentum $|p|=p_\text{on}$,
we have $g(p)=h(p)=1$.

\subsubsection{Higher partial waves}
\label{Sec:bounds_LO_higher_PW}
For $l>0$,
we can use the fact that for $m<l$,
\begin{align}
&\frac{\partial^m V_0(p\,',p)}{(\partial p)^m}\bigg|_{p=0}=
\frac{\partial^m V_0(p\,',p)}{(\partial p')^m}\bigg|_{p'=0}=0,
\end{align}
and thus
\begin{align}
&\Delta_p^{(m)}V_0(p\,',p)=\Delta_{p'}^{(m)}V_0(p\,',p)=V_0(p\,',p).
\end{align}

For the case of the ``mild'' regulator 
utilizing Eq.~\eqref{Eq:Delta_p_V0}
and performing the partial-wave projection according to Eqs.~\eqref{Eq:bound_PW_x}~and~\eqref{Eq:denominator_integrated_2a},
we derive 
\begin{align}
 g(p)=\lambda_\text{log}(p/\Lambda)/|p|^{\tilde l}, \qquad h(p)=|p|^{\tilde l},
 \label{Eq:g_h_higherPWs}
\end{align}
with $\tilde l\le l$.
Since
\begin{align}
\lambda(p/\Lambda)\le\lambda_\text{log}(p/\Lambda),
\end{align}
the same bounds can be used for the ``standard'' regulators, see Eq.~\eqref{Eq:denominator_integrated_3a}.

For the purposes of the present paper, it is sufficient to choose $\tilde l=1$.

\subsection{Bounds on the partial-wave next-to-leading-order potential}
\subsubsection{\texorpdfstring{$S$-wave}{S-wave}}
\label{Sec:bounds_PW_NLO_Swave}
For $l=0$, the bounds on the NLO partial-wave potential are the same as in Ref.~\cite{Gasparyan:2021edy}:
\begin{align}
 \left|\hat V_2(p', p)\right|\le \mathcal{\hat M}_{V_2,0} \frac{8\pi^2 }{m_N \Lambda_V}\frac{M_\pi^2}{\Lambda_{b}^2}\,,
 \label{Eq:bound_V_2_0_hat}
\end{align}
and
\begin{align}
&\left|\tilde V_2(p',p)\right|\le \mathcal{M}_{V_2,0}
\left(|p|^2+|p'|^2\right)\tilde f_\text{log}(p',p),
\label{Eq:bounds_V2_l_0}
\end{align}
when one employs the ``standard'' regulators for the LO potentials.

In the case of the ``mild'' regulator of the LO potential,
we use the partial-wave projected regularized expression, 
applying Eq.~\eqref{Eq:denominator_integrated_2a} to Eq.~\eqref{Eq:bounds_V2_regulated_full}:
\begin{align}
&\left|\tilde V_2(p',p)\right|\le \mathcal{M}_{V_2,0}
\left(|p|^2+|p'|^2\right)\tilde f_\text{log}(p',p)  \lambda_\text{log}(p'/\Lambda_\text{NLO}), \text{ or}\nonumber\\
&\left|\tilde V_2(p',p)\right|\le \mathcal{M}_{V_2,0}
\left(|p|^2+|p'|^2\right)\tilde f_\text{log}(p',p)  \lambda_\text{log}(p/\Lambda_\text{NLO}).
\label{Eq:bounds_V2_l_0_R_0}
\end{align}

\subsubsection{Higher partial waves}
For $l\ge1$, we simply adopt the bounds from Ref.~\cite{Gasparyan:2021edy}
\begin{align}
&\left|\tilde V_2(p',p)\right|\le \mathcal{M}_{V_2,\tilde l}
\left|\frac{p}{p'}\right|^{\tilde l}|p'|^2\tilde f_\text{log}(p',p),
\label{Eq:bounds_V2_l_1a}\\
&\left|\tilde V_2(p',p)\right|\le \mathcal{M}_{V_2,\tilde l}
\left|\frac{p'}{p}\right|^{\tilde l}|p|^2\tilde f_\text{log}(p',p),
\label{Eq:bounds_V2_l_1b}
\end{align}
where $0\le\tilde l \le l$.

For $\tilde l=1$, both above equations coincide:
\begin{align}
&\left|\tilde V_2(p',p)\right|\le \mathcal{M}_{V_2,1}
|p'||p|\tilde f_\text{log}(p',p).
\label{Eq:bounds_V2_l_2}
\end{align}

For the purposes of the present paper, it is sufficient to take the choice $\tilde l=1$.

\section{\texorpdfstring{Bounds on various parts of the $S$-wave NLO amplitude}
{Bounds on various parts of the S-wave NLO amplitude}}
\label{Sec:Bounds_N2tilde_nu}
In this appendix we provide bounds for various parts of the 
unrenormalized and renormalized $S$-wave NLO amplitude and their series remainders.
The unrenormalized NLO amplitude is decomposed by factoring out 
the Fredholm determinant as in 
Eq.~\eqref{Eq:decomposition_N2_pwaves}:
\begin{align}
   T_2(p',p;p_\text{on})&= N_2(p',p;p_\text{on})/D(p_\text{on})^2,\nonumber\\
   N_2 &=  V_2 D^2+ T_{2,Y} D + T_{2,\bar{Y}} D + T_{2,\bar{Y}Y},
  \label{Eq:N2_T_2Y}
\end{align}
with
\begin{align}
 & T_{2,Y}(p',p;p_\text{on})=
\int\frac{p_1^2 d p_1}{(2\pi)^3} V_2(p',p_1)Y(p_1,p;p_\text{on})\,,\nonumber\\
 & T_{2,\bar{Y}}(p',p;p_\text{on})=\int\frac{p_1'^2 d p'_1}{(2\pi)^3}
 \bar{Y}(p',p_1';p_\text{on}) V_2(p_1',p)\,,\nonumber\\ 
 & T_{2,\bar{Y}Y}(p',p;p_\text{on})=\int\frac{p_1^2 d p_1}{(2\pi)^3}\frac{p_1'^2 d p'_1}{(2\pi)^3}
 \bar{Y}(p',p_1';p_\text{on}) V_2(p_1',p_1)Y(p_1,p;p_\text{on}) .
\end{align}
Below, we derive the bounds for the quantities $T_{2,Y}$, $T_{2,\bar{Y}}$ and $T_{2,\bar{Y}Y}$
for the cases of the ``standard'' and the ``mild'' regulators of the LO potential.

\subsection{``Standard'' regulator}
For the ``standard'' regulators of the LO potential, in particular, for the local regulators
of the spin-triplet part of the one-pion-exchange potential of power $n\ge 2$,
the binding functions $g$ and $h$ have the form (see Eq.~\eqref{Eq:g_h_local_n2})
\begin{align}
 g(p_1)=\lambda(p_1/\Lambda)^2\,, \qquad h(p)=1\,, \text{ if } \ p<\Lambda\,.
\end{align}
From the bounds on $V_2$ (Eq.~\eqref{Eq:bounds_V2_l_0}) and $V_0$ (Eq.~\eqref{Eq:bounds_V0_l_0}), we obtain
\begin{align}
\left| T_{2,Y}(p',p;p_\text{on})\right|&\le
  \mathcal{M}_{V_2,0}n_\text{PW} \frac{8\pi^2 \mathcal{M}_{Y_\text{max}}}{\Lambda_V}
 \int\frac{(|p_1|^2+|p'|^2)d |p_1|}{(2\pi)^3}
\tilde f_\text{log}(p',p_1)\lambda(p_1/\Lambda)^2 \nonumber\\
&= \mathcal{M}_{V_2,0}n_\text{PW} \frac{8\pi^2 \mathcal{M}_{Y_\text{max}}}{m_N\Lambda_V^2\Lambda_b^2}
 \int\frac{d |p_1|}{\pi}(|p_1|^2+|p'|^2)
f_\text{log}(p',p_1)\lambda(p_1/\Lambda)^2 \nonumber\\
&= \mathcal{M}_{V_2,0}n_\text{PW} \frac{8\pi^2 \mathcal{M}_{Y_\text{max}}}{m_N\Lambda_V^2\Lambda_b^2}
\bigg\{\left[|p'|^2 I_{\lambda,1a} +I_{\lambda,1b}
\right]\left[1+\theta(|p'|-M_\pi)\ln\frac{|p'|}{M_\pi}  \right] \nonumber\\
&+|p'|^2 I_{\lambda,2a} +I_{\lambda,2b}\bigg\},
\label{Eq:T2Y_0}
\end{align}
where the typical integrals $I_i$ are defined and estimated in Appendix~\ref{Sec:integrals}.
Setting all external momenta on shell, $p=p'=p_\text{on}$, and
using $p_\text{on}\ll \Lambda$, gives
\begin{align}
 \left| T_{2,Y}(p_\text{on})\right|\le
 \frac{8\pi^2 \mathcal{M}_{T_2,Y}\mathcal{M}_{Y_\text{max}}}{m_N\Lambda_V^2\Lambda_b^2}\Lambda^3\ln\frac{\Lambda}{M_\pi},
\end{align}
or, assuming $\Lambda\sim\Lambda_V$,
\begin{align}
 \left| T_{2,Y}(p_\text{on})\right|\le
 \frac{8\pi^2 \mathcal{\tilde M}_{T_2,Y}\mathcal{M}_{Y_\text{max}}}{m_N\Lambda_V}
 \frac{\Lambda^2}{\Lambda_b^2}\ln\frac{\Lambda}{M_\pi}.
 \label{Eq:T2Y_standard}
\end{align}
Symmetrically, the same bound holds for $ T_{2,\bar Y}(p',p;p_\text{on})$.

Next, we consider the contribution $T_{2,\bar{Y}Y}$:
\begin{align}
\left|T_{2,\bar{Y}Y}(p',p;p_\text{on})\right|&\le
  \mathcal{M}_{V_2,0} \left(\frac{8\pi^2 n_\text{PW}\mathcal{M}_{Y_\text{max}}}{\Lambda_V}\right)^2\nonumber\\
  &\times
 \int\frac{d |p_1|}{(2\pi)^3}\frac{d |p_1'|}{(2\pi)^3}(|p_1|^2+|p_1'^2|)
\tilde f_\text{log}(p_1',p_1)\lambda(p_1/\Lambda)^2 \lambda(p_1'/\Lambda)^2 \nonumber\\
&= \mathcal{M}_{V_2,0}n_\text{PW}^2 \frac{8\pi^2 \mathcal{M}_{Y_\text{max}}^2}{m_N\Lambda_V^3\Lambda_b^2}
 \int\frac{d |p_1|d |p_1'|}{\pi^2}(|p_1|^2+|p_1'|^2)
f_\text{log}(p',p_1)\lambda(p_1/\Lambda)^2 \lambda(p_1'/\Lambda)^2\nonumber\\
&= \mathcal{M}_{V_2,0}n_\text{PW}^2 \frac{8\pi^2 \mathcal{M}_{Y_\text{max}}^2}{m_N\Lambda_V^3\Lambda_b^2}
2\left(I_{\lambda,1a} I_{\lambda,1b}+
I_{\lambda,2a} I_{\lambda,1b}+
I_{\lambda,2b} I_{\lambda,1a}
\right).
\label{Eq:T2YYbar_0}
\end{align}

Setting all external momenta on shell, $p=p'=p_\text{on}$, and
using $p_\text{on}\ll \Lambda$, we obtain
\begin{align}
 \left| T_{2,\bar{Y}Y}(p_\text{on})\right|\le
 \frac{8\pi^2 \mathcal{M}_{T_2,\bar{Y}Y}\mathcal{M}_{Y_\text{max}}^2}{m_N\Lambda_V^3\Lambda_b^2}\Lambda^4\ln\frac{\Lambda}{M_\pi},
\end{align}
or, assuming $\Lambda\sim\Lambda_V$:
\begin{align}
 \left| T_{2,\bar{Y}Y}(p_\text{on})\right|\le
 \frac{8\pi^2 \mathcal{\tilde M}_{T_2,\bar{Y}Y}\mathcal{M}_{Y_\text{max}}^2}{m_N\Lambda_V}
 \frac{\Lambda^2}{\Lambda_b^2}\ln\frac{\Lambda}{M_\pi}.
 \label{Eq:T2barYY_standard}
\end{align}

 \subsection{``Mild'' regulator}
 For the ``mild'' regulator of the LO potential, including the case when
  the spin-triplet one-pion-exchange contribution is regularized
 by the local dipole regulator,
the binding functions $g$ and $h$ have the form (see Eq.~\eqref{Eq:g_h_local_n1})
\begin{align}
 g(p_1)=\lambda_\text{log}(p_1/\Lambda)\,, \qquad h(p)=1\,, \text{ if } \ p<\Lambda\,.
\end{align}
By analogy with Eq.~\eqref{Eq:T2Y_0}
from the bounds on the regularized
$V_2$ (Eq.~\eqref{Eq:bounds_V2_l_0_R_0}) and $V_0$ (Eq.~\eqref{Eq:bounds_V0_l_0}), we obtain
\begin{align}
 \left| T_{2,Y}(p',p;p_\text{on})\right|&\le
   \mathcal{M}_{V_2,0}n_\text{PW} \frac{8\pi^2 \mathcal{M}_{Y_\text{max}}}{m_N\Lambda_V^2\Lambda_b^2}
   \int\frac{d |p_1|}{\pi}(|p_1|^2+|p'|^2)
   \nonumber\\&\times
f_\text{log}(p',p_1)\lambda_\text{log}(p_1/\Lambda) 
\lambda_\text{log}(p_1/\Lambda_\text{NLO})\nonumber\\
&= \mathcal{M}_{V_2,0}n_\text{PW} \frac{8\pi^2 \mathcal{M}_{Y_\text{max}}}{m_N\Lambda_V^2\Lambda_b^2}
\bigg\{\left[|p'|^2 I_{\lambda_\text{log},1a} +I_{\lambda_\text{log},1b}
\right]\left[1+\theta(|p'|-M_\pi)\ln\frac{|p'|}{M_\pi}  \right] \nonumber\\
&+|p'|^2 I_{\lambda_\text{log},2a} +I_{\lambda_\text{log},2b}\bigg\},
\end{align}
where the typical integrals $I_i$ are defined and estimated in Appendix~\ref{Sec:integrals}.
Setting all external momenta on shell, $p=p'=p_\text{on}$, and
using $p_\text{on}\ll \Lambda\ll \Lambda_\text{NLO}$, yields
\begin{align}
 \left| T_{2,Y}(p_\text{on})\right|\le
 \frac{8\pi^2 \mathcal{M}_{T_2,Y}\mathcal{M}_{Y_\text{max}}}{m_N\Lambda_V^2\Lambda_b^2}\Lambda^2
 \Lambda_\text{NLO}\ln\frac{\Lambda_\text{NLO}}{\Lambda}\ln\frac{\Lambda_\text{NLO}}{M_\pi},
\end{align}
or, assuming $\Lambda\sim\Lambda_V$:
\begin{align}
 \left| T_{2,Y}(p_\text{on})\right|\le
 \frac{8\pi^2 \mathcal{\tilde M}_{T_2,Y}\mathcal{M}_{Y_\text{max}}}{m_N\Lambda_V}
 \frac{\Lambda\Lambda_\text{NLO}}{\Lambda_b^2}
 \ln\frac{\Lambda_\text{NLO}}{\Lambda}\ln\frac{\Lambda_\text{NLO}}{M_\pi}.
 \label{Eq:T2Y_mild}
\end{align}

Symmetrically, the same bound holds for $ T_{2,\bar Y}(p',p;p_\text{on})$.
 
Analogously to Eq.~\eqref{Eq:T2YYbar_0}, 
the following bound holds for $T_{2,\bar{Y}Y}$:
\begin{align}
 \left|T_{2,\bar{Y}Y}(p',p;p_\text{on})\right|&\le
   \mathcal{M}_{V_2,0}n_\text{PW}^2 \frac{8\pi^2 \mathcal{M}_{Y_\text{max}}^2}{m_N\Lambda_V^3\Lambda_b^2}
   \nonumber\\&\times
 \int\frac{d |p_1|d |p_1'|}{\pi^2}
f_\text{log}(p_1',p_1)\lambda_\text{log}(p_1/\Lambda) \lambda_\text{log}(p_1'/\Lambda)
\nonumber\\&\times  
\left[|p_1|^2\lambda_\text{log}(p_1/\Lambda_\text{NLO})+|p_1'|^2\lambda_\text{log}(p_1'/\Lambda_\text{NLO})\right]\nonumber\\
&= \mathcal{M}_{V_2,0}n_\text{PW}^2 \frac{8\pi^2 \mathcal{M}_{Y_\text{max}}^2}{m_N\Lambda_V^3\Lambda_b^2}
\nonumber\\&\times
2\left(I_{\lambda_\text{log},1} I_{\lambda_\text{log},1b}+
I_{\lambda_\text{log},2} I_{\lambda_\text{log},1b}+
I_{\lambda_\text{log},2b} I_{\lambda_\text{log},1}
\right).
\end{align}
Setting all external momenta on shell, $p=p'=p_\text{on}$, and
using $p_\text{on}\ll \Lambda\ll \Lambda_\text{NLO}$, we obtain
\begin{align}
 \left| T_{2,\bar{Y}Y}(p_\text{on})\right|\le
 \frac{8\pi^2 \mathcal{M}_{T_2,\bar{Y}Y}\mathcal{M}_{Y_\text{max}}^2}{m_N\Lambda_V^3\Lambda_b^2}
 \Lambda^3\Lambda_\text{NLO}
 \ln\frac{\Lambda_\text{NLO}}{\Lambda}\ln\frac{\Lambda_\text{NLO}}{M_\pi},
\end{align}
or, assuming $\Lambda\sim\Lambda_V$:
\begin{align}
 \left| T_{2,\bar{Y}Y}(p_\text{on})\right|\le
 \frac{8\pi^2 \mathcal{\tilde M}_{T_2,\bar{Y}Y}\mathcal{M}_{Y_\text{max}}^2}{m_N\Lambda_V}
 \frac{\Lambda\Lambda_\text{NLO}}{\Lambda_b^2}
  \ln\frac{\Lambda_\text{NLO}}{\Lambda}\ln\frac{\Lambda_\text{NLO}}{M_\pi}.
  \label{Eq:T2barYY_mild}
\end{align}

\subsection{\texorpdfstring{Bounds on the function $\nu(p_\text{on})$}{Bounds on the function nu(pon)}}
In this subsection we provide bounds on the function $\nu_l(p_\text{on})$,
defined in Eq.~\eqref{Eq:definitions_N2_n}.
We introduce another function $\nu_{Y,l}$ as follows:
\begin{align}
 \nu_l(p_\text{on})=D(p_\text{on})\left[\delta_{l,0}+\nu_{Y,l}(p_\text{on}) \right],
 \label{Eq:nY_definition}
\end{align}
which equals (see Eq.~\eqref{Eq:psi_explicit})
\begin{align}
 \nu_{Y,l}(p_\text{on})=\int\frac{p_1^2 d p_1}{(2\pi)^3}Y_{0,l}(p_1,p_\text{on};p_\text{on}).
\end{align}
Using Eq.~\eqref{Eq:bound_Y_Y_max}, we derive the following bound
for the function $n_{Y,l}$
in the case of the ``standard'' regulator of the LO potential (see Appendix~\ref{Sec:LO_bounds_Swaves}):                                                                                                                                                                                 
\begin{align}
 \left|\nu_{Y,l}(p_\text{on})\right|
 &\le \frac{\mathcal{M}_{Y_\text{max}}}{\Lambda_V}
   \int\frac{d |p_1|}{\pi}
  g(p_1/\Lambda)h(p_\text{on})\nonumber\\
&  = \frac{\mathcal{M}_{Y_\text{max}}}{\Lambda_V}
   \int\frac{d |p_1|}{\pi}
  \lambda(p_1/\Lambda)^2
  = \frac{\mathcal{M}_{Y_\text{max}}}{\Lambda_V}
   I_{\lambda,1a}
=\mathcal{M}_{Y_\text{max}}\mathcal{M}_\lambda\frac{\Lambda}{\Lambda_V},
\label{Eq:nY_1}
\end{align}
where we have utilized the bounds for typical integrals provided in Appendix~\ref{Sec:integrals}.

Assuming $\Lambda\sim\Lambda_V$ yields
\begin{align}
 \left|\nu_{Y,l}(p_\text{on})\right|
 &\le
 \mathcal{M}_{Y_\text{max}}\mathcal{\tilde M}_\lambda.
 \label{Eq:nY_2}
\end{align}
For the ``mild'' regulator of the LO potential,
one should replace $\lambda(p_1/\Lambda)^2$
with $\lambda_\text{log}(p_1/\Lambda)$ and $I_{\lambda,1a}$
with $I_{\lambda_\text{log},1}$ in Eq.~\eqref{Eq:nY_1}.
Since our bounds for $I_{\lambda_\text{log},1}$ and $I_{\lambda,1a}$
are the same, see Eqs.~\eqref{Eq:I_lambda_log12} and~\eqref{Eq:I_lambda_12},
equation~\eqref{Eq:nY_2} holds also for the ``mild'' regulator.

Since the Fredholm determinant $D$ is bounded by a constant of order one
(Eq.~\eqref{Eq:MD}), the same is true for the function $\nu_l(p_\text{on})$:
 \begin{align}
  \nu_l(p_\text{on})\le\mathcal{M}_{\nu},
  \label{Eq:bound_n_l}
 \end{align}
as follows from Eqs.~\eqref{Eq:nY_2} and~\eqref{Eq:nY_definition}.
 
\subsection{Series remainders}
From the bounds on the matrix elements of the operator $Y$ ($\bar Y$) 
and its series remainders (Eqs.~\eqref{Eq:bound_Y_Y_max} and~\eqref{Eq:deltaY_max})
as well as the bounds on the Fredholm determinant $D$ and its series remainders
(Eqs.~\eqref{Eq:MD} and ~\eqref{Eq:remainder_D}),
it is straightforward to deduce also the bounds for the series remainders
of the quantities $T_{2,Y}$, $T_{2,\bar{Y}}$, $T_{2,\bar{Y}Y}$ and $\nu_Y$
by just replacing $\mathcal{M}_{Y_\text{max}}$ with $\mathcal{N}_{\delta_n Y}=\mathcal{M}_Y\delta_n Y_\text{max}$
and $\mathcal{M}_{Y_\text{max}}^2$ with $2\mathcal{M}_{Y_\text{max}} \mathcal{N}_{\delta_n Y}+\mathcal{N}_{\delta_n Y}^2$.
Being proportional to $\delta_n Y_\text{max}$ or 
$(\delta_n Y_\text{max})^2$,
$T_{2,Y}$, $T_{2,\bar{Y}}$, $T_{2,\bar{Y}Y}$ and $\nu_Y$ decrease faster than
exponential with any base, see Eq.~\eqref{Eq:deltaY_exponential}.
The series remainder of the Fredholm determinant
possesses the same property, see  Eq.~\eqref{Eq:deltaD_exponential}.
Therefore, from Eq.~\eqref{Eq:N2_T_2Y} we conclude that
$N_2$ also decreases faster than exponential
as well as the renormalized quantity
$\mathds{R}(\tilde N_{2})$ (Eq.~\eqref{Eq:R_N2_tilde}),
because those are polynomials in 
$T_{2,Y}$, $T_{2,\bar{Y}}$, $T_{2,\bar{Y}Y}$, $\nu_Y$ and $D$.

To be specific, the following bound holds:
 \begin{align}
|\delta_n [\mathds{R}(\tilde N_2)]| &=
\Big|\sum_{k_1,k_2=0}^\infty \mathds{R}(\tilde N_2)^{[k_1,k_2]}
-\sum_{k_1,k_2=0}^n \mathds{R}(\tilde N_2)^{[k_1,k_2]}\Big|\nonumber\\
&\le  \frac{8\pi^2}{m_N\Lambda_V}\mathcal{N}_{\tilde N_2} e^{-\mathcal{M}_{\delta \tilde N_2}n},\qquad\text{ for }
 n>\mathcal{\tilde M}_{\delta \tilde N_2},
 \label{Eq:delta_tildeN2_exponential}
\end{align}
where  $\mathcal{\tilde M}_{\delta \tilde N_2}$
  is of order $\mathcal{\tilde M}_{\delta \tilde N_2}\gtrsim (e\Sigma)^2$
 in the general case but is typically much smaller in realistic calculations.
The prefactors $\mathcal{N}_{\tilde N_2}$ follow from 
Eqs.~\eqref{Eq:T2Y_standard},~\eqref{Eq:T2barYY_standard},~\eqref{Eq:T2Y_mild},~\eqref{Eq:T2barYY_mild},~\eqref{Eq:bound_n_l}
and~\eqref{Eq:MD}:
\begin{align}
 \mathcal{N}_{\tilde N_2}=\frac{\Lambda^2}{\Lambda_b^2}
\ln\frac{\Lambda}{M_\pi}
\end{align}
in the case of the ``standard'' regulators of the LO potential and 
\begin{align}
 \mathcal{N}_{\tilde N_2}=\frac{\Lambda\Lambda_\text{NLO}}{\Lambda_b^2}
 \ln\frac{\Lambda_\text{NLO}}{\Lambda}\ln\frac{\Lambda_\text{NLO}}{M_\pi}
\end{align}
 in the case of the ``mild'' regulator.
 
 \section{Bounds on typical integrals}
\label{Sec:integrals}
In this appendix we provide the bounds for typical integrals that appear
in the course of evaluation of various amplitudes.

The integrals
\begin{align}
 I_{\lambda_\text{log},1}&=\int\frac{d |{p}|}{\pi}\lambda_\text{log}({p}/\Lambda),\nonumber\\
 I_{\lambda_\text{log},1a}&=\int\frac{d |{p}|}{\pi}\lambda_\text{log}({p}/\Lambda_\text{NLO})\lambda_\text{log}({p}/\Lambda),\nonumber\\
 I_{\lambda_\text{log},2}&=\int\frac{d |{p}|}{\pi}\lambda_\text{log}({p}/\Lambda)
 \theta(|{p}|-M_\pi)\ln\frac{|{p}|}{M_\pi},\nonumber\\
 I_{\lambda_\text{log},2a}&=\int\frac{d |{p}|}{\pi}\lambda_\text{log}({p}/\Lambda_\text{NLO})\lambda_\text{log}({p}/\Lambda)
 \theta(|{p}|-M_\pi)\ln\frac{|{p}|}{M_\pi},
 \end{align}
with functions $\lambda$ and $\lambda_\text{log}$ defined in Eq.~\eqref{Eq:lambda_lambdalog}
can be bounded as follows:
\begin{align}
 I_{\lambda_\text{log},1}&=\Lambda\int\frac{d \xi}{\pi}\lambda_\text{log}(\xi)\eqqcolon\mathcal{M}_\lambda \Lambda,\nonumber\\
 I_{\lambda_\text{log},1a}&<I_{\lambda_\text{log},1}=\mathcal{M}_\lambda \Lambda,\nonumber\\
 I_{\lambda_\text{log},2}&=\frac{1}{\pi}\left(2+\Lambda+2\Lambda\ln\frac{\Lambda}{M_\pi}\right)
 \le \mathcal{M}_{\lambda,2} \Lambda \ln\frac{\Lambda}{M_\pi},\nonumber\\
 I_{\lambda_\text{log},2a}&<I_{\lambda_\text{log},2}\le\mathcal{M}_{\lambda,2} \Lambda \ln\frac{\Lambda}{M_\pi}.
 \label{Eq:I_lambda_log12}
\end{align}

Analogously, for the integrals 
\begin{align}
 I_{\lambda,1}&=\int\frac{d |{p}|}{\pi}\lambda({p}/\Lambda),\nonumber\\
 I_{\lambda,1a}&=\int\frac{d |{p}|}{\pi}\lambda({p}/\Lambda)^2,\nonumber\\
 I_{\lambda,1b}&=\int\frac{|{p}|^2d |{p}|}{\pi}\lambda({p}/\Lambda)^2,\nonumber\\
 I_{\lambda,2}&=\int\frac{d |{p}|}{\pi}\lambda({p}/\Lambda)
 \theta(|{p}|-M_\pi)\ln\frac{|{p}|}{M_\pi},\nonumber\\
 I_{\lambda,2a}&=\int\frac{d |{p}|}{\pi}\lambda({p}/\Lambda)^2
 \theta(|{p}|-M_\pi)\ln\frac{|{p}|}{M_\pi}\nonumber\\
 I_{\lambda,2b}&=\int\frac{|{p}|^2d |{p}|}{\pi}\lambda({p}/\Lambda)^2,
 \theta(|{p}|-M_\pi)\ln\frac{|{p}|}{M_\pi},
 \end{align}
we obtain the following bounds:
\begin{align}
 I_{\lambda,1}&=\Lambda\int\frac{d \xi}{\pi}\lambda(\xi)
 < \Lambda\int\frac{d \xi}{\pi}\lambda_\text{log}(\xi)=\mathcal{M}_\lambda \Lambda,\nonumber\\
 I_{\lambda,1a}&<I_{\lambda,1}\le\mathcal{M}_\lambda \Lambda,\nonumber\\
 I_{\lambda,1b}&=\Lambda^3\int\frac{\xi^2d \xi}{\pi}\lambda(\xi)^2
 <\Lambda^3\int\frac{d \xi}{\pi}\lambda(\xi)
 <\Lambda^3\int\frac{d \xi}{\pi}\lambda_\text{log}(\xi)=\mathcal{M}_\lambda \Lambda^3,\nonumber\\
 I_{\lambda,2}&<I_{\lambda_\text{log},2}\le \mathcal{M}_{\lambda,2} \Lambda \ln\frac{\Lambda}{M_\pi},\nonumber\\
 I_{\lambda,2a}&<I_{\lambda,2}\le \mathcal{M}_{\lambda,2} \Lambda \ln\frac{\Lambda}{M_\pi},\nonumber\\
 I_{\lambda,2b}&<\Lambda^2 I_{\lambda,2}\le \mathcal{M}_{\lambda,2} \Lambda^3 \ln\frac{\Lambda}{M_\pi}\,.
 \label{Eq:I_lambda_12}
\end{align}
 
Next, we estimate the integral
\begin{align}
 I_{\lambda_\text{log},1b}&=\int\frac{|{p}|^2 d |{p}|}{\pi}\lambda_\text{log}({p}/\Lambda_\text{NLO})\lambda_\text{log}({p}/\Lambda).\nonumber\\
 \end{align}
Direct estimation under the assumption $\Lambda_\text{NLO}\gg \Lambda$ gives
\begin{align}
 I_{\lambda_\text{log},1b}&=\frac{2}{\pi}\Lambda^2\Lambda_\text{NLO}\ln\frac{\Lambda_\text{NLO}}{\Lambda}
 +O(\Lambda_\text{NLO})\le \mathcal{M}_{\lambda,1a}\Lambda^2\Lambda_\text{NLO}\ln\frac{\Lambda_\text{NLO}}{\Lambda}.
\end{align}
Finally, we derive a bound for the integral
\begin{align}
 I_{\lambda_\text{log},2b}&=\int\frac{|{p}|^2 d |{p}|}{\pi}\lambda_\text{log}({p}/\Lambda_\text{NLO})\lambda_\text{log}({p}/\Lambda)
 \theta(|{p}|-M_\pi)\ln\frac{|{p}|}{M_\pi}.
\end{align}
Direct calculation yields
\begin{align}
 I_{\lambda_\text{log},2b}&=\frac{1}{\pi}\Lambda^2\Lambda_\text{NLO}\ln\frac{\Lambda_\text{NLO}}{\Lambda}
 \ln\frac{\Lambda_\text{NLO}}{M_\pi}+O(\Lambda_\text{NLO}\ln\Lambda_\text{NLO}/M_\pi)\nonumber\\
 &\le \mathcal{M}_{\lambda,1a}\Lambda^2\Lambda_\text{NLO}\ln\frac{\Lambda_\text{NLO}}{\Lambda}
 \ln\frac{\Lambda_\text{NLO}}{M_\pi}.
\end{align}

\bibliography{5.1}
\bibliographystyle{apsrev}

\end{document}